\documentclass[aps,twocolumn,amsmath,amssymb,preprintnumbers]{revtex4}
\usepackage{amsmath} \usepackage{amsfonts} \usepackage{amssymb}
\usepackage{bbm}
\usepackage{epsfig}
\usepackage{graphics}
\usepackage{graphicx}
\begin{document}

\title[ ]{Quantum mechanics from classical statistics}

\author{C. Wetterich}
\affiliation{Institut  f\"ur Theoretische Physik\\
Universit\"at Heidelberg\\
Philosophenweg 16, D-69120 Heidelberg}

\begin{abstract}
Quantum mechanics can emerge from classical statistics. A typical quantum system describes an isolated subsystem of a classical statistical ensemble with infinitely many classical states. The state of this subsystem can be characterized by only a few probabilistic observables. Their expectation values define a density matrix if they obey a ``purity constraint''. Then all the usual laws of quantum mechanics follow, including Heisenberg's uncertainty relation, entanglement and a violation of Bell's inequalities. No concepts beyond classical statistics are needed for quantum physics - the differences are only apparent and result from the particularities of those classical statistical systems which admit a quantum mechanical description. Born's rule for quantum mechanical probabilities follows from the probability concept for a classical statistical ensemble. 
In particular, we show how the non-commuting properties of quantum operators are associated to the use of conditional probabilities within the classical system, and how a unitary time evolution reflects the isolation of the subsystem. As an illustration, we discuss a classical statistical implementation of a quantum computer.
\end{abstract}

\maketitle

\section{Introduction}
\label{Introduction}
A realization of quantum mechanics as a classical statistical system may shed new light on the conceptual interpretation of experiments based on entanglement, as teleportation or quantum cryptography \cite{Ze}. One may even speculate that steps in quantum computing \cite{Zo} could be realized by exploiting classical statistics. Recently, classical statistical ensembles that are equivalent to four-state and two-state quantum mechanics have been constructed explicitly \cite{CW1,CW2}.  This constitutes a proof of equivalence of few-state quantum statistics and particular classical statistical ensembles with infinitely many degrees of freedom. In view of the particular manifolds of classical states for these examples one may wonder if quantum statistical systems are very special cases of classical statistics, or if they arise rather genuinely under certain conditions. In this paper we argue that quantum statistics can indeed emerge rather generically if one describes small ``isolated'' subsystems of classical ensembles with an infinity of states.

An atom in quantum mechanics is an isolated system with a few degrees of freedom. This contrasts with quantum field theory, where an atom is described as a particular excitation of the vacuum. The vacuum in quantum field theory is a complicated object, involving infinitely many degrees of freedom. In a fundamental theory of particle physics, which underlies the description of atoms, collective effects, such as spontaneous symmetry breaking, are crucial for its understanding. Our treatment of atoms in the context of classical statistics is similar to the conceptual setting of quantum field theory. A classical statistical system with infinitely many states describes the atom and its environment or the atom and the vacuum. The quantum statistical features become apparent if one concentrates on a subsystem that describes the isolated atom. 

Only a small part of the information contained in the probability distribution for the classical statistical ensemble is used for the description of the properties of the subsystem. Statistical subsystems are also relevant if the description of the atom does not use all the possible ``microscopic information'' which could, in principle, be available on length scales many orders of magnitude smaller than the size of the atom.  The subsystem corresponds then to a ``coarse grained approach'', for example ignoring the constituents of the atomic nucleus. Furthermore, the probability distribution at a given time can be interpreted as a subsystem of a distribution for probabilities of events at different times. The part of the information which is contained in the probability distribution for the classical statistical ensemble but not used for the subsystem will generically be called ``environment''. The detailed properties and physical meaning of the environment will not be important for the emergence of quantum structures. 

Our classical statistical description of quantum systems has four crucial ingredients. (1) {\em Probabilistic observables} have in a given quantum state only a probabilistic distribution of possible measurement values, rather than a fixed value as for classical observables in a given state of the classical ensemble. Probabilistic observables obtain from classical observables by ``integrating out'' the environment degrees of freedom. The fact that this map is not invertible avoids conflicts with the Kochen-Specker theorem \cite{KS,Str}.  The probabilistic nature of the observables can be understood as a result of ``coarse graining of the information'', starting from classical observables on a suitable level. Alternatively, this concept may be used as a basic definition of observables \cite{POB},\cite{BB}.

(2) {\em  Incomplete statistics} characterizes the subsystem. This means that the joint probability for finding a measurement value $a$ for the ``system observable'' $A$, and $b$ for a second system observable $B$, cannot be computed from the state of the subsystem alone for all pairs $(A,B)$. Typically, joint probabilities require information about the precise state of the environment.

(3) {\em Conditional probabilities} are used for a computation of the probabilities for the possible outcomes of two measurements of observables $A$ and $B$. In particular, if $B$ is measured after $A$, the outcome of the measurement of $B$ depends on the previous measurement of $A$. The classical correlations $\langle A\cdot B\rangle$ are not uniquely defined for the quantum system - they depend on detailed properties of the environment. We argue that the appropriate conditional correlations for measurements of properties of the isolated subsystem should only require information which characterizes the subsystem. With this requirement the conditional probabilities induce the concept of {\em quantum correlations}, and we propose that quantum correlations rather than classical correlations should be used for a description of measurements of pairs of two observables. This avoids conflicts with Bell's inequalities \cite{Bell,Z} which apply for complete statistics if the classical correlation is used \cite{CC}. 

(4) {\em The unitary time evolution} of quantum mechanics is a special case of a more general classical evolution which can also describe the phenomena of decoherence and syncoherence. We propose that the unitary time evolution reflects the isolation of the subsystem and corresponds to a partial fixed point (or better ``fixed manifold'') of the more general classical time evolution of the probability distribution.

A classical statistical description of quantum mechanics should not be confounded with a deterministic description. The no go theorems for large classes of local deterministic ``hidden variable theories'' remain valid. We rather take the attitude that a probabilistic approach is appropriate for the basic setting of both the quantum and the classical world. Deterministic behavior is a particular case (albeit rather genuine) which can originate from the collective properties of many degrees of freedom, as the laws of thermodynamics, or the motion of planets which are composed of many atoms. Other deterministic features arise from the discreteness of the spectrum of quantum observables, as the energy levels in atoms. The classical statistical ensembles of this paper should therefore not be reduced to the concept of a finite large number of point particles for which locations and momenta are given, in principle, in a deterministic way. `Classical statistics'' means the description in terms of an ensemble of classical states for which observables take fixed values, and a classical probability distribution for these states which changes in time by a ``deterministic'' evolution equation. 

We demonstrate in this paper that all features of quantum mechanics can be described by a classical statistical ensemble if appropriate observables for the subsystem are selected, if sequences of measurements are described by correlations compatible with the subsystem, and if the probabilities for the classical states follow a suitable time evolution. The strategy of our approach is the following. Out of the infinitely many classical observables that can, in principle, be measured in the system and its environment, we select subclasses of system and quantum observables which only measure properties of the system. For suitable quantum observables and their measurement correlations we show that they have all the properties of the observables associated to the non-commuting operators in quantum mechanics.  

We begin the discussion in sect. \ref{Classicalquantumcomputer} with a simple example of a classical realization for the qubits of a quantum computer. While this does not yet reveal all features of quantum physics, we present in the following sections a systematic discussion how all characteristics of quantum mechanics are realized in classical statistical ensembles with appropriate properties. In sect. \ref{Observablesand} we describe the concept of probabilistic observables for the subsystem and we present explicit classical ensembles which realize all aspects of two-state and four-state quantum systems in sect. \ref{Simplequantum}. Sect. \ref{Correlations} discusses the correlations between measurements of ``quantum observables'' which can be computed from the information contained in the state of the subsystem. In sect. \ref{Quantumtimeevolution} we turn to the unitary time evolution of the state of an isolated subsystem. Sect. \ref{Propertiesofquantum} is devoted to a detailed discussion of properties of quantum observables within our classical statistical setting. In particular, we discuss the issue of entanglement and Bell's inequalities in sect. \ref{Entanglement}. After a detailed discussion of sequences of measurements and their relation to the quantum mechanical commutator of two operators in sect. \ref{Sequenceofmeasurements}, we conclude in sect. \ref{Conclusions}.

\section{Classical statistical quantum computer}
\label{Classicalquantumcomputer}

In this section we discuss simple examples of classical statistical ensembles that realize certain aspects of quantum mechanics. The demonstration that all properties of quantum mechanics can emerge from a classical setting, and a more systematic discussion under what circumstances this happens, will be given in the following sections. Here we describe the operations of a quantum computer purely within the framework of a time evolution of probabilities for the states of a classical statistical ensemble.

\medskip\noindent
{\bf Qubit}

We start with a two level observable $A$ which can take only two values $+1$ and $-1$. Equivalently, we can identify $A=1$ with the value $1$ for a bit, while $A=-1$ corresponds to the value $0$. We consider families of statistical ensembles, where a particular ensemble is specified by a particular probability distribution for  the ``classical states'' of the ensemble. For every classical state $\tau$ the observable has one of the values $A_\tau=1$ or $A_\tau=-1$, and the probability for the state $p_\tau$ obeys $0\leq p_\tau\leq 1~,~\sum_\tau p_\tau=1$. The time evolution of the system is described by the time dependence of the probabilities $p_\tau(t)$, defining a trajectory within the family of ensembles. 

In order to compute the expectation value $\langle A\rangle$ we do not need to know the complete probability distribution of the ensemble $\{p_\tau\}$. It is sufficient to know the probability $w_+$ for $A$ having the value $+1$, and similarly $w_-$ for $A=-1$. Since $w_++w_-=1$, one of the probabilities $w_\pm$, or the relative ratio of probabilities $w_+/w_-$, is sufficient to compute
\begin{equation}\label{A1}
\langle A\rangle=w_+-w_-.
\end{equation}

In the following we assume that $w_\pm$ can be computed in terms of three real numbers $\rho_k$ obeying $\sum_k\rho^2_k\leq 1$. In other words, the complete statistical information contained in the probability distribution $\{p_\tau\}$ is not necessary for a determination of $\langle A\rangle$, which only requires the part of the information contained in the three numbers $\rho_k$. This is a typical situation for an isolated subsystem, whose behavior can be described by the state of the subsystem which is specified by $(\rho_k)$, independently of the detailed state of the environment. The statistical information which characterizes the probability distribution of the ensemble beyond the three numbers $\rho_k$ describes the state of the environment.

We will not need the detailed properties of the environment. As a concrete example we  may consider a classical ensemble with eight states labeled by $\tau=1,\dots, 8$, characterized by the probabilities $p_\tau\geq 0$ for each state $(\sum_\tau p_\tau=1)$. Instead of $\tau$ we can use a triple-index $(\sigma_1,\sigma_2,\sigma_3)$, with $\sigma_j=\pm 1$. If we define
\begin{equation}\label{A2}
\rho_j=\sum_{\sigma_1,\sigma_2,\sigma_3}\sigma_j p(\sigma_1,\sigma_2,\sigma_3),
\end{equation}
three linear combinations of the probabilities $p(\sigma_1,\sigma_2,\sigma_3)$ specify the state of the subsystem, while the remaining four independent linear combinations characterize the environment. We may use
\begin{equation}\label{AS1}
p(\sigma_1,\sigma_2,\sigma_3)=p_s(\sigma_1,\sigma_2,\sigma_3)+\delta p_e(\sigma_1,\sigma_2,\sigma_3)
\end{equation}
with
\begin{equation}\label{AS2}
p_s(\sigma_1,\sigma_2,\sigma_3)=\frac18(1+\sigma_1\rho_1)(1+\sigma_2\rho_2)
(1+\sigma_3\rho_3)
\end{equation}
and $\delta p_e$ obeying
\begin{equation}\label{AS3}
\sum_{\sigma_1,\sigma_2,\sigma_3}\delta p_e(\sigma_1,\sigma_2,\sigma_3)=0~,~
\sum_{\sigma_1,\sigma_2,\sigma_3}\sigma_j\delta p_e(\sigma_1,\sigma_2,\sigma_3)=0.
\end{equation}
Different choices of $\delta p_e$ differ then only in properties of the environment, while $\rho_j$ is independent of $\delta p_e$. As far as only the properties of the subsystem are concerned all $\delta p_e$ obeying the condition \eqref{AS3} and $0\leq p_s+\delta p_e\leq 1$ are equivalent. Typically, the environment may be far more extended, with (infinitely) many states $\tau$. For example, $\tau$ could be characterized by other labels beyond $(\sigma_1,\sigma_2,\sigma_3)$ which do not influence the $\rho_j$ and are summed over in eq. \eqref{A2}.  We note that in eq. \eqref{AS1} the split between the environment and the system cannot be done by simply associating some of the ``classical bits'' $\sigma_j$ to the system and others to the environment.

We will consider a simple dependence $w_+(\rho_k)$ given by
\begin{equation}\label{A3}
\langle A\rangle =\rho_3~,~w_+=\frac12(1+\rho_3).
\end{equation}
In order to describe the time evolution of $\langle A\rangle$ we need to specify the time evolution of the state of the subsystem. Let us investigate rotations
\begin{equation}\label{A4}
\rho_k(t,t')=\hat S_{kl}(t,t')\rho_l(t')~,~\hat S\hat S^T=1,
\end{equation}
as implemented by an evolution equation
\begin{equation}\label{A5}
\frac{\partial}{\partial t}\rho_k=T_{kl}\rho_l~,~(T)^T=-T.
\end{equation}
(Sums over repeated indices are implied.)
As a particular example we may choose a rotation in a plane diagonal to the $1-2$ and $2-3$ planes with $\varphi$ linearly increasing with $t$
\begin{eqnarray}\label{A6a}
&&\hat S=\\
&&\left(-\begin{array}{ccccc}
\cos^2\varphi&,&\sqrt{2}\sin\varphi\cos\varphi&,&\sin^2\varphi\nonumber\\
\sqrt{2}\sin\varphi\cos\varphi&,&1-2\sin^2\varphi&,&\sqrt{2}\sin\varphi\cos\varphi\\
\sin^2\varphi&,&-\sqrt{2}\sin\varphi\cos\varphi&,&\cos^2\varphi
\end{array}\right)
\end{eqnarray}
where
\begin{eqnarray}\label{A6aa}
T&=&\sqrt{2}\dot\varphi\left(\begin{array}{rrrr}
0,&1,&~&0\\-1,&0,&~&1\\0,&-1,&~&0
\end{array}\right).
\end{eqnarray}
With $\rho_{k,0}=\rho_k(\varphi=0)$ eq. \eqref{A4} yields
\begin{equation}\label{A7a}
\langle A(t)\rangle=\sin^2\varphi\rho_{1,0}-\sqrt{2}\sin\varphi\cos\varphi\rho_{2,0}+
\cos^2\varphi\rho_{3,0}.
\end{equation}
For $\sum_k\rho^2_{k,0}=1$ this describes a typical precession pattern of a spin in quantum mechanics in an appropriately chosen homogeneous  magnetic field. While $\langle A\rangle$ only requires the knowledge of $\rho_3$, the time evolution of $\rho_3$ also involves the two other variables characterizing the state, $\rho_1$ and $\rho_2$. These variables will contain the information about the correlations which are characteristic for a quantum system. 

Rotations with arbitrary angles $\varphi$ can be realized on the classical level by a time evolution of the probabilities $p_\tau$ according to eq. \eqref{A2}. For this purpose it is sufficient that $p_s(\sigma_1,\sigma_2,\sigma_3)$ in eqs. \eqref{AS1}, \eqref{AS2} evolves according to the rotations of $\rho_j$. This may be accompanied by an arbitrary evolution of $\delta p_e(\sigma_1,\sigma_2,\sigma_3)$ which ensures that the condition \eqref{AS3} and $0\leq p_s+\delta p_e\leq 1$ continues to hold. The simplest example is $\delta p_e=0$ for all $t$. However, the precise time evolution of $\delta p_e$ has no influence on the evolution of $\rho_k(t)$ and $\langle A(t)\rangle$. The time evolution of the subsystem is decoupled from the environment. As a special example we may consider $\varphi(t=\Delta)=\pi/2$ with
\begin{equation}\label{A8}
\rho_3(t)=\rho_{1,0}~,~\rho_1(t)=\rho_{3,0}~,~\rho_2(t)=-\rho_{2,0},
\end{equation}
represented by
\begin{eqnarray}\label{10A}
p_s(\sigma_1,\sigma_2,\sigma_3;t)&=&p_s(\sigma_3,\sigma_2,\sigma_1;0),\nonumber\\
p_s(\sigma_1,\sigma_2,\sigma_3;t)&=&p_s(\sigma_3,-\sigma_2,\sigma_1;0),
\end{eqnarray}
independently of $\delta p_e(t)$ or $\delta p_e(0)$. 

For $\rho_2\rho_k\leq 1$ the three numbers $\rho_k$ can be represented by a $2\times 2$ hermitean density matrix in terms of the Pauli matrices $\tau_k$
\begin{equation}\label{10B}
\rho=\frac12(1+\rho_k\tau_k).
\end{equation}
In turn, a ``pure state density matrix'', which obeys $\rho_k\rho_k=1$, can be expressed in terms of a normalized two-component complex vector $\psi_\alpha$ - a ``quantum state'' or ``qubit'' - as
\begin{equation}\label{10C}
\rho_{\alpha\beta}=\psi_\alpha\psi^*_\beta~,~\psi^*_\alpha\psi_\alpha=1,
\end{equation}
such that the expectation value of $A$ obeys the quantum rule $(\hat A=\tau_3)$
\begin{equation}\label{10D}
\langle A\rangle=\psi^*_\alpha(\tau_3)_{\alpha\beta}\psi_\beta=\langle\psi|\hat A|\psi\rangle.
\end{equation}
Rotations of the vector $(\rho_k)$ act on $\psi_\alpha$ as unitary transformations
\begin{equation}\label{10E}
\psi_\alpha(t)=U_{\alpha\beta}(t)\psi_\beta(0).
\end{equation}

Inversely, for any Hamiltonian of a two-state quantum system we can construct the evolution operator $U$ and the associated evolution of $\rho_k(t)$ by using eq. \eqref{10B} and 
\begin{equation}\label{10F}
\rho_{\alpha\beta}(t)=U_{\alpha\gamma}(t)\rho_{\gamma\delta}(0)
U^\dagger_{\delta\beta}(t).
\end{equation}
With eqs. \eqref{AS1}, \eqref{AS2} we can infer the classical probability distributions $p_\tau(t)$ whose time evolution precisely reproduces the time evolution of the expectation value of the observable $A$, which is associated in the quantum system to the operator $\hat A=\tau_3$. For example, we can interprete this observable as the third component of the spin in appropriate units, $A=S_3$. Simultaneously, we can also describe by eq. \eqref{10D}, or more generally by
\begin{equation}\label{10G}
\langle A\rangle=\text{tr}(\rho\hat A),
\end{equation}
the time evolution of the orthogonal spin components associated to the operators $\tau_1$ and $\tau_2$.

\medskip\noindent
{\bf Classical statistical realization of quantum gates}

The evolution \eqref{A8}, \eqref{10A} realizes the ``Hadamard gate'' for a qubit, with associated unitary evolution operator
\begin{equation}\label{10H}
U=\frac{1}{\sqrt{2}}\left(\begin{array}{cc}
1,&1\\1,&-1\end{array}\right).
\end{equation}
(The overall phase of $U$ is without physical significance, since the overall phase of $\psi$ drops out in eq. \eqref{10C}.) One could realize the Hadamard transformation by a continuous time evolution of a classical ensemble with probabilities $p_\tau(t)$ changing according to eq. \eqref{A6a}, with 
\begin{equation}\label{10I}
\varphi(t)=\frac{\pi t}{2\Delta},
\end{equation}
such that for a ``read out'' at $t=\Delta$ the angle $\varphi=\frac\pi2$ realizes eq. \eqref{A8}. Alternatively, we may associate the probabilities $p_\tau(t=\Delta)$ with a distribution of random numbers, which are obtained from  random numbers at $t=0$ by eq. \eqref{A8}.

Other quantum gates can be realized similarly. For example, the $\frac\pi4$-phase shift
\begin{equation}\label{10J}
U=\left(\begin{array}{ll}
1,&0\\0,&e^{i\frac\pi4}
\end{array}\right)
\end{equation}
can be realized by a rotation of $(\rho_k)$ in the $1-2$ plane 
\begin{equation}\label{A6}
T=\dot\varphi\left(\begin{array}{rrr}
0,&+1,&0\\-1,&0,&0\\0,&0,&0
\end{array}\right)~,~
\hat S=
\left(\begin{array}{rrr}
\cos\varphi,&\sin\varphi,&0\\
-\sin\varphi,&\cos\varphi,&0\\
0,&0,&1
\end{array}\right)
\end{equation}
and
\begin{equation}\label{A7}
\varphi=\frac{\pi t}{4\Delta}.
\end{equation}
For $t=\Delta$ this amounts to
\begin{eqnarray}\label{23A}
\rho_1(t)&=&\frac{1}{\sqrt{2}}\big(\rho_1(0)+\rho_2(0)\big),\\
\rho_2(t)&=&\frac{1}{\sqrt{2}}\big(-\rho_1(0)+\rho_2(0)\big),\nonumber\\
\rho_3(t)&=&\rho_3(0).\nonumber
\end{eqnarray}

We can also construct classical ensembles realizing quantum gates acting on two qubits, as the ``controlled not gate'' or CNOT-gate represented by
\begin{equation}\label{F1a}
U=\left(\begin{array}{llll}
1,&0,&0,&0\\0,&1,&0,&0\\
0,&0,&0,&1\\0,&0,&1,&0
\end{array}\right).
\end{equation}
For this purpose we enlarge our classical ensemble and consider a subsystem characterized by $15$ real numbers $\rho_k$ obeying $\rho_k\rho_k\leq 3$. They can be grouped in to a $4\times 4$ density matrix
\begin{equation}\label{F2a}
\rho=\frac14(1+\rho_kL_k)~,~\text{tr}(L_kL_l)=4\delta_{kl},
\end{equation}
with $L_k$ appropriately normalized generators of $SU(4)$ given in a direct product basis by \cite{CW1}
\begin{eqnarray}\label{FXA}
L_1&=&\tau_3\otimes 1~,~L_2=1\otimes \tau_3~,~L_3=\tau_3\otimes\tau_3,\\
L_4&=&1\otimes\tau_1~,~L_5=1\otimes\tau_2~,~L_6=\tau_3\otimes\tau_1,\nonumber\\
L_7&=&\tau_3\otimes\tau_2~,~L_8=\tau_1\otimes 1~,~L_9=\tau_2\otimes 1,\nonumber\\
L_{10}&=&\tau_1\otimes\tau_3~,~L_{11}=\tau_2\otimes\tau_3~,~
L_{12}=\tau_1\otimes\tau_1,\nonumber\\
L_{13}&=&~\tau_1\otimes\tau_2,~L_{14}=-\tau_2\otimes\tau_2~,~L_{15}
=\tau_2\otimes\tau_1.\nonumber
\end{eqnarray}

Pure states obey $\rho^2=\rho$. They can be associated to a complex four-component vector $\psi$ according to eq. \eqref{10C}, which describes the quantum states of four-state quantum mechanics  or two qubits. The unitary transformation \eqref{F1a} is associated to a transformation of the density matrix $\rho$ and the fifteen-component real vector $(\rho_k)$ according to 
\begin{equation}\label{F3a}
\rho'=U\rho U^\dagger~,~\rho'_k=\text{tr}(L_k\rho'),
\end{equation}
resulting for the density matrix $\rho$ in the exchange of elements
\begin{eqnarray}\label{F4a}
&&\rho_{13}\leftrightarrow\rho_{14}~,~\rho_{23}\leftrightarrow\rho_{24}~,~
\rho_{31}\leftrightarrow\rho_{41}~,~\rho_{32}\leftrightarrow\rho_{42},\nonumber\\
&&\rho_{33}\leftrightarrow\rho_{44}~,~\rho_{34}\leftrightarrow \rho_{43},
\end{eqnarray}
and for the vector $(\rho_k)$ in
\begin{eqnarray}\label{F5a}
\rho_2\leftrightarrow\rho_3~,~\rho_5\leftrightarrow\rho_7~,~\rho_8\leftrightarrow\rho_{12},\nonumber\\
\rho_9\leftrightarrow\rho_{15}~,~\rho_{10}\leftrightarrow\rho_{14}~,~\rho_{11}\leftrightarrow\rho_{13}
\end{eqnarray}
Eq. \eqref{F5a} corresponds to a particular rotation of the $15$-component vector $(\rho_k)$.

For a realization in terms of classical statistics we may employ an ensemble with $2^{15}$ states $\tau$. They can be labeled as $\{\sigma_k\}=(\sigma_1,\sigma_2,\dots,\sigma_{15})$ with $k=1,\dots,15$ and $\sigma^2_k=1$. We can again use a classical probability distribution of the type \eqref{AS1} with
\begin{eqnarray}\label{F6aa}
p_s\big(\{\sigma_k\}\big)=2^{-15}\prod_k(1+\sigma_k\rho_k),
\end{eqnarray}
and
\begin{eqnarray}\label{F6a}
\sum_{\{\sigma_k\}}\delta p_e\big(\{\sigma_k\}\big)&=&0~,~\sum_{\{\sigma_k\}}\sigma_j\delta p_e\big(\{\sigma_k\}\big)=0.
\end{eqnarray}
Again, one finds
\begin{equation}\label{30Aa}
\rho_j=\sum_{\{\sigma_k\}}\sigma_j p\big(\{\sigma_k\}\big)
\end{equation}
and the rotations among the $\rho_k$ are realized by an evolution of $p_s$ with time varying $\rho_k$, while the evolution of the ``environmental information'' $\delta p_e\{\sigma_k\}$ is arbitrary as long as the constraints \eqref{F6a} are obeyed and $0\leq p_s+\delta p_e\leq 1$. This demonstrates that the time evolution of classical probabilities can realize the CNOT gate, even though the number of classical states $2^{15}$ is very high. (We do not attempt here to discuss possible classical realizations of the CNOT gate with a smaller number of classical states.)

Also the Hadamard gate and the $\pi/4$-phase shift can operate separately on qubit $1$ and qubit $2$. For qubit $1$ we replace in eqs. \eqref{A8}, \eqref{23A}: $(\rho_1,\rho_2,\rho_3)\to (\rho_8,\rho_9,\rho_1)$, while for qubit $2$ the replacement is $(\rho_1,\rho_2,\rho_3)\to(\rho_4,\rho_5,\rho_2)$. The different gates can operate one after the other if we switch the evolution of probabilities after a time step $\Delta$ to the evolution characteristic for the next gate. A sequence of two gates corresponds to a sequence of two rotations of the vector $(\rho_k)$ or the associated classical probabilities $p\big(\{\sigma_k\}\big)$. Obviously, the order of the sequence matters since in general rotations are not commutative.

\medskip\noindent
{\bf Classical statistical realization of a quantum \\computer}

We have now most elements for a description of a quantum computer in terms of the time evolution of probabilities for the states of a classical statistical ensemble. Our purpose is here purely an existence proof on the conceptual side, rather than a proposal for a practical realization which could speed up computations for certain types of problems. It is meant as an introductory example how classical statistics can be linked to quantum mechanics. Beyond the operation of quantum gates we still need a description of the initialization of some algorithm, the readout of results of a computation and the scaling to a larger number of qubits.

Initialization means the preparation of an initial state for the qubits on which quantum gates can operate consecutively in discrete time steps $\Delta$. A convenient initial state has $0$ for all qubits. In our two-qubit example this means that the two observables associated to the generators $L_1$ and $L_2$ should have expectation value $-1$, i.e. $\rho_1=\rho_2=-1$. We note that the generators $L_1$ and $L_2$ commute, $L_1L_2=L_2L_1=L_3$. We want to start with a pure state which requires $\rho_k\rho_k=3$ and therefore choose $\rho_3=1$. For all other $\rho_k$ vanishing, $\rho_{k\geq 4}=0$, the density matrix
\begin{equation}\label{30B}
\rho=\frac14(1-L_1-L_2+L_3)=diag(0,0,0,1)
\end{equation}
describes indeed a pure state, with associated wave function $\psi=(0,0,0,1)$. The associated initial classical probability distribution is
\begin{equation}\label{30C}
p\big(\{\sigma_k\}\big)=2^{-15}(1-\sigma_1)(1-\sigma_2)(1+\sigma_3)+\delta p_e\big(\{\sigma_k\}\big).
\end{equation}

For a certain number of time steps $\Delta$ the classical probabilities $p\big(\{\sigma_k\}\big)$ are assumed to change in a way such that a sequence of CNOT, Hadamard or phase shift quantum gates are operated on the two qubits. After this sequence of gates the classical probability distribution has reached a ``final value'' $p_f\big(\{\sigma_k\}\big)$ which one wants to ``read out'' by a measurement of the two qubits. The read out is done by measuring two two-level observables $A^{(1)},A^{(2)}$ which can take the values $\pm 1$. The respective probabilities of finding $+1$ or $-1$ are related to the expectation values similar to eq. \eqref{A1}, with
\begin{equation}\label{30D}
\langle A^{(1)}\rangle=\rho_1(t_f)~,~\langle A^{(2)}\rangle=\rho_2(t_f),
\end{equation}
and $\rho_{1,2}(t_f)$ determined from $p_f\big(\{\sigma_k\}\big)$ according to eq. \eqref{30Aa}. In other words, the probability for finding for qubit $1$ the value $1$ is given by $w^{(1)}_+=\frac12\big(1+\rho_1(t_f)\big)$, and for qubit $2$ by $w^{(2)}_+=\frac12\big(1+\rho_2(t_f)\big)$. 

As appropriate for two qubits there are four possible outcomes of the read out $(1,1),(1,-1),(-1,1)$, and $(-1,-1)$. (We use here the notation $-1$ for the value $0$ of the bit, according to the values of the two-level-observables $A^{(k)}$.) For a given initial state and a well defined sequence of quantum gates, one may wonder if it is possible to compute the joint probabilities $p_{++},p_{+-},p_{-+},p_{--}$ for the observables $A^{(1)}$ and $A^{(2)}$ having values according to these four possibilities. Since $p_{++}+p_{+-}+p_{-+}+p_{--}=1$ this requires three independent numbers. The values $\rho_1(t_f)$ and $\rho_2(t_f)$ provide only two of them. They are not enough for a computation of the joint probabilities that qubit $1$ has the value $+1$ and qubit $2$ the value $-1$, etc.. The missing piece of information is
\begin{equation}\label{30E}
C_{12}=\sum_{\{\sigma_k\}}\sigma_1\sigma_2p\big(\{\sigma_k\}\big)
=p_{++}+p_{--}-p_{+-}-p_{-+}.
\end{equation}
Together with $\rho_1=p_{++}+p_{+-}-p_{-+}-p_{--}$ and $\rho_2=p_{++}-p_{+-}+p_{-+}-p_{--}$ this would yield the necessary information for the computation of the four joint probabilities $p_{++},p_{+-},p_{-+}$ and $p_{--}$ separately.

In general, the quantity 
\begin{equation}\label{30F}
C_{ij}=\sum_{\{\sigma_k\}}\sigma_i\sigma_jp\big(\{\sigma_k\}\big)
\end{equation}
depends on $\delta p_e$ and therefore on the detailed properties of the environment. It is then not a property of the subsystem alone and therefore not available for a prediction of ``readouts'' by subsystem data only. For the initial state \eqref{30B} we have $p_{++}=p_{+-}=p_{-+}=0$, $p_{--}=1$ and we can identify for $t=0$
\begin{equation}\label{30G}
C_{12}=\rho_3=\rho_1\rho_2.
\end{equation}
Eq. \eqref{30G} holds for $t_f$ only if the ``read-out'' state is precisely a simultaneous eigenstate of $A^{(1)}$ and $A^{(2)}$. Only for this special case we can use the ``system data'' $\rho_3$ in order to predict the joint probabilities, while for general states this is not possible.

We will argue that the joint probabilities $p_{++}$ etc. are actually not what is needed for a prediction of the outcome of the four possibilities of the readout. The readout of qubit $1$ is a measurement, and in general measurements change the ensemble. What is needed is the probability $w_{++}$ to find for $A^{(2)}$ the value $+1$ if $A^{(1)}$ is measured to have the value $+1$. (We will see in sects. \ref{Correlations}, \ref{Sequenceofmeasurements} that the order of the measurements does not matter in our case.) Thus the {\em conditional probabilities} $w_{++}$ and $w_{+-}$ for finding $A^{(2)}=\pm 1$ under the condition $A^{(1)}=1$ are needed, and similarly $w_{-+}$ and $w_{--}$ for a measurement of $A^{(2)}$ if $A^{(1)}=-1$. We will see in sect. \ref{Correlations} that the conditional probabilities can be expressed by system data and therefore predicted without knowing details of the environment. In particular, one finds
\begin{equation}\label{40A}
w_{++}-w_{+-}-w_{-+}+w_{--}=\rho_3.
\end{equation}
Together with $w^{(1)}_+=w_{++}+w_{+-}~,~w^{(2)}_+=w_{++}+w_{-+}$, which involve $\rho_1$ and $\rho_2$, all conditional probabilities can indeed be expressed in terms of $\rho_1,\rho_2$ and $\rho_3$. The probabilities for the four different readout results can therefore be predicted using the subsystem data alone.

The predictability of probabilities $w_{++}$ etc. is crucial for performing quantum computations for which the $2^Q$ alternatives associated to the readout of $Q$ quantum bits can be used. If quantum computing only involves the state of the system independently of the unknown state of the environment, the computability of the conditional probabilities in terms of $\rho_k$ is a key ingredient. This brings us to the important issue which type of correlations should be used for measurements of properties of the subsystem. The conditional probabilities fixed by eq. \eqref{40A} define a ``measurement correlation''
\begin{equation}\label{41A}
\langle A^{(1)}A^{(2)}\rangle_m=\rho_3.
\end{equation}
This is distinct from  the ``classical'' or ``ponintwise'' correlation
\begin{equation}\label{41B}
\langle A^{(1)}\cdot A^{(2)}\rangle=C_{12}.
\end{equation}
While the measurement correlation is computable in terms if the state of the subsystem, the classical correlation is not, since it involves details of the environment. The outcome of a ``good measurement'' in an isolated subsystem should only depend on the state of the subsystem and not on the state of the environment. This excludes the use of the classical correlation. We will discuss this important issue for the understanding of quantum systems in more detail in sect. \ref{Correlations}.

Finally, we briefly discuss the scaling for our classical statistical implementation of a quantum computer for an arbitrary number of $Q$ qubits. One can associate $Q$ qubits to $M$-state quantum mechanics with $M=2^Q$. In turn, the most general density matrix involves $K=M^2-1$ components of the vector $\rho_k$. All quantum gates can be realized by rotations of $(\rho_k)$. Generalizing our construction of associated classical probabilities $p_\tau$ involves $S=2^K$ classical states. We observe that the number of classical states
\begin{equation}\label{41C}
S=2^{(2^{2Q}-1)}
\end{equation}
increases very rapidly with $Q$. Large numbers of classical states are characteristic for a classical statistics implementation of quantum mechanics. Realizing spin observables with an arbitrary direction involves already an infinity of classical states \cite{CW2}. We emphasize, however, that most of the information contained in the probability distribution for the classical ensemble $\{p_\tau\},\tau=1\dots S$, only concerns properties of the environment. The state of the quantum system involves only $K$ real numbers $\rho_k$ or $M$ complex numbers $\psi_\alpha$ in case of pure states.

\section{Observables and expectation values}
\label{Observablesand}

The quantum features encountered in the preceding section can be generalized by addressing systematically the embedding of a subsystem in a more extended classical ensemble  that also includes the environment or the vacuum. We discuss in the following sections systems that correspond to a finite number of quantum states $M$. The generalization to $M\to\infty$, needed for quantum observables with a continuous spectrum, is straightforward, as shown for a quantum particle in a potential in ref. \cite{CWQP}.

\medskip
\noindent
{\bf Probabilistic observables}

The most crucial effect of the embedding of the subsystem into a classical statistical ensemble, typically for infinitely many classical states, is the appearance of probabilistic observables for the description of the subsystem. For a given state of the subsystem - which will be associated with a quantum state - they have a probability distribution of values rather than a fixed value as for the standard classical observables in a classical state.  For subsystems that are equivalent to $M$-state quantum mechanics the spectrum of the possible outcomes of measurements for the probabilistic observables contains at most $M$ different real values $\gamma_a$. In a given quantum state the probabilistic observable is characterized by probabilities $w_a$ to find $\gamma_a$, where $w_a$ depends on the state. The simplest example are two-level observables, which can resolve only one bit, such that $\gamma_1=1~,~\gamma_2=-1$.

We may use the example of the observable $A$ in the preceeding section, where for a given state of the subsystem characterized by $(\rho_1,\rho_2,\rho_3)$ only the probability $w_+(\rho_k)$ for finding the value $A=1$ is known. In fact, the emergence of probabilistic observables from a reduction of effective degrees of freedom is a generic feature in statistical physics. As another example, we may group classical states $\tau=(\sigma,\beta)$ into collective states $\sigma$ by summing over $\beta$, with probabilities for the collective states given by $p_\sigma=\sum_\beta p_{(\sigma,\beta)}$. Then a classical observable $A$ which has a sharp value $\pm 1$ for every state $(\sigma,\beta)$ will have only a probability distribution of values $\pm 1$ in the state $\sigma$, where the probability $w_{\sigma,+}$ to find the value $+1$ in the state $\sigma$ obtains by summing the probabilities $p_{(\sigma,\beta)}$ over all $\beta$ for which $A_{(\sigma,\beta)}=1$. 

Let us consider a subsystem that can be described by $n$ discrete classical two-level-observables $A^{(k)}$. We assume that these observables form a basis in a sense to be specified later. The simplest quantum mechanical analogue for $n=3$ is two-state quantum mechanics, with $A^{(k)}$ corresponding to three orthogonal ``spins'' in an appropriate normalization. This may be viewed as an atom with spin one half where only the spin degree of freedom is resolved, as for example in Stern-Gerlach type experiments. 

The subsystem is embedded into a classical statistical ensemble with infinitely many states labeled by $\tau$. On this level the observables $A^{(k)}$ are standard classical or deterministic observables. They can only take fixed values $A^{(k)}_\tau=\pm 1$ for any state $\tau$ of the classical statistical ensemble. For arbitrary $n$, we denote the expectation value or average of $A^{(k)}$ by $\rho_k$, 
\begin{equation}\label{1}
\rho_k=\langle A^{(k)}\rangle=\sum_\tau p_\tau A^{(k)}_\tau~,~-1\leq \rho_k\leq 1.
\end{equation}
The classical probabilities obey $p_\tau\geq 0$ for all states $\tau$ of the classical statistical ensemble. As usual, one has $\sum_\tau p_\tau=1$. 
However, not all possible classical probability distributions $\{p_\tau\}$ obeying these conditions correspond to quantum systems, and we will discuss restrictions below.

The probabilistic observables associated to these classical observables have probabilities $w^{(k)}_\pm=(1\pm\rho_k)/2$ to find $\gamma_{1,2}=\pm 1$. We can compute $w^{(k)}_+$ by summing the probabilities $p_\tau$ over all classical states for which $A^{(k)}_\tau=1$, and similarly for $w^{(k)}_-$. This maps the classical observable, as characterized by the values $A^{(k)}_\tau$ in every classical state $\tau$, to a probabilistic observable characterized by $\gamma^{(k)}_a$ and $w^{(k)}_a$. 

The map from the classical observables to the probabilistic observables is not invertible. A different classical observable $A'^{(k)}$, with different values $A'^{(k)}_\tau$ in the classical states, may be mapped to the same probabilistic observable $A^{(k)}$. For this it is sufficient that the probabilities $w^{(k)}_\pm$ are the same for every probability distribution $\{p_\tau\}$ which corresponds to a quantum system. This lack of invertibility avoids conflicts of our classical statistical description with the Kochen-Specker theorem \cite{KS}, as we will discuss in sect. \ref{Propertiesofquantum}. Since the classical observables $A^{(k)}_\tau$ contain much more information than the associated probabilistic observables, they ``measure'' properties of both the quantum system and its environment. The transition to probabilistic observables ``integrates out'' the environment degrees of freedom \cite{CW2}.

\medskip\noindent
{\bf Quantum system and system observables}

We will assume that the $n$ numbers $\rho_k$ are the only information that is needed and available for a computation of expectation values for the ``system observables'' of the subsystem. In this sense, the state of the subsystem is characterized by the $n$ expectation values of the basis observables $A^{(k)}$. Only a very limited amount of the information contained in the probability distribution $\{p_\tau\}$ for the total system is needed for the subsystem. We will refer to a subsystem with these properties as the ``quantum system'', even though in certain special cases it can also describe a classical statistical system.

For a given quantum system the system observables are those classical observables that lead to probabilistic observables $A$ for which the probabilities $w_a$ can be computed in terms of $\{\rho_k\}$. Then the expectation values of functions $f(A)$ can also be computed from $\{\rho_k\}$, $\langle f(A)\rangle=\sum_a f(\gamma_a)w_a(\rho_k)$. We will assume that the relation between $w_a$ and $\rho_k$ is linear. 

Our first question concerns a classification of possible system observables for the subsystem. It is straightforward to define rescaled observables $c A^{(k)}$ by $(c A^{(k)})_\tau=c A^{(k)}_\tau,\langle c A^{(k)}\rangle=c\rho_k$.  Furthermore, we can trivially shift the observable by a piece $e_0$ proportional to the unit observable. The rescaled and shifted observables $A$ obey
\begin{equation}\label{2}
\langle A\rangle=\rho_ke^{(A)}_k+e^{(A)}_0,
\end{equation}
where repeated indices are summed. Here we associate to each $A^{(k)}$ an $n$-dimensional unit vector $e^{(k)}$ with components $e^{(k)}_m=\delta^k_m$.  Then the vector $e^{(A)}$ reads $e^{(A)}=c e^{(k)}$ if $A=c A^{(k)}$. One may use $c=\hbar/2$ if $A$ describes a spin with standard units of angular momentum. Other units may be employed for alternative interpretations, as for example occupation number $n=(1+A^{(3)})/2$ which equals one for occupied and zero for empty. (Contrary to widespread belief, the value of $\hbar$ is not a genuine property of quantum mechanics, but rather an issue of units.) One could include $e^{(A)}_0$ into an $n+1$-dimensional vector $e^{(A)}$. We will not do so since in the following we mainly discuss observables with $e^{(A)}_0=0$.

For an arbitrary system observable $A$ we may write the expectation value 
\begin{equation}\label{42A}
\langle A\rangle=\sum_a\gamma_a w_a(\rho_k),
\end{equation}
with $w_a$ depending linearly on $\rho_k$, as a linear combination $\langle A\rangle=\sum_kc_k\langle A^{(k)}\rangle$. (If necessary, we substract an appropriate constant shift.) This should hold for all probability distributions $\{p_\tau\}$ which describe the quantum system.  We can associate to each such observable the vector $e^{(A)}=\sum_kc_k e^{(k)}~,~e^{(A)}_k=c_k$, such that eq. \eqref{2} remains valid. Thus the vector $e^{(A)}$ can be used for a partial characterization of a system observable. Explicit constructions of such probabilistic observables can be found in \cite{CW2}, where observables of this type correspond to rotated spins in the quantum mechanical analogue. The possibility to write the expectation value in the form \eqref{2} is a necessary, albeit not sufficient condition for a classical observable to be a system observable. Beyond the vector $e^{(A)}$ a system observable also needs the specification of the explicit form of $w_a(\rho_k)$.

Probabilistic observables are characterized by the spectrum $\{\gamma_a\}$ of possible measurement values, and the associated probabilities $w_a$. The multiplication of the observable by a constant $c$ and the addition of a piece proportional to the unit observable are always defined by $\gamma_a\to c\gamma_a+e_0$. However, the sum and the product of two probabilistic observables $A,B$ are not defined a priori. At this stage we can only give a necessary condition that a possible linear combination C of two observables $A$ and $B$ can be defined as a system observable: it has to be represented by linear combinations of the associated vectors $e$. If $C=c_AA+c_BB$, one should have $e^{(C)}=c_Ae^{(A)}+c_Be^{(B)}$, such that $\langle C\rangle=c_A\langle A\rangle+c_B\langle B\rangle$ can again be computed from eq. \eqref{2}.

\medskip\noindent
{\bf Incomplete statistics}

In general, the joint probabilities for pairs of two observables $A^{(k)}$ and $A^{(l)}$ cannot be computed from the information which specifies the state of the subsystem. Indeed, for each of the $n(n-1)/2$ pairs of unequal basis observables one would have to specify the probabilities $p_{++},p_{+-},p_{-+}$ and $p_{--}$  that they have simultaneously the values $A^{(k)}=+1,A^{(l)}=+1$, or $A^{(k)}=+1,A^{(l)}=-1$ etc. Since $p_{++}+p_{+-}+p_{-+}+p_{--}=1$, this requires three numbers for each pair, or a total of $3n(n-1)/2$ numbers. This is more than the $n$ numbers $\rho_k$ which characterize the state of the subsystem. 

The joint probabilities are available only at the level of the classical statistical ensemble that characterizes the subsystem and its environment. We may pick a particular classical representation for the probabilistic observable $A^{(1)}$, as well as for $A^{(2)}$, specified by $A^{(1)}_\tau$ and $A^{(2)}_\tau$. The joint probability $p_{++}$ obtains then by summing the classical probabilities $p_\tau$ over all states $\tau$ for which both $A^{(1)}_\tau=1$ and $A^{(2)}_\tau=1$. However, the joint probability $p_{++}$ involves properties of the environment and is not a property of the subsystem alone. Two different classical observables $A^{(1)}_\tau$ and $A'^{(1)}_\tau$, which are mapped to the same probabilistic observable $A^{(1)}$ for the subsystem, will lead to different results for $p_{++}$. Since $A^{(1)}_\tau$ and $A'^{(1)}_\tau$ differ only by properties of the environment, this clearly demonstrates that joint probabilities involve precise knowledge of the state of the environment.

From the point of view of the subsystem this ``environment information'' is no longer available - it has been ``integrated out'' by the coarse graining of the information. We refer to a probabilistic setting for which joint probabilities are not available or not used for the correlation of measurements as ``incomplete statistics''. Hence the subsystem and the associated quantum mechanics is described by incomplete statistics in the sense of ref. \cite{3}. 

\medskip\noindent
{\bf Purity}

Next we are interested in some general properties of the basis observables. For example, one typical question may ask if two of them can have simultaneously a sharp value.  A ``classical eigenstate'' of a probabilistic observable $A$ is an ensemble for which $A$ has a ``sharp value'' with vanishing dispersion, $\langle A^2\rangle -\langle A\rangle^2=0$. For example, the eigenstate of the observable $A^{(k)}$ with ''classical eigenvalue'' $\langle A^{(k)}\rangle=1$ is characterized by $p_\tau=0$ whenever $A^{(k)}_\tau=-1$. The maximal number of sharp ``basis observables'' $A^{(k)}$ can be characterized by the ``purity'' $P$ of the ensemble,
\begin{equation}\label{3}
P=\rho_k\rho_k.
\end{equation}
(Note that $P$ depends on the set of basis observables that characterize the subsystem.)  For $P=0$ one finds equipartition with $\langle A\rangle=0$ for all two level observables. Obviously, $\tilde M$ sharp observables require $P\geq \tilde M$, since at least for $\tilde M$ values of $k$ one needs $\rho_k=\pm 1$. For an ensemble with $P=1$ at most one observable $A^{(k)}$ can be sharp. Typical classical ensembles that describe isolated quantum systems will have a maximal purity smaller than $n$, such that not all $A^{(k)}$ can have sharp values simultaneously. We recall that the purity \eqref{3} is a statistical property involving expectation values. For a given classical state $\tau$ all observables $A^{(k)}$ have a sharp value.

More precisely, the purity is a statistical property of the subsystem, with a conceptual status somewhat similar to the entropy. It is a measure for the size of the fluctuations - systems with larger fluctuations have a smaller purity. A measure for the fluctuations of the basis observables for the subsystem is given by
\begin{eqnarray}\label{48A}
G_k&=&\langle(A^{(k)}-\langle A^{(k)}\rangle)^2\rangle\nonumber\\
&=&\langle(A^{(k)})^2\rangle-\langle A^{(k)}\rangle^2=1-\rho^2_k,
\end{eqnarray}
with $0\leq G_k\leq 1$. The purity can then be expressed by the ``measure of  total fluctuations'' $G=\sum_kG_k$ as 
\begin{equation}
P=n-G=n-\sum^n_{k=1}G_k.
\end{equation}
For maximal fluctuations one has $G_k=1,G=n$ and therefore $P=0$, whereas minimal fluctuations with $G_k=0~,~G=0$ lead to the maximal possible value of the purity $P=n$.

\medskip
\noindent
{\bf Density matrix}

For $M$ an integer obeying $M\geq P+1$ we may represent the $\rho_k$ by an $M\times M$ hermitean ``density matrix''$\rho_{\alpha\beta}$:
\begin{eqnarray}\label{3a}
\rho=\frac 1M(1+\rho_kL_k).
\end{eqnarray}
The matrices $L_k$ \cite{CW1} are $SU(M)$-generators $k=1..M^2-1$, obeying
\begin{eqnarray}\label{3b}
\text{tr}L_k&=&0~,~L^2_k=1~,~\text{tr}(L_kL_l)=M\delta_{kl},\\
\{L_k,L_l\}&=&2\delta_{kl}+2d_{klm}L_m~,~[L_k,L_l]=2if_{klm}L_m.\nonumber
\end{eqnarray}
The matrix $\rho$ has properties of a density matrix in quantum mechanics, 
\begin{eqnarray}
\text{tr}\rho=1~,~\text{tr}\rho^2=\frac 1M(1+P)\leq 1.
\end{eqnarray}
Furthermore, we require that $\rho$ is a positive matrix in the sense that all its eigenvalues are positive or zero. This implies the positivity of all diagonal elements,
\begin{equation}\label{53A}
\rho_{\alpha\alpha}\geq 0.
\end{equation}

For a given $M$ the positivity of $\rho$ imposes constraints on the allowed values $\rho_k$ for which the subsystem can describe $M$-state quantum mechanics. We will discuss these ``positivity conditions'' in more detail below. We note that the condition for $\rho$ being a density matrix can always be realized for $M=n+1$ by choosing only diagonal and therefore mutually commuting $L_k$. We are interested, however, in the minimal $M$ for which $\rho$ is positive and the ``purity constraint'' $P\leq M-1$ holds. For a typical value $M=P+1=\sqrt{n+1}$ not all $L_k$ can commute.

In analogy to quantum mechanics, a ``classical pure state'' obeys $\rho^2=\rho$ and therefore requires $P=M-1$. The density matrix can be diagonalized by a unitary transformation. In consequence, any pure state density matrix can be written in the form $\rho=U\hat\rho_mU^\dagger$ for a suitable $U,$ with $UU^\dagger=1$. Here $(\hat\rho_m)_{\alpha\beta}=\delta_{m\alpha}\delta_{m\beta}$ obeys trivially $\hat\rho^2_m=\hat\rho_m$, and this property is preserved by unitary transformations. The positivity condition is obeyed obviously for all pure state density matrices, since the eigenvalues are one or zero.

\medskip\noindent
{\bf Quantum operators}

We can associate to any system observable $A$ an operator $\hat A$ such that the quantum mechanical rule for the computation of expectation values holds $(e_k\equiv e^{(A)}_k)$ 
\begin{equation}\label{5}
\hat A=e_kL_k~,~\langle A\rangle =\rho_ke_k=\text{tr}(\rho\hat A).
\end{equation}
We will concentrate on the minimal $M$ needed for a given maximal purity of the ensemble and assume that a suitable positivity constraint for the $\rho_k$ holds such that $\rho$ is a positive matrix. For $n=M^2-1$ the operators for the basis variables $A^{(k)}$ are given by the full set of all $SU(M)$ generators $L_k$. If $n<M^2-1$ only part of the $L_k$ are used as a basis for the observables.

Many characteristic features of the system observables $A$ can now be inferred from standard quantum mechanics, as demonstrated by a few examples. For $M=2$ at most one of the three possible two-level-observables $A^{(k)}$ can have a sharp value. This occurs for an ensemble where the density matrix describes a quantum mechanical pure state, $\rho=\frac12(1\pm \hat A)$, tr$\rho^2=\frac12+\frac14$tr$\hat A^2=1$, with $\hat A^{(k)}=L_k=\tau_k$. For such an ensemble the expectation value of the two orthogonal two-level-observables must vanish, $\langle A^{(l)}\rangle=0$ for $l\neq k$. Thus, whenever one basis observable is sharp, the two others have maximal uncertainty, as for the spin one-half system in quantum mechanics. 

Another example for $M=4$ describes two different two-level-observables (say the $z$-direction of two spins $S^1_z,S^2_z$) by $L_1=$ diag $(1,1,-1,-1)$ and $L_2=$ diag $(1,-1,1,-1)$. We may also consider a combined observable for measurements of the two spins $S^1_z$ and $S^2_z$. It has the value $+1$ whenever the signs of the two spins are found to be equal, and $-1$ if they are opposite. This observable is represented by $L_3=$ diag $(1,-1,-1,1)$. Consider an ensemble characterized by $\rho_3=-1~,~\rho_1=\rho_2=0$. For this ensemble one has $\langle A^{(1)}\rangle=\langle A^{(2)}\rangle =0$ such that for both two-level-observables the values $+1$ and $-1$ are randomly distributed in the ensemble. Nevertheless, $\langle A^{(3)}\rangle=-1$ indicates that the two ``spins'' are maximally anticorrelated. Whenever the first spin takes the value $+1$, the second one necessarily assumes $-1$ and vice versa. In sect. \ref{Entanglement} we discuss that pure states of this type show the characteristics of an entangled quantum state. Our third example considers the observable $S$ corresponding to the sum $\hat S=L_1+L_2$. For the particular pure state density matrices $(\hat\rho_m)_{\alpha\beta}=\delta_{m\alpha}\delta_{m\beta}$ one has $\langle S\rangle=2$ (for $m=1$), $\langle S\rangle=0$ (for $m=2,3)$ and $\langle S\rangle=-2$ (for $m= 4)$. Thus $S$ has the properties of a total spin, composed of two half integer spins (say $S_z=S^1_z+S^2_z)$. 

\medskip\noindent
{\bf Pure states and wave function}

The general form $\rho=U\hat\rho_mU^\dagger$ allows us to ``take the root'' of a pure state density matrix by introducing the quantum mechanical wave function $\psi_\alpha$ as an $M$-component complex normalized vector, $\psi^\dagger\psi=1$,
\begin{eqnarray}\label{6}
\rho_{\alpha\beta}&=&\psi_\alpha\psi^*_\beta~,~\psi_\alpha=
U_{\alpha\beta}(\hat\psi_m)_\beta~,~\nonumber\\
(\hat\psi_m)_\beta&=&\delta_{m\beta}~,~\langle A\rangle=\psi^\dagger\hat A\psi.
\end{eqnarray}
All the usual rules for expectation values in quantum mechanical pure states apply. 

Pure states play a special role since they describe classical ensembles with minimal uncertainty for a given integer $M$.  For $M=4$ a pure state has purity $P=3$ and three different observables can have sharp values, corresponding to the maximum number of three commuting quantum mechanical operators. For $M>2,P=M-1$ and $\{L_k,L_l\}=2\delta_{kl}+2d_{klm}L_m$, the condition for a pure state, $\rho^2=\rho$ or $\rho_k[\rho_l d_{klm}-(M-2)\delta_{km}]=0$, is not automatically obeyed for all $\rho_k$ with $\rho_k\rho_k=P$. Pure states have to obey additional restrictions.

For a pure state, the ``copurity'' 
\begin{equation}\label{6a}
C= tr[(\rho^2-\rho)^2]
\end{equation}
must also vanish. While the purity $P$ is conserved by all orthogonal $SO(n)$ transformations of the vector $(\rho_k)$, pure states are transformed into pure states only by the subgroup of $SU(M)$ transformations. The $SU(M)$ transformations are realized as unitary transformations of the wave function $\psi$, where the overall phase of $\psi$ remains unobservable since it does not affect $\rho$ in eq. \eqref{6}. In our classical statistical description of quantum phenomena, the particular role of the classical pure states constitutes the basic origin for the unitary transformations in quantum mechanics. Just as in quantum mechanics, we can write the density matrix $\rho$ for an arbitrary ensemble as a linear combination of appropriate pure state density matrices. 

The pure states form the manifold $SU(M)/SU(M-1)\times U(1)$ \cite{CW1}. This homogeneous space forms a submanifold of $S^{M^2-2}$. In other words, the copurity constraint imposes on the $M^2-2$ independent real numbers $\rho_k=f_k$, $f_kf_k=M-1$, a number of $M(M-2)$ additional constraints, corresponding to the $2(M-1)$-dimensional space $SU(M)/SU(M-1)\times U(1)$. A simple way to realize these constraints is the condition that the allowed $f_k$ have to obey
\begin{equation}\label{54A}
f_k=\psi^\dagger L_k\psi~,~\psi^\dagger \psi=1
\end{equation}
for some arbitrary normalized wave function $\psi$ \cite{CW1}. 

A linear combination of two complex wave functions
\begin{equation}\label{56B}
\psi=c_1\psi^{(1)}+c_2\psi^{(2)}
\end{equation}
defines again a possible pure state if the coefficients $c_1$ and $c_2$ obey the restriction that $\psi$ is normalized, $\psi^\dagger\psi=1$. There exists a pure state density matrix \eqref{6} for every possible pure state. A classical ensemble, for which the probability distributions can realize arbitrary density matrices, can therefore also realize arbitrary (normalized) superpositions of two wave functions. This allows us to describe all the interference phenomena characteristic for quantum mechanics in terms of appropriate classical ensembles \cite{CW1}. 

\medskip\noindent
{\bf Mixed states}

A mixed quantum state has purity $P<M-1$. For example, the elements $\rho_k$ may obey
$\rho_k=\left({P}/(M-1)\right)^{1/2}f_k$. The general positivity condition for a mixed quantum state can be formulated by using the possibility to diagonalize any hermitean matrix $\rho$ by a unitary transformation
\begin{equation}\label{56Aa}
\rho=Udiag(p_\alpha)U^\dagger.
\end{equation}
In addition to $P\leq M-1$ we require
\begin{equation}\label{56Ba}
p_\alpha\geq 0.
\end{equation}
In other words, classical statistical ensembles can describe $M$-state quantum systems, if the expectation values $\rho_k$ obey the purity constraint
\begin{equation}\label{56A}
\sum_k\rho_k\rho_k=P=M\sum_\alpha p^2_\alpha-1\leq M-1,
\end{equation}
and further obey the positivity constraint \eqref{56Aa}, \eqref{56Ba}. We will often refer to the combination of eqs. \eqref{3a}, \eqref{56Aa}-\eqref{56A} collectively as the ``purity constraint'', as mentioned in the abstract. Actually, the purity constraint \eqref{56A} follows from the positivity \eqref{56Ba} and the normalization $\sum_\alpha p_\alpha=1$ - the generalized purity constraint \eqref{56Aa}-\eqref{56A} is therefore equivalent to the positivity constraint for the density matrix. The use of the wording ``purity constraint'' is motivated by the basic observation that typical quantum systems are characterized by a lower bound for the fluctuations and therefore an upper bound for the purity. The positivity of $\rho$ \eqref{56Ba} will be crucial for many steps in our construction. For example, together with the normalization of $\rho$ it guarantees the important inequality for the diagonal elements of $\rho$
\begin{equation}\label{61A}
0\leq \rho_{\alpha\alpha}\leq 1
\end{equation}
which follows directly from the properties of the unitary matrices in eq. \eqref{56Aa}
\begin{equation}\label{61B}
\rho_{\alpha\alpha}=\sum_\beta p_\beta|U_{\alpha\beta}|^2.
\end{equation}

\medskip
\noindent
{\bf Probabilistic quantum observables}

The vector $e^{(A)}$ is sufficient for a determination of the expectation value of a system observable $A$ in any state of the subsystem (characterized by $\rho_k$). However, the typical observables for the subsystem are probabilistic observables and we further have to specify the probability distribution for the possible outcomes of measurements for every state $\{\rho_k\}$. The probabilities $w_a(\rho_k)$ are needed for a computation of expectation values $\langle A^p\rangle$ of powers of $A$. Inversely, knowledge of all $\langle A^p\rangle (\rho_k)$ allows the reconstruction of the probability distribution $w_a(\rho_k)$. We will next concentrate on ``quantum observables''. They constitute a subclass of the system observables for which the operators associated to $A^p$ are given by $\hat A^p$. Since a quantum operator $\hat A$ is uniquely characterized by the vector $e^{(A)}$, we can also express $\hat A^p$ in terms of $e^{(A)}$. Thus for a quantum observable the vector $e^{(A)}$ is sufficient for the computation of $\langle A^p\rangle$. It specifies the quantum observable uniquely.

Not all system observables are quantum observables. As an example for a system observable that is not a quantum observable we may take the random two level observable $R$ with spectrum $\gamma_\alpha=\pm 1$ and $w_+=w_-=1/2$ for every state $(\rho_k)$ of the subsystem. It obeys $\langle R\rangle=0~,~\langle R^2\rangle=1$ for all $\rho_k$. From $\langle R\rangle=0$ we infer $e^{(R)}=0,~\hat R=0$. However, $\langle R^2\rangle$ differs from tr$(\rho\hat R^2)=0$. The operator associated to $R^2$ is the unit operator and not the zero operator. We infer that for $p\geq 2$ the property
\begin{equation}\label{61C}
\langle A^p\rangle=\text{tr}(\rho\hat A^p)
\end{equation}
is not generic for all system observables. We can therefore define in a quantum system probabilistic observables for which all expectation values $\langle A^p\rangle$ are computable in terms of the state $(\rho_k)$, while the operator relation \eqref{61C} does not hold. They only play a minor role in quantum systems and we will concentrate in the following on the subclass of quantum observables for which the relation \eqref{61C} holds. 

For a given $M$ we consider observables with a spectrum of at most $M$ different values $\gamma_a$. We first concentrate on a non-degenerate spectrum of $M$ different $\gamma_a$ and identify $a=\alpha$. The probabilities to find $\gamma_\alpha$ in the state $\rho_k$ of the subsystem are denoted by $w_\alpha(\rho_k)\geq 0~,~\sum_\alpha w_\alpha(\rho_k)=1~$, $\langle A\rangle=\sum_\alpha w_\alpha(\rho_k)\gamma_\alpha=\rho_ke_k$. The expectation value $\langle A\rangle=\textup{tr}(\rho\hat A)$ is invariant under a change of basis by unitary transformations, $\rho\to \rho'=U\rho U^\dagger~,~\hat A\to \hat A'=U\hat A U^\dagger$. We may choose a basis with diagonal $\hat A'=diag(\lambda_1,\dots,\lambda_M)~,~\langle A\rangle=\sum_\alpha\rho'_{\alpha\alpha}\lambda_\alpha$, suggesting that the spectrum $\gamma_\alpha$ can be identified with the eigenvalues $\lambda_\alpha$ of the operator $\hat A$, and $w_\alpha(\rho_k)=\rho'_{\alpha\alpha}$. 

Among the system observables we therefore consider the subclass of quantum observables the spectrum  $\{\gamma_\alpha\}$ and probabilities
\begin{equation}\label{8A}
w_\alpha(\rho_k)=\rho'_{\alpha\alpha}
=(U_A\rho U^\dagger_A)_{\alpha\alpha}.
\end{equation}
The purity constraint for $\rho_k$ guarantees $0\leq w_\alpha(\rho_k)\leq 1$. For a given observable $A$ the unitary matrix $U_A$ is fixed. Thus $A$ is characterized by $\gamma_\alpha$ and $U_A$, with expectation value 
\begin{equation}\label{62A}
\langle A\rangle=\sum_\alpha w_\alpha(\rho_k)\gamma_\alpha=\text{tr}(\rho\hat A).
\end{equation}
To each quantum observable, we can associate a hermitean quantum operator, with a spectrum of possible measurement values given by the eigenvalues $\lambda_\alpha=\gamma_\alpha$ of the operator. The classical probability for the outcome of the measurement in a given state is the corresponding diagonal element of the density matrix in a basis where $\hat A$ is diagonal. We describe the mathematical structures related to quantum observables in more detail in sect. \ref{Propertiesofquantum}, where we also deal with degenerate spectra of less than $M$ different $\gamma_a$. There we will construct an  explicit classical realization for the quantum observables.  

We can now discuss powers of the probabilistic observable $A,~\langle A^p\rangle=\sum_\alpha w_\alpha(\rho_k)\gamma^p_\alpha$. The observable $A^p$ should belong to the observables of the subsystem, since it can be associated with $p$ measurements of $A$, multiplying the $p$ measurement results that must be identical. We can therefore associate an operator $\tilde A_p$ to the observable $A^p~,~\langle A^p\rangle=\textup{tr}(\rho\tilde A_p)$. For quantum observables this is realized by $\tilde A_p=\hat A^p$ and we conclude $\langle A^p\rangle=\textup{tr}(\rho\hat A^p)$. For $\langle A^2\rangle=$tr$(\rho\hat A^2)$ a classical eigenstate of $A$ obeys [tr$(\rho\hat A)]^2=$tr$(\rho\hat A^2)$. The possible classical eigenvalues are the eigenvalues of the operator $\hat A$. If a pure state is an eigenstate of $A$ one has $\hat A\psi=\lambda\psi$ with $\lambda\equiv \lambda_\alpha=\gamma_\alpha$ one of the eigenvalues of $\hat A$. 

\medskip\noindent
{\bf Classical quantum observables}

Consider now the classical ensemble of states $\tau$ which describe the subsystem together with its environment. It is straightforward to characterize the properties of the classical observables $A_\tau$ that are mapped to the probabilistic quantum observables and the corresponding quantum operators $\hat A$. First, the spectrum of possible outcomes of individual measurements equals the (sharp) values of the classical observable  $A_\tau$ in the classical states $\tau$. It consists of the eigenvalues of $\hat A$. Second, for all allowed probability distributions $\{p_\tau\}$ for the states  $\tau$ it must be possible to express $w_\alpha$ as a linear combination of $\rho_k=\langle A^{(k)}\rangle$ according to eq. \eqref{8A}. More precisely, these statements have to hold for all states obeying eqs. \eqref{56Aa}, \eqref{56Ba} with $P\leq M-1$. Classical observables $A_\tau$ with these properties are called classical quantum observables.

At first sight the second requirement for a classical observable to be a quantum observable may seem rather special. However, in many circumstances the quantum observables are related to the basis observables by simple ``physical operations''. For example, the spin observable in an arbitrary direction obtains from the three basis observables $A^{(k)}(M=2)$ by rotation, such that the relation \eqref{8A} arises naturally \cite{CW2}. Other simple operations are the addition and multiplication of ``commuting observables'', as we will explain in sect. \ref{Propertiesofquantum}. At this point it may be worthwhile to pause. We have selected a set of classical observables with a remarkable property: whenever the ensemble obeys a simple purity constraint for the expectation values of some basis observables, all the laws of quantum mechanics apply for these classical observables, as for example Heisenberg's uncertainty relation based on the commutator of the associated operators. 

In summary, we define the quantum observables as a subclass of the classical observables with a spectrum of at most $M$ different possible measurement values $\gamma_a$. Furthermore, the associated probabilities $w_a$ should be given by the relation \eqref{8A} which is linear in $\rho_k$, with coefficients depending only on the observable. This implies the relation \eqref{2} and defines $e^{(A)}_k$. (We may add a piece proportional to the  unit observable.) It is then also possible to compute the expectation values $\langle A^p\rangle$ in terms of the information contained in the quantum state through an expression linear in the $\rho_k$. Therefore $\langle A^p\rangle$ does not involve any properties of the environment - these expectation values are completely determined by the subsystem. To any quantum observable we can associate a unique vector $(e_k)$ and therefore a unique quantum operator. In turn, to each hermitean quantum operator we can also associate a unique probabilistic quantum observable, with $w_\alpha$ given by eq. \eqref{8A}. On the level of probabilistic observables the quantum observables can therefore be fully characterized by the vector $(e_k)$. The map from the classical quantum observables to the vectors $e^{(A)}$ is not invertible, however.

\medskip\noindent
{\bf Quantum product of quantum observables}

The correspondence between probabilistic observables and hermitean quantum operators allows for the introduction of a symmetric ``quantum product'' $(AB)_s=(BA)_s$ between two probabilistic quantum observables, which is associated to the operator product $\frac12\{\hat A,\hat B\}$. Similarly, we can now define linear combinations of quantum observables. The sum $A+B$ is the quantum observable which is associated to the sum of operators $\hat A+\hat B$. Together with linear combinations, the quantum product $(AB)_s$ defines an algebra for the probabilistic quantum observables. The algebra of probabilistic quantum observables is isomorphic to the algebra of quantum operators with product $\frac12\{\hat A,\hat B\}$. For both, it can be expressed as operations in the space of $(e_k)$. In particular, we observe $A^2=(A A)_s$. We will see in sect. \ref{Correlations} that a particular use of the ``quantum product'' $(A B)_s$ arises from an investigation of conditional probabilities. 

In general, a probabilistic quantum observable $A$ describes an equivalence class of classical observables that all lead to the same quantum operator $\hat A$. The product $\frac12\{\hat A,\hat B\}$ then induces a product structure between equivalence classes. If a representative for each class is selected one can also define $(AB)_s$ on the level of classical observables \cite{3}. We emphasize that the product $(AB)_s$ is {\em not} the classical or pointwise product $A\cdot B$ where $(A\cdot B)_\tau=A_\tau B_\tau$.

\medskip\noindent
{\bf Statistical interpretation of quantum mechanics}

We obtain the laws of quantum mechanics from a classical statistical ensemble. Therefore the postulates concerning the relation between the quantum calculus and statistical statements about the outcome of observations can be derived from the probability concept for a classical statistical ensemble. Consider Born's rule for a pure state wave function $\psi_\alpha$. Let us assume that the basis vectors $(\hat\psi_m)_\beta=\delta_{m\beta}$ denote eigenstates of an observable $A$ with eigenvalues $\gamma_m$, i.e. $\hat A\hat\psi_m=\gamma_m\hat\psi_m$. We further assume that the spectrum is non-degenerate. Then Born's rule states that the probability for finding in a measurement of $A$ the value $\gamma_\alpha$ is given by the absolute square of the corresponding component of the wave function, $w_\alpha=|\psi_\alpha^2|$. 

This statement is one of the axioms of quantum mechanics. It can be easily derived for our implementation of quantum systems as classical statistical ensembles. Since $\hat A=diag(\gamma_\alpha)$ is diagonal in the chosen basis we find for the quantum observable associated to $\hat A$ that the expectation values of arbitrary powers of $A$ obey
\begin{equation}\label{57A}
\langle A^p\rangle=\text{tr}(\hat A^p\rho)=\sum_\alpha(\gamma_\alpha)^p\rho_{\alpha\alpha}.
\end{equation}
The classical statistical rule for expectation values then implies 
\begin{equation}\label{57B}
w_\alpha=\rho_{\alpha\alpha}.
\end{equation}
In turn, for a pure state density matrix the diagonal elements obey
\begin{equation}\label{57C}
\rho_{\alpha\alpha}=|\psi_\alpha|^2,
\end{equation}
thus establishing Born's rule. The statistical interpretation of a mixed state density matrix follows along the same lines, since eqs. \eqref{57A}, \eqref{57B} do not need the assumption of a pure state. The basic ingredients are the statistical interpretation of $\langle A^p\rangle$ as an expectation value of an observable in a classical statistical ensemble, and the relation for quantum observables $\langle A^p\rangle=\text{tr}(\hat A^p\rho)$. Born's rule follows for all classical quantum observables. The generalization for the case of a degenerate spectrum implies sums over all $\alpha$ for which the eigenvalue in the corresponding basis state has the same value $\gamma_\alpha$. 

In summary of this section we have considered subsystems of a classical statistical ensemble whose state is characterized by the expectation values $\rho_k$ of basis observables. If the $\rho_k$ obey a purity constraint they define a quantum density matrix $\rho$ \eqref{3a}. The positivity of $\rho$ is equivalent to the purity constraint. We have discussed system observables for which $\langle A^p\rangle$ is computable in terms of $(\rho_k)$, and concentrated on the subclass of quantum observables for which eq. \eqref{61C} holds. For quantum observables all the rules of quantum mechanics for the outcome of measurements, as Born's rule or Heisenberg's uncertainty relation, can be inferred from the probabilities of the classical statistical ensemble. 

\section{Simple quantum systems}
\label{Simplequantum}

In this section we present explicit examples which demonstrate the general considerations of the preceding section. We discuss classical ensembles that realize all aspects of quantum-operators and their expectation values for two-state and four-state quantum mechanics. These ensembles involve infinitely many classical states, while the subsystems are characterized by only a few real numbers. We give explicitly the classical observables that are represented by quantum operators and comment on the associated equivalence classes of classical observables that are mapped to the same probabilistic observables and quantum operators. The issue of correlations for measurements of pairs of such observables will be addressed in the next section. 

\medskip
\noindent
{\bf Two-state quantum mechanics}

As an instructive example we may consider a classical ensemble corresponding to two-state quantum mechanics \cite{CW2}. For this purpose we generalize the setting of section \ref{Classicalquantumcomputer} by considering spins in arbitrary directions. In sect. \ref{Classicalquantumcomputer} the three cartesian spins $A^{(1)},A^{(2)},A^{(3)}$ are realized on the classical level by the observables $\sigma_1,\sigma_2,\sigma_3$. We now denote the direction of the spin by a unit vector $(e_1,e_2,e_3),e_ke_k=1$, such that $A(e_k)$ is again a two level observable, $A^2(e_k)=1$, and $A^{(1)}=A(1,0,0)~,~A^{(2)}=A(0,1,0)~,~A^{(3)}=A(0,0,1)$. Next we choose for every angular direction on $S^2/Z_2$ (denoted now by $g_k$) a discrete variable $\gamma(g_k)=\pm 1~,~g_kg_k=1$, with special cases $\sigma_1=\gamma(1,0,0)~,~\sigma_2=\gamma(0,1,0)~,~\sigma_3=\gamma(0,0,1)$ treated in sect. \ref{Classicalquantumcomputer}. Here it is sufficient to have a variable $\gamma$ for every point on the half-sphere $S^2/Z_2$ since we can identify $\gamma(-g_k)=-\gamma(g_k)$. We may consider the infinitely many classical bits $\gamma(g_k)$ as the limit of a sequence where an increasing number of points on $S^2/Z_2$ is included. The classical probability distribution depends now on sequences of an infinite number of classical bits $\gamma(g_k)~,~\tau=\big\{\gamma(g_k)\big\}$ , $p_\tau=p\Big(\big\{\gamma(g_k)\big\}\Big)$. We again use the form $(3)$
\begin{equation}\label{V1}
p\Big(\big\{\gamma(g_k)\big\}\Big)=p_s\Big(\big\{\gamma(g_k)\big\}\Big)+\delta p_e\Big(\big\{\gamma(g_k)\big\}\Big)
\end{equation}
with 
\begin{equation}\label{V2}
p_s\Big(\{\gamma(g_k)\big\}\Big)=\prod_{g_k}
\left[\frac12 \big(1+\rho_kg_k\gamma(g_k)\big)\right]
\end{equation}
and 
\begin{equation}\label{V3}
\sum_{\big\{\gamma(g_k)\big\}}p_s\Big(\big\{\gamma(g_k)\big\}\Big)=1~,~
\sum_{\big\{\gamma(g_k)\big\}}=\prod_{g_k}\sum_{\gamma(g_k)=\pm 1}.
\end{equation}

The classical representation of the spin in the direction $e_k$ involves precisely one of the discrete variables
\begin{equation}\label{V4}
A(e_k)=A(e_k)\Big(\big\{\gamma(g_k)\big\}\Big)=\gamma(e_k).
\end{equation}
In other words, the observable $A(e_k)$ ''reads out'' the classical bit with $g_k=\pm e_k$. 
Its expectation value is given by
\begin{equation}\label{V5}
\langle A(e_k)\rangle=\sum_{\big\{\gamma(g_k)\big\}}
\gamma(e_k)p\Big(\big\{\gamma(g_k)\big\}\Big)=\rho_ke_k,
\end{equation}
where we assume as in sect. \ref{Classicalquantumcomputer}
\begin{equation}\label{V6}
\sum_{\big\{\gamma(g_k)\big\}}\delta p_e=0~,~\sum_{\big\{\gamma(g_k)\big\}}
\gamma(e_k)\delta p_e=0.
\end{equation}
Correspondingly, the probability to find in the ensemble the value $+1$ for the observable $A(e_k)$ reads
\begin{equation}\label{V7}
w_+(e_k)=\frac12\big(1+\langle A(e_k)\rangle\big)=\frac12(1+\rho_ke_k).
\end{equation}
It can be expressed in terms of the vector $(\rho_k)$ which characterizes the state of the subsystem and corresponds to the expectation values of the three basis observables
\begin{equation}\label{V8}
\rho_1=\langle A(1,0,0)\rangle~,~\rho_2=\langle A(0,1,0)\rangle~,~\rho_3=
\langle A(0,0,1)\rangle.
\end{equation}
The probabilities $w_+(e_k)$ (with $w_-(e_k)=1-w_+(e_k)$) characterize the probabilistic two-level observables $A(e_k)$. For $P=\rho_k\rho_k\leq 1$ this classical statistical ensemble realizes two-state quantum mechanics.

\medskip\noindent
{\bf Four-state quantum mechanics}

We next generalize the ensemble in order to describe four-state quantum mechanics. As long as only two-level observables are concerned this is straightforward. Instead of three basis observables we now consider $15$ two-level basis observables $A^{(m)}~,~m=1\dots 15$, whose expectation values characterize the quantum state
\begin{equation}\label{V9}
\langle A^{(m)}\rangle=\langle A(\hat e^{(m)}_k)\rangle=\rho_m.
\end{equation}
The associated quantum operators are the fifteen $SU(4)$ generators $L$ given in eq. \eqref{FXA}. The directions of general two-level observables can again be associated with a fifteen-component vector $(e_k)$, with basis vectors $\hat e^{(m)}_k=\delta^m_k$. We impose the restrictions
\begin{equation}\label{V10}
e_ke_k=1~,~e_ke_md_{kml}=0,
\end{equation}
with $d_{kml}=d_{mkl}$ the symmetric ``structure constants'' in eq. \eqref{3b} which obey for every $k~d_{kkl}=0$. The manifold parameterized by $(e_k)$ with the condition \eqref{V10} is the homogenous space $SU(4)/SU(3)\times U(1)$. 

For every point $\bar g_k$ of $SU(4)/SU(3)\times U(1)\times Z_2$ we introduce again a discrete variable $\gamma(\bar g_k)~,~\gamma^2(\bar g_k)=1$, $\gamma(-\bar g_k)=-\gamma(\bar g_k)$. The split of an arbitrary classical probability distribution describing an isolated subsystem is done similarly as before
\begin{equation}\label{V11}
p\Big(\big\{\gamma(\bar g_k)\big\},\zeta\Big)=p_s\Big(\big\{\gamma(\bar g_k)\big\}\Big)\bar p_s(\zeta)+\delta p_e
\Big(\big\{\gamma(\bar g_k)\big\},\zeta\Big)
\end{equation}
with $\zeta$ a further collective variable characterizing additional degrees of freedom of the classical states. With $p_s$ given again by eq. \eqref{V2}, $\sum_\zeta\bar p_s(\zeta)=1$,
\begin{equation}\label{V12}
\sum_\tau\delta p_e=0~,~\sum_\tau\gamma(e_k)\delta p_e=0,
\end{equation}
we find for all two level observables $A(e_k)$
\begin{equation}\label{V13}
\langle A(e_k)\rangle=\rho_ke_k=\text{tr}(\rho\hat A).
\end{equation}
The last part holds for $P=\rho_k\rho_k\leq 3$ and $\rho$ the associated density matrix \eqref{3a}, provided we represent $A(e_k)$ by the operator $\hat A=e_kL_k$. We observe that $\hat A^2=1$ holds only if the condition \eqref{V10} is obeyed. This condition states that instead of arbitrary $SO(15)$ - rotations of the basis observables $A^{(m)}$ we only consider $SU(4)$ transformations thereof. 

Four state quantum mechanics admits probabilistic observables with a spectrum with up to four different values $\gamma_a$. The associated classical statistical ensemble should describe this situation. In particular, the part of the classical probability distribution $\{p_\tau\}$ which describes the subsystem should contain the information about the probabilities $w_a(\rho_k)$ to find the measurement values $\gamma_a$ for observables with three or four different values $\gamma_a$. This is the role of $\bar p_s(\zeta)$ in eq. \eqref{V11}. The description of observables with three or four different values in the spectrum will be associated to quantum operators $\hat A=e_kL_k$ with values of $(e_k)$ not obeying the second restriction in eq. \eqref{V10}. 

We can restrict the discussion to observables for which the spectrum $(\gamma_1,\gamma_2,\gamma_3,\gamma_4)$ obeys
\begin{equation}\label{M1}
\sum_\alpha\gamma_\alpha=0~,~\sum_\alpha\gamma^2_\alpha=4.
\end{equation}
Observables with a different normalization can be obtained by multiplicative scaling $A\to\lambda A$, and the first condition in eq. \eqref{M1} can always be achieved by subtracting an appropriate part proportional to the unit observable. If the state of the subsystem can be described by $(\rho_k)$ with $\rho_k\rho_k\leq 3$ and therefore by an associated density matrix $\rho$, and if the probabilistic observable $A$ can be associated to a quantum operator $\hat A$ such that 
$\langle A\rangle=\text{tr}(\rho\hat A)$, the conditions \eqref{M1} read
\begin{equation}\label{M2}
\text{tr}\hat A=0~,~\text{tr}\hat A^2=4.
\end{equation}
This is obeyed for $\hat A=e_kL_k$ if $(e_k)$ is a unit vector on $S^{15},e_ke_k=1$. The conditions \eqref{M1} fix $\gamma_3$ and $\gamma_4$ as a function of $\gamma_1$ and $\gamma_2$, such that the spectrum of normalized observables can be characterized by two parameters $(\gamma_1,\gamma_2)$, while
\begin{eqnarray}\label{M3}
\gamma_3+\gamma_4&=&-(\gamma_1+\gamma_2),\nonumber\\
\gamma^2_3+\gamma^2_4&=&4-\gamma^2_1-\gamma^2_2,\nonumber\\
\gamma_3-\gamma_4&=&\pm \sqrt{8-3\gamma^2_1-3\gamma^2_2-2\gamma_1\gamma_2}.
\end{eqnarray}
This determines $\gamma_3$ and $\gamma_4$ uniquely up to an exchange $\gamma_3\leftrightarrow\gamma_4$. We observe the restriction
\begin{equation}\label{M4}
\gamma^2_1+\gamma^2_2+\frac12(\gamma_1+\gamma_2)^2\leq 4.
\end{equation}

The two level observables correspond to the special case where $\gamma_1=\pm 1~,~\gamma_2=\pm 1$. They obey the second condition \eqref{V10} for $(e_k)$, which defines a submanifold of $S^{15}$. This submanifold is the homogeneous space $SU(4)/SU(2)\times SU(2)$, as can be seen by the following argument. We can characterize the $(e_k)$ obeying eq. \eqref{V10} by the properties of the associated operator $\hat A=e_kL_k$ as tr$\hat A=0~,~\hat A^2=1~,~\hat A^\dagger=\hat A$. The spectrum of a hermitean matrix remains conserved by unitary $SU(4)$ transformations $\hat A\to U\hat AU^\dagger$. On the other hand, for any pair $(\hat A, \hat B)$ of hermitean matrices with identical spectrum there exists a unitary transformation such that $\hat B=U\hat AU^\dagger$. (The spectrum defines the orbits of the $SU(4)$ transformations.) We can therefore obtain $\hat A$ by a unitary transformation from the generator $L_1,\hat A=UL_1U^\dagger$. Since $L_1$ is invariant under the subgroup $SU(2)\times SU(2)$ we can use the transformations in $SU(4)/SU(2)\times SU(2)$ in order to describe the two-level operators $\hat A$ and the associated $(e_k)$. 

Let us next specify a classical statistical ensemble that can describe four-state quantum mechanics. Consider unit vectors $(g_k)$ on $S^{15}~,~g_kg_k=1~,~k=1\dots 15$. To every $g_k$ we associate a discrete label $\alpha(g_k)$ which can take the value $\alpha(g_k)=1,2,3,4$. The classical states are sequences of these discrete variables
\begin{equation}\label{M5}
\tau=\big\{\alpha(g_k)\big\}~,~\sum_\tau=\prod_{g_k}\sum^4_{\alpha(g_k)=1}.
\end{equation}
This construction generalizes the one for $M=2$: instead of $(g_k)\in S^3$ we now have $(g_k)\in S^{15}$, and the discrete variable $\gamma(g_k)$, which can take two values for $M=2$, is replaced by $\alpha(g_k)$ which can assume four discrete values. The classical probability distribution is again written in the form
\begin{eqnarray}\label{M6}
p_\tau&=&(p_s)_\tau+(\delta p_e)_\tau,\nonumber\\
(p_s)_\tau&=&\prod_{g_k}w_\alpha(\rho_k,g_k),
\end{eqnarray}
which $(\delta p_e)_\tau$ characterizes the environment and $(p_s)_\tau$ the system. The probabilities $w_\alpha\geq 0$ obey $\sum_\alpha w_\alpha=1$ and are defined as follows. We diagonalize the hermitean matrix $G=g_kL_k$ by a unitary transformation
\begin{equation}\label{M7}
G=g_kL_k~,~\tilde G=U(g_k)GU^\dagger(g_k)=diag\big(\lambda_\alpha(g_k)\big),
\end{equation}
where $U(g_k)$ is specified such that $\lambda_1\geq \lambda_2\geq \lambda_3\geq \lambda_4$. From the density matrix $\rho=(1+\rho_kL_k)/4$ we obtain
\begin{equation}\label{M8}
w_\alpha(\rho_k,g_k)=\big[U(g_k)\rho(\rho_k)U^\dagger(g_k)\big]_{\alpha\alpha}.
\end{equation}
This construction implies $\sum_\tau(p_s)_\tau=1$ such that $\delta p_e$ has to obey
\begin{equation}\label{M9}
\sum_\tau(\delta p_e)_\tau=0.
\end{equation}

A classical realization of quantum observables $A(e_k)$ can be implemented by
\begin{equation}\label{M10}
\big(A(e_k)\big)_\tau=\gamma_{\alpha(g_k=e_k)}.
\end{equation}
Here $\gamma_{\alpha(g_k=e_k)}$ is defined by $\hat A=e_kL_k,e_ke_k=1$, and 
\begin{equation}\label{M11}
\gamma_\alpha=\big[U(g_k)\hat AU^\dagger(g_k)\big]_{\alpha\alpha}.
\end{equation}
In other words, the observable $A(e_k)$ again ``reads out'' from the sequence $\big\{\alpha(g_k)\big\}$ the element for $g_k=e_k$. Then $\hat A(e_k)=G(g_k)$, and therefore $U(g_k)\hat AU^\dagger(g_k)$ is diagonal, such that $\gamma_\alpha$ equals the eigenvalue $\lambda_\alpha(g_k)$ of $\tilde G$ in eq. \eqref{M7}. In all classical states $\tau$ the observable $A(e_k)$ has a fixed value which is one of the eigenvalues of the associated operator $\hat A$. 

The sequence $\big\{\alpha(g_k)\big\}$ may be constructed by adding consecutively different ``angles'' on $S^{15}$. For illustration we may consider a reduced system where only three different angles $(g^{(m)}_k)$ are included, such that $\alpha_m=\alpha(g^{(m)}_k)$ and the $4^3$ classical states $\tau$ obey
\begin{equation}\label{M12}
\tau=(\alpha_1,\alpha_2,\alpha_3)~,~(p_s)_\tau=w_{\alpha_1}w_{\alpha_2}w_{\alpha_3}.
\end{equation}
In this system we may define three classical quantum observables
\begin{equation}\label{M13}
A^{(m)}=A(e_k=g^{(m)}_k)~,~A^{(m)}_\tau=\lambda^{(m)}_{\alpha_m},
\end{equation}
with $\lambda^{(m)}$ an eigenvalue of $G^{(m)}=g^{(m)}_kL_k$. 

The two level observables obtain as a special case for those $e_k$ for which $\gamma_{\alpha(e_k)=1}=\gamma_{\alpha(e_k)=2}=1$ , $\gamma_{\alpha(e_k)=3}=\gamma_{\alpha(e_k)=4}=-1$. The associated $g_k$ are those for which $\tilde G=diag(1,1,-1,-1)=L_1$. We infer that the additional index $\zeta$ in eq. \eqref{V11} corresponds to the sequence $\big\{\alpha(\tilde g_k)\big\}$ for those $\tilde g_k$ for which $d_{klm}\tilde g_k\tilde g_l\neq 0$. In other words, we can write the sequence $\big\{\alpha(g_k)\big\}=\big\{\gamma(\bar g_k),\alpha(\tilde g_k)\big\}$. Here we observe that those $g_k$ for which the spectrum of $G(g_k)$ consists only of $\pm 1$ need only two classical possibilities $\alpha(g_k)=1~,~\alpha(g_k)=2$ corresponding to $\gamma=1$ and $\gamma=-1$. (Similarly, if the spectrum of $G(g_k)$ has only three distinct values, three values for $\alpha(g_k)$ are sufficient.)

For the part of the classical probability distribution characterizing the environment we impose in addition to eq. \eqref{M9} the conditions
\begin{equation}\label{M14}
\sum_\tau\big(A(e_k)\big)^p_\tau(\delta p_e)_\tau=0
\end{equation}
for $p=1,2,3$. We can then evaluate the expectation values
\begin{equation}\label{M15}
\langle\big(A(e_k)\big)^p\rangle=\sum_\tau\big(A(e_k)\big)^p_\tau(p_s)_\tau
=\sum_\alpha(\gamma_\alpha)^pw_\alpha(\rho_k,e_k),
\end{equation}
where $(\gamma_1,\gamma_2,\gamma_3,\gamma_4)$ is the ordered spectrum of eigenvalues of $\hat A=e_kL_k$. The last expression in eq. \eqref{M15} can be interpreted as
\begin{equation}\label{M16}
\langle\big(A(e_k)\big)^p\rangle=\text{tr}(\rho\hat A^p)
\end{equation}
in a basis where $\hat A$ is diagonal with ordered eigenvalues. The trace is invariant under a change of basis by unitary transformations and we have therefore realized the quantum rule for the computation of expectation values of powers of $A(e_k)$. In particular, one finds 
\begin{equation}\label{M16A}
\langle A(e_k)\rangle=\text{tr}\big(\rho\hat A(e_k)\big)=\rho_ke_k.
\end{equation}
The discussion shows that $\big(A(e_k)\big)_\tau$ describes a classical quantum observable. It can be mapped to a probabilistic quantum observable with spectrum $\big\{\gamma_\alpha(e_k)\big\}$ and associated probabilities $w_\alpha(\rho_k,e_k)$. 

We emphasize that the four conditions \eqref{M9} \eqref{M14} for $p=0,1,2,3$ are actually sufficient in order to ensure eqs. \eqref{M15} \eqref{M16} for arbitrary $p$. The reason is that $\big(A(e_k)\big)_\tau$ can take at most four different classical value $\gamma_\alpha(e_k)$. We can define the probabilities of their occurrence $w_\alpha$ by the sum of the classical probabilities for all states $\tau$ for which $A_\tau=\gamma_\alpha(e_k)$
\begin{equation}\label{M17}
w_\alpha(\rho_k,e_k)=\sum_\tau p_{\tau|A_\tau=\gamma_\alpha(e_k)}.
\end{equation}
Eq. \eqref{M15} for $p=1,2,3$ fixes $w_\alpha(\rho_k,e_k)$ uniquely for any given $\rho_k$ and $e_k$. Then arbitrary functions $f(\big(A(e_k)\big)$ must have expectation values given by
\begin{equation}\label{M18}
\langle f\big(A(e_k)\big)\rangle=\text{tr}(\rho f\big(\hat A(e_k)\big).
\end{equation}

\medskip\noindent
{\bf Equivalence classes of probabilistic observables}

Two classical observables $A_\tau,B_\tau$ are equivalent from the point of view of the subsystem  if they have even formally the same spectrum $\{\gamma_a\}$ and if for all allowed classical probability distributions $\{p_\tau\}$ the probabilities $w_\alpha$ are equal. This defines equivalence classes associated to probabilistic observables which are characterized by four pairs $\big(\gamma_\alpha,w_\alpha(\rho_k)\big)$. For two equivalent observables the expectation values of arbitrary functions $f(A),f(B)$ coincide
\begin{equation}\label{M19}
\langle f(A)\rangle=\langle f(B)\rangle.
\end{equation}
If the allowed $\{p_\tau\}$ realize all possible $(\rho_k)$ with $\rho_k\rho_k\leq 3$, and if $A$ and $B$ are quantum observables, the associated operators have to coincide, $\hat A=\hat B$. Nevertheless, the classical observables $A_\tau,B_\tau$ are distinct if
\begin{eqnarray}\label{M20}
\langle(A-B)^2\rangle&=&\sum_\tau(A_\tau-B_\tau)^2p_\tau\nonumber\\
&=&2\langle A^2\rangle-2\sum_\tau A_\tau B_\tau p_\tau>0
\end{eqnarray}
for some allowed probability distribution $\{p_\tau\}$. In this respect the crucial observation is that in general the classical correlation
\begin{equation}\label{M21}
\sum_\tau A_\tau B_\tau p_\tau=\langle A\cdot B\rangle
\end{equation}
depends on the environment $(\delta p_e)_\tau$. It is often smaller then $\langle A^2\rangle$ such that $A_\tau$ and $B_\tau$ are distinct classical observables. 

As an example, we consider two ``diagonal quantum observables'' $B$ and $C$ such that $A=f(C)$ is equivalent to $B$ for some appropriate function $f$. The spectra of $B$ and $C$, namely $(\gamma^{(B)}_1,\gamma^{(B)}_2,\gamma^{(B)}_3,\gamma^{(B)}_4)$ and $(\gamma^{(C)}_1,\gamma^{(C)}_2,\gamma^{(C)}_3,\gamma^{(C)}_4)$, obey both eq. \eqref{M3} and are related by $\gamma^{(B)}_\alpha=f(\gamma^{(C)}_\alpha)$. The associated $e^{(B)}_k,e^{(C)}_k$ vanish for $k\geq 4$ and obey 
\begin{eqnarray}\label{M22}
e^{(B,C)}_1&=&\frac12(\gamma^{(B,C)}_1+\gamma^{(B,C)}_2)~,~\nonumber\\
e^{(B,C)}_2&=&\frac12(\gamma^{(B,C)}_1+\gamma^{(B,C)}_3),\nonumber\\
e^{(B,C)}_3&=&\frac12(\gamma^{(B,C)}_1+\gamma^{(B,C)}_4).
\end{eqnarray}
They are distinct if the spectra of $B$ and $C$ differ. The observables $B$ and $C$ therefore read out different generalized classical bits $\alpha_B=\alpha(e^{(B)}_k)$ and $\alpha_C=\alpha(e^{(C)}_k)$ from the sequence $\big\{\alpha(g_k)\big\}$. The classical correlation function receives contribution from $p_s$ and $\delta p_e$,
\begin{equation}\label{M23}
\langle f(C)\cdot B\rangle=\langle f(C)\cdot B\rangle_s+\langle f(C)\cdot B\rangle_{env},
\end{equation}
with
\begin{eqnarray}\label{M24}
&&\langle f(C)\cdot B\rangle_s\nonumber\\
&&\qquad=\sum_{\alpha_B}\sum_{\alpha_C}w_{\alpha_B}(\rho_k,e^{(B)}_k)
w_{\alpha_C}(\rho_k,e^{(C)}_k)
\gamma^{(B)}_{\alpha_B}f(\gamma^{(C)}_{\alpha_C})\nonumber\\
&&\qquad=\langle B\rangle\langle f(C)\rangle.
\end{eqnarray}

The contribution form the environment
\begin{equation}\label{M25}
\langle f(C)\cdot B\rangle_{env}=\sum_\tau(\delta p_e)_\tau B_\tau f(C_\tau)
\end{equation}
is not fixed by the conditions \eqref{M14} for $A(e^{(B)}_k)$ and $A(e^{(C)}_k)$. Its value depends on the detailed choice of $\big\{(\delta p_e)_\tau\big\}$. As a particular environment we may take $(\delta p_e)_\tau=0$ for all $\tau$, which trivially obeys eq. \eqref{M14} and implies $\langle f(C)\cdot B\rangle_{env}=0$. It is now easy to find allowed probability distributions for which
\begin{equation}\label{M26}
\langle \big( f(C)-B\big)^2\rangle=2\langle B^2\rangle-2\langle B\rangle\langle f(C)\rangle
\end{equation}
differs from zero. For example, this will be the case if $\langle B\rangle=e^{(B)}_k\rho_k=0$. We conclude that the classical observables represented by $B_\tau$ and $f(C_\tau)$ are distinct. As we have mentioned already, the concept of equivalence classes is crucial in order to avoid conflicts with the Kochen-Specker theorem \cite{KS,Str}. Contradiction would arise if for two commuting operators $\hat B$ and $\hat C$, with $\hat B=f(\hat C)$, the corresponding classical observables $B$ and $f(C)$ would always be identical. 

The size of the equivalence classes, or the ensemble of those classical observables which are mapped to a given probabilistic observable, depends on the notion of allowed probability distributions $\{p_\tau\}$. For the split \eqref{M6} and assuming that all $\rho_k$ with $\rho_k\rho_k\leq 3$ are allowed, this depends on the allowed distributions $\big\{(\delta p_e)_\tau\big\}$ characterizing the environment. For a fixed $\big\{(\delta p_e)_\tau\big\}$, as for example $(\delta p_e)_\tau=0$, the equivalence classes are very large. For an observable with a given spectrum of at most four different values $\gamma_\alpha$ only three conditions of the type \eqref{M17} must be met. This can be achieved by a huge number of possible choices for values $A_\tau$ among the four values $\gamma_\alpha$. Even though these conditions have to hold for arbitrary values of $\rho_k$, there are many different possible choices of $A_\tau$ - in general not of the simple form \eqref{M10} - which can obey these conditions. If we take a different fixed choice of $\big\{(\delta p_e)_\tau\big\}$ the equivalence class corresponding to a given probabilistic observable will contain different classical observables as for the case $(\delta p_e)_\tau=0$. If we admit both $\big\{(\delta p_e)_\tau\big\}$ for the allowed probability distributions, only the classical observables which fulfill eq. \eqref{M17} for both choices of $(\delta p_e)_\tau$ belong to the equivalence class. Observables for which $\langle A^p\rangle$ differs for the two different $\big\{(\delta p_e)_\tau\big\}$ are not system observables. As we increase the number of allowed $\big\{(\delta p_e)_\tau\big\}$ the ensemble of system observables shrinks, and so do the equivalence classes associated to a given probabilistic observable. If arbitrary $\big\{(\delta p_e)_\tau\big\}$ are admitted each equivalence class would contain only one classical observable. For $\big\{(\delta p_e)_\tau\big\}$ obeying the constraints \eqref{M14} we have shown by explicit construction that the equivalence classes contain distinct classical observables.

We end this section by noting that the construction of a classical ensemble representing four-state quantum mechanics can be generalized in a straightforward way to $M$-state quantum mechanics with arbitrary $M$. Quantum mechanics for observables with a continuous spectrum can be obtained for appropriate limits $M\to\infty$. On the other hand, if one is not aiming for a description of all possible quantum observables, as for the classical statistical implementation of a quantum computer in sect. \ref{Classicalquantumcomputer}, the space of classical states $\tau$ can be reduced by restriction to a subset of $g_k$. 

\section{Correlations}
\label{Correlations}

Beyond a rule for the computation of expectation values of observables, any  theory must provide a prediction for the outcome of two consecutive measurements. After a first measurement of the observable $A$ the result of a subsequent measurement of another observable $B$ is, in general, influenced by the first measurement. In a statistical system two measurements are typically correlated and one has to specify the ``measurement correlation''. It is, a priori, not always obvious which correlation should be chosen, since the measurement of $A$ may have changed the ensemble or the knowledge of the observer. 

\medskip\noindent
{\bf Conditional probability}

For simplicity we concentrate in this section on two-level-observables, $\langle A^2\rangle=\langle B^2\rangle=1,~\hat A^2=\hat B^2=1$. The probability of finding $B=1$ after a measurement $A=1$ amounts to the {\em conditional probability} $(w^B_+)^A_+$. There are, in principle, different ways to specify the conditional probability. A valid definition should be appropriate for the properties of a given measurement. For a ``good measurement'' we know that after the measurement $A=1$ the ensemble must be an eigenstate to the eigenvalue $\gamma^{(A)}=1$ - otherwise a subsequent measurement of $A$ would not necessarily yield the same value as the first one. Then $B$ is measured under this condition. 

We take here the attitude that there is only one given reality, but physicists can at best give a statistical description of it. The ``fundamental laws'' are genuinely of a statistical nature \cite{GenStat} and only establish relations within different possibilities for the history of the real world. Measuring for an observable $A$ in a given state the value $\gamma_{\bar\alpha}$ simply eliminates the other possible alternatives (which may have nonvanishing probabilities $w_{\alpha\neq \bar\alpha}$). After the measurement of $A$ it makes only sense to ask what are the outcomes of other measurements under the condition that $A$ has been measured to have the value $\gamma_{\bar\alpha}$. 

On the level of the classical statistical system with infinitely many degrees of freedom, which describes the system and its environment, the elimination of the possible histories which are not compatible with the first measurement of $A$ is not unique. Many different classical probability distributions $p_\tau$ can be eigenstates of $A$ with $\gamma^{(A)}=1$. One will have to specify how this elimination is done. Indeed, for  a given classical representation $A_\tau$ one may ``eliminate'' after the first measurement all states $\tau$ for which $A_\tau=-1$. Setting $p_\tau=0$ if $A_\tau=-1$ leaves an eigenstate with $\langle A\rangle=1$, independently of how the probabilities for the states with $A_\tau=1$ are distributed after the first measurement. One possibility would be to keep the relative probabilities of all states $\tau$ for which $A_\tau=1$ the same as before the measurement. However, one could apply the same procedure to a second representation $A'_\tau$, where now $p_\tau=0$ for all $\tau$ with $A'_\tau=-1$. If one keeps again the relative probabilities of the states with $A'_\tau=1$, the results would differ from applying this prescription to $A_\tau$. With such a prescription the state of the ensemble after the measurement $A=1$ would therefore depend on the precise choice of $A_\tau$ or $A'_\tau$, whose difference concerns only properties of the environment. Such a prescription (which is actually behind the use of the classical correlation for pairs of measurements) can only make sense if the measurement can resolve the details of the environment. It is clearly inappropriate for a measurement of the subsystem whose outcome does not depend on the environment. 

\medskip\noindent
{\bf Measurement correlation}

The measurement correlation $\langle BA\rangle_m$ describes the outcome of measurements of pairs of two observables $A$ and $B$. As a criterion for a  measurement that preserves the isolation of the subsystem and only measures its properties we postulate that it should be possible to determine the measurement correlation $\langle BA\rangle_m$ by using only information which is available for the subsystem. It must be possible to compute $\langle BA\rangle_m$ from the $\rho_k$ characterizing the original state of the subsystem before the measurement of $A$. No information about the details of the environment should be needed. 

The {\em measurement correlation} or {{\em conditional correlation}  $\langle BA\rangle_m$ multiplies the measured values of $A$ and $B$, weighed with the probabilities that they occur
\begin{eqnarray}\label{7}
\langle BA\rangle_m&=&(w^B_+)^A_+w^A_{+,s}-(w^B_-)^A_+w^A_{+,s}\nonumber\\
&&-(w^B_+)^A_-w^A_{-,s}+(w^B_-)^A_-w^A_{-,s}.
\end{eqnarray}
Here $w^A_{\pm,s}$ denotes the probability that $A$ is measured as $\pm 1$ in the state $s$, with $\langle A\rangle=w^A_{+,s}-w^A_{-,s}=$ tr $(\rho\hat A)$, $w^A_{+,s}+w^A_{-,s}=1$. The conditional probabilities must obey  $(w^B_+)^A_\pm +(w^B_-)^A_\pm=1$. The conditional correlation needs a specification of the conditional probabilities as $(w^B_\pm)^A_+$. For their computation we use the prescription that after the first measurement $A=1$ the density matrix $\rho_{A+}$ must describe an eigenstate of $A$, tr$(\hat A\rho_{A+})=1$. This effect of a first measurement may be called {\em state reduction}. The subsequent measurement of $B$ then involves this state,
\begin{equation}\label{8}
(w^B_+)^A_+-(w^B_-)^A_+=\text{tr}(\hat B\rho_{A+}).
\end{equation}

The relation \eqref{8} is based on the property that the quantum observable $B$ obeys eqs. \eqref{2}, \eqref{5} for an arbitrary quantum state of the subsystem. Our assumption is therefore that after the measurement of $A$ the classical ensemble still describes a quantum system. This seems reasonable for appropriate measurements since otherwise the first measurement destroys the isolation of the subsystem instead of only changing its state. This assumption has far reaching consequences, however. It necessarily implies eq. \eqref{8} and excludes the option of using the classical correlation for the general description of subsequent measurements, as we will see below.

\medskip\noindent
{\bf Quantum correlation}

For $M=2$ the matrix $\rho_{A+}$ is unique, $\rho_{A+}=\frac12(1+\hat A)$, such that 
\begin{equation}\label{9}
(w^B_\pm)^A_+=\frac12\pm \frac14\text{tr}
(\hat B\hat A)~,~(w^B_\pm)^A_-=\frac12\mp\frac14\text{tr}(\hat B\hat A).
\end{equation}
However, for $M>2$ one has tr$(\hat A\rho_{A+})=1,\text{tr}\rho_{A+}=1$ for 
\begin{eqnarray}\label{10}
\rho_{A+}&=&\frac 1M(1+\hat A+X)~,~\text{tr}(\hat A X)=0~,~\nonumber\\
\text{tr}X&=&0,
\end{eqnarray}
where the purity after the measurement obeys
\begin{equation}\label{14A}
P=M\text{tr}(\rho^2_{A+})=1+\frac1M\text{tr}X^2.
\end{equation}
With
\begin{equation}\label{11}
\rho^2_{A+}-\rho_{A+}=\frac{1}{M^2}(X^2+\{\hat A,X\})-\left(1-\frac 2M\right)\rho_{A+}
\end{equation}
a necessary condition for $\rho_{A+}$ describing a pure state is $~P=M-1$,  tr$X^2=M(M-2)$, which implies $X=0$ only for $M=2$.  

We may distinguish between a ``maximally destructive measurement'' where all information about the original ensemble except for the value of $A$ is lost, and a ``minimally destructive measurement'' for which an original pure state remains a pure state after the measurement. A maximally destructive measurement is described by $X=0$ in eq. \eqref{10}, leading to 
\begin{equation}\label{12}
\langle B\rangle_{A+}=(w^B_+)^A_+-(w^B_-)^A_+=\frac1M\text{tr}(\hat B\hat A)=\langle BA\rangle_{\max}.
\end{equation}
Here we denote by $\langle BA\rangle_{\max}$ the conditional correlation for maximally destructive measurements and use that $\rho_{A-}$ obtains from $\rho_{A+}$ by changing the sign of $\hat A$ in eq. \eqref{10} (with $X=0$). We can use $\langle BA\rangle_{\max}$ for the definition of a scalar product between the observables $B$ and $A$, since it does not depend on the initial ensemble. The two-level observables $A^{(k)}$ form an orthogonal basis in this sense, $\langle A^{(k)}A^{(l)}\rangle_{\max}=\delta_{kl}$. 

A minimally destructive measurement of $A=1$ projects out all states with $A=-1$, without further changes of the original ensemble and associated density matrix $\rho$,
\begin{equation}\label{13}
\rho_{A+}=\frac{1}{2(1+\langle A\rangle)}(1+\hat A)\rho(1+\hat A).
\end{equation}
Similarly, a first measurement $A=-1$ maps $\rho$ on $\rho_{A-}$, which obtains from eq. \eqref{13} by changing on the r.h.s. all $+$ signs to $-$ signs. We note that $X$ in eq. \eqref{10} depends on $\rho$. Due to the normalization factor and since $\langle A\rangle$ depends on $\rho$, the map $\rho\to\rho_{A_+}$ is not linear in $\rho$.  If $\rho$ is the density matrix corresponding to a pure state $\psi$, this also holds for $\psi_{A+}$, with
\begin{equation}\label{17Aa}
\psi_{A+}=\big[2(1+\langle A\rangle)\big]^{-1/2}(1+\hat A)\psi.
\end{equation}
For $M=2$ eq. \eqref{13} yields $\rho_{A+}=\frac12(1+\hat A)$. In sect. \ref{Sequenceofmeasurements} we will discuss the map $\rho\to \rho_{A+}$ in more detail when we address the issue of sequences of measurements.

With
\begin{equation}\label{16A}
\langle BA\rangle_m =\text{tr}(\hat B\rho_{A+})w^A_{+,s}-\text{tr}(\hat B\rho_{A-})w^A_{-,s}, 
\end{equation}
and 
\begin{equation}\label{14a}
w^A_{\pm,s}=(1\pm\langle A\rangle)/2
\end{equation}
our prescription for $\rho_{A_+}$ yields for the measurement correlation
\begin{equation}\label{14}
\langle BA\rangle_m=\frac12\text{tr}(\{\hat A, \hat B\}\rho). 
\end{equation}
Thus the conditional correlation for minimally destructive measurements in the classical statistical ensemble corresponds precisely to the expression of this correlation in quantum mechanics. It involves the anticommutator and is therefore related to the quantum mechanical operator product. On the level of probabilistic observables we can express the conditional correlation $\langle BA\rangle_m$ in terms of the  expectation value of the quantum product $(BA)_s$
\begin{equation}\label{17A}
\langle BA\rangle_m=\langle (BA)_s\rangle,
\end{equation}
demonstrating the close connection between the quantum product and the conditional correlation.

The two point correlation is commutative, $\langle BA\rangle_m=\langle AB\rangle_m$. We will postulate that the two point correlation \eqref{17A} describes in general the correlation between two measurements for quantum systems and call it ``{\em quantum correlation}''. We have motivated its use by two subsequent measurements, but the order of the measurements does actually not matter. It seems therefore natural to use this correlation for any measurement of pairs of observables, independently of the time order. 

\medskip\noindent
{\bf Classical correlation}

At first sight, a possible alternative choice may be the ``classical correlation'' which is based on the classical product $A\cdot B$, as defined on the level of the classical ensemble, $(A\cdot B)_\tau=A_\tau B_\tau$. We will see in sect. \ref{Propertiesofquantum} however, that $A\cdot B$ is usually not a quantum observable and can therefore not be determined from the information characterizing a quantum state, i.e. from $\{\rho_k\}$. Using the classical correlation $\langle B\cdot A\rangle=\sum_\tau p_\tau(B\cdot A)_\tau=\sum_\tau p_\tau B_\tau A_\tau$ would therefore need information which relates to the environment, but not only to the subsystem. In other words, the use of the classical product corresponds to a state reduction after the first measurement where substantial information about the relation between the subsystem and the environment is retained. This is not what a good measurement in an isolated subsystem does. The classical correlation can therefore not serve for the description of such measurements. For any measurement where the outcome (including the state reduction) can be expressed in terms of information available for the subsystem, the choice of the quantum product seems natural. It retains a maximum of the information which is available in the subsystem.

Since the correct choice of the correlation for a description of two measurements is crucial we may describe the issue in some more detail. In any statistical setting one should distinguish the probability $w_{++}$ that the two level observables $A$ and $B$ are {\em measured} with values $A=1,B=1$ from the probability $p_{++}$ that they ``have'' the values $A=B=1$ in the classical statistical ensemble before the first measurement. On the classical statistical level one can express
\begin{equation}\label{19A}
p_{++}=\frac14(1+\langle A\rangle+\langle B\rangle+\langle A\cdot B\rangle)
\end{equation}
in terms of the classical correlation
\begin{equation}\label{19B}
\langle A\cdot B\rangle=\sum_\tau p_\tau A_\tau B_\tau.
\end{equation}
The probability $p_{++}$ does not specify a priori the conditional information relating two subsequent measurements, which is necessary for $w_{++}$. In general, one needs a separate prescription how $w_{++}$ should be computed from the available statistical information. Only under particular circumstances, one may be able to identify $w_{++}$ with $p_{++}$. In other words, $w_{++}=p_{++}$ is an additional basic assumption which does not hold true in general. This contrasts to the case of a single measurement for $A$, where $w_+=p_+=(1+\langle A\rangle)/2$ by definition.

If we try the identification $w_{++}=p_{++}$ for measurements in an isolated subsystem we run into severe problems. In general, the classical correlation $\langle A\cdot B\rangle$ is not computable in the subsystem. It is a property of the system and its environment and cannot be obtained from the information characterizing the quantum state, i.e. from $\rho_k$. This problem is closely linked to the fact that the mapping from the classical observables described by $A_\tau$ to the probabilistic observables described by $\gamma^{(A)}_a=\pm 1$ and $w^{(A)}_a=w^{(A)}_\pm$ is not invertible. The classical observable $A_\tau$ describes properties of the subsystem and its environment, while the characterization of the environment is only lost on the level of the probabilistic observable $A$. The use $w_{++}=p_{++}$ corresponds to an implicit definition of the conditional probability for two measurements where after the first measurement $A=1$ all classical states $\tau$ for which $A_\tau=-1$ are eliminated without changing the relative probabilities of those states where $A_\tau=1$. This elimination process depends, however, on the particular observable $A_\tau$ and therefore also reflects properties of the environment, not only of the subsystem. Two observables $A_\tau$ and $A'_\tau$, which lead to the {\em same} probabilistic observable $A$, produce, in general, {\em different} classical products $\langle A\cdot B\rangle\neq \langle A'\cdot B\rangle$. (These issues are discussed in more detail in sect. \ref{Propertiesofquantum}.)

\medskip\noindent
{\bf Choice of measurement correlation}

If the state of an isolated subsystem can be described by $\{\rho_k\}$, this information must also be sufficient for a prediction of the outcome of two measurements. The probability $w_{++}$ must be computable in terms of $\{\rho_k\}$. For this reason we employ the quantum correlation $\langle AB\rangle_m$ \eqref{14}, \eqref{17A} and postulate
\begin{equation}\label{19C}
w_{++}=\frac14(1+\langle A\rangle+\langle B\rangle+\langle AB\rangle_m),
\end{equation}
as advocated already before for subsequent measurements. (A different motivation for the use of the quantum correlation is given in \cite{3}.) Indeed, now $w_{++}$ can be expressed in terms of $\rho_k$
\begin{equation}\label{19D}
w_{++}=\frac14\Big(1+e^{(A)}_ke^{(B)}_k+\rho_k
[e^{(A)}_k+e^{(B)}_k+d_{mlk}e^{(A)}_me^{(B)}_l]\Big)
\end{equation}
The other probabilities $w_{+-}$ etc. obtain from eqs. \eqref{19C}, \eqref{19D} by appropriate changes of relative signs.

The prescription for the probabilities of the outcome of two measurements influences strongly the statistical properties of correlations. For example, one may ask if a ``hidden variable theory'' is possible, where there exist discrete functions $\tilde A(v)=\pm 1~,~\tilde B(v)=\pm 1$ such that $\langle AB\rangle_m=\int dv\tilde p(v)\tilde A(v)\tilde B(v)$ with some probability distribution $\tilde p(v)$. Just as in quantum mechanics, this can be excluded by the use of Bell's inequalities \cite{Bell}, \cite{CC}. In our classical statistical setting the correlation function \eqref{14}, or the probabilities for the outcome of two measurements, are exactly the same as in quantum mechanics. On the other hand, Bell's inequalities apply to the classical correlation $\langle A\cdot B\rangle$. Besides theoretical arguments we have therefore also experimental evidence that in general the classical correlation function should not be used for the description of the outcome of two measurements.

\medskip\noindent
{\bf Complete sets of measurements}

Consider the special case of two two-level-observables $A$ and $B$ for which the associated quantum operators $\hat A$ and $\hat B$ commute,
\begin{equation}\label{114A}
[\hat A,\hat B]=0.
\end{equation}
After a sequence of two measurements, where first $A$ is measured to have the value $\gamma_A=\pm1$, and subsequently $B$ is measured to have the value $\gamma_B=\pm 1$, the state of the system is given by
\begin{equation}\label{114B}
\rho_{BA,\gamma_B\gamma_A}=
\frac{{\cal N}}{16}
(1+\gamma_B\hat B)(1+\gamma_A\hat A)\rho(1+\gamma_A\hat A)(1+\gamma_B\hat B),
\end{equation}
where the normalization factor ${\cal N}$ assures $\text{tr}\rho_{BA,\gamma_B\gamma_A}=1$. Due to the vanishing commutator \eqref{114A} the order of the measurements of $A$ and $B$ does not matter and we may say that $A$ and $B$ are measured ``simultaneously''. The state of the system after the simultaneous measurements of $A$ and $B$ depends only on the initial state and the ``outcome of the measurement'', i.e. the four possible values $(\gamma_A,\gamma_B)=(+,+),(+,-),(-,+),(-,-)$. We can consider the combined measurements of $A$ and $B$ as a single measurement. 

The state after the combined measurement can be written in terms of the four projectors
\begin{equation}\label{114C}
P_{\gamma_B\gamma_A}=\frac{1}{4}(1+\gamma_B\hat B)(1+\gamma_A\hat A)
\end{equation}
as
\begin{equation}\label{114D}
\rho_{\gamma_B\gamma_A}={\cal N}P_{\gamma_B\gamma_A}\rho P_{\gamma_B\gamma_A}.
\end{equation}
We observe that $\rho_{\gamma_B\gamma_A}$ is automatically an eigenstate of any quantum observable $D$ for which the associated operator obeys $\hat D=\hat A\hat B$ (and therefore also commutes with $\hat A$ and $\hat B$). The associated eigenvalue $\gamma_D$ is given by the product $\gamma_D=\gamma_A\cdot\gamma_B$. A simultaneous measurement of $D,A$, $B$ is therefore possible, but the measurement of $D$ does not yield any new information about the system. For four-state quantum systems $(M=4)$ the simultaneous measurement of two ``commuting two-level observables'' constitutes a ``complete set of measurements''. After the measurement of the complete set, the system is in a pure state. (We assume here that ${\cal N}$ is finite and postpone a more detailed discussion of the limiting case ${\cal N}\to \infty$ to sect. \ref{Sequenceofmeasurements}.) The same must actually happen after the measurement of a quantum observable with a non-degenerate spectrum of four different eigenvalues. After the measurement of a given eigenvalue the system is projected to the corresponding eigenstate which yields in this case a unique pure state density matrix. There is a close connection between measurements of non-degenerate quantum observables (and also general quantum observables) and complete sets of measurements for two-level quantum observables. We will not discuss this in detail in this paper. 

It is straightforward to generalize these concepts to general $M$-state quantum systems. A complete set of measurements of two-level observables projects the state after the measurement to a pure state density matrix. We will assume a minimal set in the sense that ``redundant observables'', whose values can be predicted uniquely after measurements of the minimal set, are removed from the set. (They are the analogue of the observable $D$ in the preceding paragraph.) Instead of a set of commuting two-level observables one may also use ``commuting observables'' with a spectrum of more than two distinct eigenvalues. After a complete set of measurements the system is in a pure state which is a simultaneous eigenstate of a ``maximal set of commuting operators'', as familiar from quantum mechanics. 

\section{Quantum time evolution}
\label{Quantumtimeevolution}

We have seen how quantum structures can arise from the description of subsystems where the ``state of the system'' is described by $n$ expectation values of ``basis observables''. For $P<n$ the appearance of ``non-commuting structures'' is mandatory. The question remains why such quantum systems are omnipresent in nature, in contrast to ``commuting structures'' for $P=n$. The answer may be rooted in stability properties of the time evolution. We discuss in this section the emergence and particularities of the unitary time evolution which is characteristic for quantum mechanics. 

\medskip\noindent
{\bf Time evolution of the subsystem}

Let us consider some continuous time evolution of the classical probability distribution $\{p_\tau\}$. It relates the ensemble at time $t_2$ to the ensemble at some earlier time $t_1$, and induces a transition from $\rho_k(t_1)$ to $\rho_k(t_2)$, 
\begin{equation}\label{15}
p_\tau(t_2)=\tilde S_{\tau\rho}(t_2,t_1)p_\rho(t_1)~,~\rho_k(t_2)=S_{kl}(t_2,t_1)\rho_l(t_1).
\end{equation}
We may decompose the transition matrix $S_{kl}$ into the product of an orthogonal matrix $\hat S_{kl}$, which preserves the length of the vector $(\rho_1\dots,\rho_n)$ and therefore the purity, and a scaling $d,~S_{kl}=\hat S_{kl}d$. For an infinitesimal evolution step this implies
\begin{eqnarray}\label{16}
\partial_t\rho_k(t)&=&T_{kl}\rho_l(t)+D\rho_k(t)~,~D=\partial_t\ln d(t,t_1),\nonumber\\
T_{kl}&=&-T_{lk}=\partial_t\hat S_{km}(t,t_1)\hat S_{lm}(t,t_1).
\end{eqnarray}

For a given maximal purity during the evolution, eq. \eqref{16} can be rewritten as an equation for the density matrix $\rho$,
\begin{eqnarray}\label{17}
\partial_t\rho_{\alpha\beta}=&-&i[H,\rho]_{\alpha\beta}+R_{\alpha\beta\gamma\delta}
\left(\rho_{\gamma\delta}-\frac 1M\delta_{\gamma\delta}\right)\nonumber\\
&+&D(\rho_{\alpha\beta}
-\frac1M\delta_{\alpha\beta}).
\end{eqnarray}
This corresponds to a split of the infinitesimal $SO(n)$ transformation $\delta\rho_k=T_{kl}\rho_l$ into a unitary part corresponding to the subgroup $SU(M)$ and represented by the hermitean Hamiltonian $H=H_kL_k+H_0$, and remaining rotations of $SO(n)/SU(M)$ represented by $R$ or $\tilde T_{kl}$, 
\begin{eqnarray}\label{18}
T_{kl}&=&-2f_{klm}H_m+\tilde T_{kl}~,~[L_k,L_l]=2if_{klm}L_m,\nonumber\\
\tilde T_{kl}&=&\frac1M
R_{\alpha\beta\gamma\delta}(L_k)_{\beta\alpha}(L_l)_{\gamma\delta}. 
\end{eqnarray}
In general, $H,R$ and $D$ may depend on $\rho_k$. 

\medskip\noindent
{\bf Unitary time evolution}

We are interested in possible partial fixed points of the evolution for which $R=0$ and $D=0$, while $H$ is independent of $\rho_k$. (Partial fixed points of this type have been found explicitly in the classical time evolution of non-relativistic boson fields \cite{4}.) Then eq. \eqref{17} reduces to the linear von-Neumann equation for the density matrix. In case of a pure state density matrix this implies the Schr\"odinger equation $i\partial_t\psi=H\psi$. One recovers the unitary time evolution of quantum mechanics. The evolutions with $R=D=0$ are singled out by the property that a pure state of the subsystem remains a pure state during the evolution. It will be interesting to find out how this property is related precisely to the notion of the isolation of the subsystem.

The more general evolution equation away from the ``unitary partial fixed point'' can describe ``decoherence'' \cite{DC} as a decrease of purity for $D<0$, or ``syncoherence'' as the approach to the pure state partial fixed point with increasing purity for $D>0$. The latter typically accounts for a situation where the subsystem described by the observables $A^{(k)}$ can exchange energy with the environment. An example is the evolution from a mixed state of an atom in different energy states to a pure state of an atom in the ground state by virtue of radiative decay of the excited states. A static pure state density matrix obtains as usual as a solution of the quantum mechanical eigenvalue problem $H\psi=E_j\psi$. We suggest that the omnipresence of quantum systems in nature is due to the existence of such partial fixed points which reflect the isolation of the subsystem. 

The change of the purity is related to $D$ in eq. \eqref{16}, 
\begin{equation}\label{118A}
\partial_tP=2\rho_k\partial_t\rho_k=2DP.
\end{equation}
For $D<0$ the purity decreases - this describes decoherence. Decoherence is not time reversible - an arrow of time is singled out by the ``direction'' of the approach to equipartition. This also holds for the opposite process of an increase of purity, i.e. syncoherence. If the time evolution of the subsystem is time-reversal invariant, $D$ must vanish.

The remaining rotations described by $T_{kl}$ in eq. \eqref{16} could, in principle, be equivalent in both time directions. The unitary subgroup corresponding to $\tilde T_{kl}=0$ in eq. \eqref{18} may be singled out by the observation that only such an evolution is compatible with the ``principle of equivalent state and observable transformations'' (PESOT), which states that instead of a time evolution of the probability distribution (or the state of the subsystem) one may equivalently describe the time evolution by time dependent observables \cite{CW1}. Only unitary transformations preserve the spectrum of quantum observables. Then PESOT corresponds to the well known equivalence of the Schr\"odinger and Heisenberg pictures in quantum mechanics. 

\medskip\noindent
{\bf Hamilton operator}

If $H$ is independent of $\rho_k$ it can be considered as an observable of the subsystem, $H=H_kL_k+H_0$, with fixed coefficients $H_k,H_0$. By Noether's theorem it is associated with the energy of the subsystem, where $E_j$ denotes the possible energy eigenvalues. (If one wants to use standard energy units one replaces $H\to H/\hbar$.) On the other hand, $R$ and $D$ account for the interactions of the subsystem with its environment. They vanish in the limit of ``perfect isolation'' of the subsystem. If the interactions with the environment are strong enough, the subsystem is typically not evolving towards the equipartition fixed point,  $\rho_{\alpha\beta}=\frac1M\delta_{\alpha\beta}$, but rather towards a Boltzmann type density matrix $\rho\sim \exp[-\beta(H+\mu_i N_i)]$ (for conserved quantities $N_i$ and chemical potentials $\mu_i$), which is close to a pure state density matrix if the temperature $T=\beta^{-1}$ is small as compared to the typical separation of the two lowest energy eigenvalues $E_j$. In contrast, if the isolation from the environment becomes efficient fast enough, the subsystem follows subsequently its own unitary time evolution, as well known from quantum mechanics. 

Such a behavior would correspond to the approach to a partial fixed point at $D=0$, $R_{\alpha\beta\gamma\delta}=0$, as described in \cite{CW2}. Consider the case where the lowest eigenvalue of $H$ is not degenerate. Energy exchange with the environment would induce an increase of $P$ until the  maximal purity $P=M-1$ is reached. For small enough $T$ one ends in the unique ground state of the system - as characteristic for many atoms for the temperature of the earth. This scenario could provide a simple explanation why subsystems with the behavior of an isolated quantum system are omnipresent in nature. As well known from quantum mechanics, the uniqueness of the lowest energy state explains that the isolated subsystems are all identical, i.e. the identity of the atoms. 

At this point we may recapitulate what we have achieved. Starting form a classical statistical ensemble we have identified a class of classical observables which have the same expectation values and admit the same algebra as the operators in a corresponding $M$-state quantum system. This holds provided the ``purity constraint''  $P\leq M-1$ (together with eqs. \eqref{54A}, \eqref{56B}) is obeyed. Also the conditional correlations which describe measurements of two such observables are the same as in the quantum system. Furthermore, we have found the criteria for the time evolution of the classical probability distribution that ensure a unitary evolution of the density matrix for the corresponding quantum system. Obviously, these properties are related to the specific subset of classical observables that can describe the subsystem, in the sense that no further information about the environment is needed for a prediction of the outcome of measurements in the subsystem. In the remaining part of this paper we will discuss the properties of these specific quantum observables in some more detail.

\section{Properties of quantum observables}
\label{Propertiesofquantum}

The quantum structures for probabilistic observables discussed so far do not need any specification of the representation as classical observables. Many classical systems with different states $\tau$, classical probabilities $p_\tau$ and classical values of the observable in a given state, $A_\tau$, may describe the same state of the subsystem according to eq. \eqref{1} and provide for classical realizations of quantum observables. Nevertheless, the implementation of quantum observables implies certain restrictions on the possible classical realizations. We will discuss those in the following, mainly for the purpose of conceptual foundations. One possible realization has been presented in sect. \ref{Simplequantum}. Here we will discuss the issue in a more general context.

\medskip\noindent
{\bf Classical and probabilistic quantum observables}

Classical observables are maps from the set of probability distributions $\Omega$, with elements $\{p_\tau\}=(p_1\dots,p_S)$, to real numbers, $\Omega{\stackrel {A^{(C)}}{\to}}{\mathbbm R}$, with $\{p_\tau\}\to\langle A^{(C)}\rangle=\sum_\tau p_\tau A^{(C)}_\tau$. (For simplicity we employ a language with a finite number $S$ of classical states, which can be extended to an infinite set at the end in some specified limiting procedure.) We will restrict the discussion to those elements of $\Omega$ which correspond to ``quantum states'', i.e. which obey the bound for the purity of the ensemble. In general, the classical observables $A^{(C)}$  describe the system and its environment. (We use ``system'' for the (isolated) subsystem or quantum system from now on.) 

We are interested in the subclass of quantum observables $A^{(Q)}$ whose expectation values and quantum correlations can be computed in the system. (We often use in this section the upper index $A^{(Q)}$ or $A^{(C)}$ in order to underline that we deal with observables on the classical level, as specified by $A^{(Q)}_\tau,A^{(C)}_\tau$ in the classical states $\tau$.) On the classical level, quantum observables are classical observables with special properties. First, for a $M$-state quantum system, a quantum observable has at most $M$ different classical values $A^{(Q)}_\tau=\gamma_a$. (For a non-degenerate spectrum we can identify $a=\alpha=1\dots M.)$ For any given quantum observable we can classify the classical states $\tau$ according to the value of $A^{(Q)}_\tau~,~\tau=(\gamma_a,\sigma_{\gamma_a})$, where for each given $\gamma_a$ one typically has a large degeneracy of classical states, labeled by $\sigma_{\gamma_a}$. We can define the probability $w_a$ for the occurrence of a possible measurement value $\gamma_a$ as 
\begin{equation}\label{W1}
w_a=\sum_{\sigma_{\gamma_a}} p(\gamma_a,\sigma_{\gamma_a})~,~\langle A^{(Q)}\rangle=\sum_a\gamma_a w_a.
\end{equation}

As a crucial ingredient, $w_a$ must be computable from the quantities which specify the quantum state, i.e. from the expectation values of the ``basis observables'' 
$\langle A^{(k)}\rangle=\rho_k$. We assume a linear relation 
\begin{equation}\label{W2}
w_a{(\rho_k)}=\sum_{k} c_{ak}\rho_k+c_{a0}.
\end{equation}
The probabilities must be normalized, $\sum_aw_a(\rho_k)=1$, for arbitrary $\rho_k$, which implies the conditions
\begin{equation}\label{30X}
\sum_ac_{ak}=0~,~\sum_ac_{a0}=1.
\end{equation}
For quantum observables the coefficients $c_{ak},c_{a0}$ are restricted further since $w_a(\rho_k)$ must obey eq. \eqref{8A}. For non-degenerate eigenvalues one needs
\begin{equation}\label{30A}
c_{\alpha k}=\frac1M (UL_kU^\dagger)_{\alpha\alpha}~,~c_{\alpha 0}=\frac1M,
\end{equation}
for some suitable unitary matrix $U$. In this case the ``quantum determination'' of probabilities \eqref{W2} amounts to a condition for the classical probabilities, namely that a unitary matrix $U$ exists such that 
\begin{equation}\label{33A}
\sum_{\sigma_{\gamma_\alpha}}p(\gamma_\alpha,\sigma_{\gamma_\alpha})=\frac1M
(1+\sum_k\rho_k(UL_kU^\dagger)_{\alpha\alpha}).
\end{equation}

Thus quantum observables are defined by two properties:\\
(i) the restriction of the spectrum to at most $M$ different values, (ii) the  ``quantum determination'' of probabilities $w_a(\rho_k)$. While (i) only involves a property of the classical observable, the second restriction (ii) depends on relations to the basis observables and on the selection of possible quantum states out of the most general probability distributions $\{p_\tau\}$. For the specification of a quantum observable we need at least $\gamma_a$ and $c_{ak},c_{a0}$. A quantum observable $A^{(Q)}$ has the important property that its classical product $A^{(Q)}\cdot A^{(Q)}$ (defined by $(A^{(Q)}\cdot A^{(Q)})_\tau=(A^{(Q)}_\tau)^2)$ is again a quantum observable, with spectrum $(\gamma^2_a)$ and the same $w_a$ as for $A^{(Q)}$. This extends to higher polynomials and arbitrary functions $f(A^{(Q)})$.

We can now associate to any classical quantum observable $A^{(Q)}$ a probabilistic quantum observable, $A^{(Q)}\to A$, which is characterized by the spectrum of possible measurement values $(\gamma_a)$ and the associated probabilities $w_a$. Only this information will be needed for a computation of expectation values $\langle(A^{(Q)})^p\rangle$ in a ``quantum state'' of the system, while the detailed form of $\{p_\tau\}$ is not relevant. On the level of classical observables the quantum observables are characterized by a distribution of values $A^{(Q)}_\tau=\gamma_a$ for the classical states $\tau$. This distribution still contains much more information than the spectrum $\gamma_a$ and the associated probabilities $w_a$. Therefore $A^{(Q)}_\tau$ still describes the system and partly the environment. Only on the level of probabilistic observables $A$ the parts of $A^{(Q)}_\tau$ relevant for the environment are projected out, such that $A$ only ``measures'' properties of the system.

We also can associate to every $A^{(Q)}$ a quantum operator $\hat A$ by a map $A^{(Q)}\to \hat A$. It is constructed from eq. \eqref{5} by observing
\begin{equation}\label{33B}
\langle A^{(Q)}\rangle=\sum_a\gamma_a(c_{a,0}+\sum_k c_{a,k}\rho_k)
=e^{(A)}_0+\sum_k e^{(A)}_k\rho_k.
\end{equation}
This identifies
\begin{eqnarray}\label{W3}
e^{(A)}_0&=&\sum_a\gamma_ac_{a,0}~,~e^{(A)}_k=\sum_a\gamma_a c_{a,k},\nonumber\\
\hat A&=&e^{(A)}_0+\sum_k e^{(A)}_kL_k~,~\langle A^{(Q)}\rangle=\textup{tr} (\hat A\rho).
\end{eqnarray}
We note that the map \eqref{W3} is possible for arbitrary probabilistic system observables obeying eq. \eqref{W2}. Without the restriction of the type \eqref{30A} for quantum observables, however, $(A^{(Q)})^2$ will, in general, not be mapped to $\hat A^2$. A simple example is the random two-level observable $R$, with $\gamma_1=1~,~\gamma_2=-1~,~c_{ak}=0~,~c_{10}=c_{20}=1/2$. It is mapped to $\hat A=0$, while $R^2=1$. 

In turn, we have a map from the space of quantum operators ${\cal O}$ to the space of probabilistic quantum observables ${\cal P}$, since for every operator $\hat A$ the spectrum $\{\gamma_a\}$ is defined, and the probabilities $w_a$ can be computed for all quantum states $\{\rho_k\}$ or density matrices $\rho$. The latter obtain from the diagonal elements $(U\rho U^\dagger)_{\alpha\alpha}$, with $U$ the unitary matrix used for the diagonalization of $\hat A$. The map from the space of classical quantum observables ${\cal Q}$ to the probabilistic quantum observables ${\cal P}$ is equivalent to the sequence of maps 
${\cal Q} \to {\cal O},{\cal O}\to{\cal P}$. 

However, the map from the ensemble of classical quantum observables ${\cal Q}$ to the space of quantum operators ${\cal O}$ is not invertible. The classical observables $A^{(Q)}$ involve a specification of $A^{(Q)}_\tau$ for every classical state $\tau$, which is much more information than contained in the coefficients $c_{a,k}~,~c_{a,0}$. We may encounter situations where a quantum observable $B^{(Q)}$ is mapped to an operator $\hat B$, while also a function $f(A^{(Q)})$ of a different quantum observable $A^{(Q)}$ is mapped to the same operator, $f(\hat A)=\hat B$. (Here $f(\hat A)$ is an operator valued function, while $f(A^{(Q)})$ is based on the classical product $A^{(Q)}\cdot A^{(Q)}$.) Such a situation does not imply an identification of the quantum observables at the classical level, i.e. in general one has $B^{(Q)}\neq f(A^{(Q)})$. This lack of invertibility of the map ${\cal Q}\to {\cal O}$ constitutes an important difference between our approach and many alternative attempts of a ``classical formulation of quantum mechanics'', which associate to each $\hat A$ a unique classical observable. For example, this is typically assumed for ``hidden variable theories''. Also for the Kochen-Specker theorem \cite{KS} the existence of a map $\hat A\to A^{(Q)}$ is a crucial hypothesis, which is not obeyed in our setting.

\medskip\noindent
{\bf Algebra of observables}

On the level of classical observables we always can define a linear combination, $C=\lambda_AA^{(Q)}+\lambda_BB^{(Q)}$, and the pointwise product, $D=A^{(Q)}\cdot B^{(Q)}$, of two quantum observables $A^{(Q)},B^{(Q)}$, where $C_\tau=\lambda_AA^{(Q)}_\tau+\lambda_BB^{(Q)}_\tau,D_\tau=A^{(Q)}_\tau B^{(Q)}_\tau$. However, in general neither $C$ nor $D$ are quantum observables. Consider the simplest case, $M=2$, and the two basis observables $A^{(1)}$ and $A^{(2)}$ with spectrum $\gamma^{(1)}_\alpha=\gamma^{(2)}_\alpha=\pm 1$. A linear combination $C=\cos \vartheta A^{(1)}+\sin\vartheta A^{(2)}$ has a spectrum $\gamma^{(C)}_\alpha=\pm \cos \vartheta\pm \sin\vartheta$. This observable has four different possible measurement values. It can therefore not be a quantum observable of the system with $M=2$, even though $\langle C\rangle$ can be computed in terms of $\rho_{1,2}=\langle A^{(1),(2)}\rangle$. We conclude that the ``rotated spin'', which corresponds to the operator $\hat A(\vartheta)=\cos \vartheta\hat A^{(1)}+\sin\vartheta\hat A^{(2)}$, has to be described by a quantum observable $A^{(Q)}_{(\vartheta)}$ that is again a two level observable with spectrum $\gamma_\alpha=\pm 1$, rather than by a linear combination of $A^{(1)}$ and $A^{(2)}$ of the type $C$. This necessity arises for each  value of the angle $\vartheta$ and we have discussed in detail in \cite{CW2} that this needs a classical ensemble with infinitely many classical states $\tau$. The reader should note that one can define two types of linear combinations. On the level of classical observables one can define combinations of the type $C$, while on the level of operators or the associated probabilistic quantum observables a natural definition is $\hat A(\vartheta)$ or $A^{(Q)}(\vartheta)$. In general, a projection on the subsystem does not map $C$ to $\hat A(\vartheta)$ or the probabilistic observable $A(\vartheta)$.

The classical product $D=A^{(1)}\cdot A^{(2)}$ has a spectrum $\gamma^{(D)}_\alpha=\pm 1$. The condition (i) for a quantum observable is obeyed by $D$. However, the probability $w^{(D)}_\alpha$ for finding $\gamma^{(D)}_\alpha=1$ needs knowledge of the joint probability to find $A^{(1)}=1,A^{(2)}=1$ or $A^{(1)}=-1,~A^{(2)}=-1$. This information cannot be extracted from $\rho_1$ and $\rho_2$, which only yield the probabilities for finding $A^{(1)}=\pm 1$ (namely $w^{(1)}_\pm =(1\pm \rho_1)/2)$ or for finding $A^{(2)}=\pm 1$ (namely $w^{(2)}_\pm=(1\pm\rho_2)/2)$. Nor is it contained in the expectation value $\rho_3$ of the third basis variable for $M=2$. We conclude that the classical observable $D$ does not obey the condition (ii) for a quantum observable. 

We conclude that different algebras can be formulated on the levels of classical observables and probabilistic observables. The map ${\cal Q}\to{\cal P}$ defines equivalence classes of classical quantum observables. The algebra of classical observables, as defined by $D=A\cdot B~,~C=A+B$, is an algebra defined within the space of all classical observables, but the operations of addition and multiplication do not remain within the restricted space of classical quantum observables ${\cal Q}$. They can therefore not be transported to the space of probabilistic quantum observables ${\cal P}$. On the other hand, we have seen in sect. \ref{Observablesand} that a new algebra can be defined, acting in the space of the probabilistic observables ${\cal P}$, i.e. involving the product $(AB)_s$. This new structure is closely related to the operator algebra in quantum mechanics. It is possible to transport the algebra to the space of classical quantum observable ${\cal Q}$ by selecting a fixed classical representative for each possible outcome of multiplications and additions. It is not clear, however, if this is useful and we will not need such a construction for our purposes. 

\bigskip
\noindent
{\bf Representation of quantum observables as classical observables}

After these general remarks we now present an explicit classical ensemble and classical quantum observables for a system with given $M$. We recall that the classical quantum observables $A^{(Q)}$ which are mapped to a given $\hat A$ are not unique. Also the specification of the classical states $\tau$ and the corresponding construction of classical observables is not supposed to be unique. At the end, all measurable information of the system can be expressed in terms of the expectation values of quantum operators, such that the details of the classical observables do not matter. Only the existence of the classical observables in a setting free of contradictions is therefore needed in order to demonstrate a realization of quantum mechanics as a classical statistical ensemble. We have already given such an example in sect. \ref{Simplequantum}. Here we discuss a classical realization closely related to it which is, however, somewhat more general since restricted sets of quantum observables are covered as well. 

It is sufficient to determine at least one classical quantum observable for every operator $\hat A$, i.e. for every hermitean $M\times M$ matrix, which is needed for the description of the system. The quantum observable $\lambda A^{(Q)}$, with $\lambda\in {\mathbbm R}$, is mapped to the operator $\lambda\hat A$. We will therefore restrict our discussion to operators with unit norm, say tr $\hat A^2=M$. Also the addition of a part proportional to the unit observable translates for operators to the addition of a corresponding piece proportional to the unit operator, $A+c\to \hat A+c$. We can therefore restrict the discussion to traceless operators, tr$\hat A=0$. We follow a simple construction principle. Consider first a single operator $\hat A$ with a spectrum of $m(\hat A)\leq M$ distinct eigenvalues $\lambda_{a(\hat A)}(\hat A)$. We associate to it $m(\hat A)$ discrete classical states, labeled by $a(\hat A)=1\dots m(\hat A)$. In these states the classical observable $A^{(Q)}$, which is mapped to $\hat A$, takes the values $A^{(Q)}_{a(\hat A)}=\lambda_{a(\hat A)}(\hat A)$. Add now a second operator $\hat B$ with $m(\hat B)$ distinct eigenvalues $\lambda_{a(\hat B)}(\hat B)$. If this operator is ``independent'' we construct the direct product space with states $\tau$ labeled by the double index $\tau=\big(a(\hat A),a(\hat B)\big)$, and $A^{(Q)}_\tau=\lambda_{a(\hat A)}(\hat A),B^{(Q)}_\tau=\lambda_{a(\hat B)}(\hat B)$. This is continued until all independent operators are included. As stated above, the resulting ensemble has infinitely many classical states $\tau$, if the number of independent operators is infinite. (A well defined sequence of subsequently included operators induces a well defined limit process for the construction of the ensemble \cite{CW2}.) Our construction yields explicitly a classical quantum observable for every independent operator. We recall that many further classical observables that do not obey the restrictions for quantum observables can be defined in the ensemble.

Details of the construction will depend on the notion of independent operators. As a simple criterion we may call two operators independent if tr$(\hat A-\hat B)^2\geq\epsilon$, and take the limiting process $\epsilon\to 0$. Other more restrictive definitions of ``independent'' may be possible. Different contradiction-free definitions of ``independent'' lead to different classical realizations, which all result in the same quantum properties of the system. The explicit construction above has demonstrated that such classical realizations exist. Of course, there are also classical ensembles with ``many more'' states than those used in our explicit construction. If one is only interested in the quantum observables the states $\tilde \tau$ of such a larger ensemble can be mapped to the states $\tau$ of the ensemble used in the construction by summing the probabilities of all states which have the same $A_\tau$ for all quantum observables. 

For our discussion of a classical realization of a quantum computer in sect. \ref{Classicalquantumcomputer} we have used this type of construction for the states of the classical ensemble. However, we have taken there only the operators $\tau_k$ or $L_k$, which are associated to the basis observables $A^{(k)}$, as independent operators. This was sufficient for the purpose of resenting initialization, gate operations and readout of a quantum computer. If we want to describe further quantum  observables, as arbitrarily rotated spins, we have to include further labels for the classical states. Since infinitely many ``rotation directions'' exist, one needs an infinity of classical states \cite{CW2}. This has been implemented in sect. \ref{Simplequantum}.

\medskip
\noindent
{\bf Classical sum and product for observables}

We next turn to the conditions under which linear combinations or classical products of two classical quantum observables can again be quantum observables. We will find that this requires that the associated quantum operators commute. For $C^{(Q)}=\lambda_AA^{(Q)}+\lambda_BB^{(Q)}$ the expectation value can always be expressed in terms of $\rho_k$. We can therefore compute $e^{(C)}_k=\lambda_Ae^{(A)}_k+\lambda_Be^{(B)}_k$ such that $\langle C\rangle$ obeys eq.\eqref{2}. This allows for the construction of an operator $\hat C$ \eqref{5} which obeys for all $\rho$
\begin{equation}\label{W4}
\langle C\rangle=\textup{tr}(\hat C\rho)=\lambda_A\langle A\rangle+\lambda_B\langle B\rangle=\textup{tr}
\Big[(\lambda_A\hat A+\lambda_B\hat B)\rho\Big].
\end{equation}
We therefore can identify $\hat C=\lambda_A\hat A+\lambda_B\hat B$. At this step, however, the spectrum of possible measurement values for $C$ does not necessarily coincide with the spectrum of eigenvalues of $\hat C$, which is a necessary condition for a quantum observable. On the classical level the spectrum of $C$ consists of all linear combinations
\begin{equation}\label{Z1}
\gamma^{(C)}_c=\gamma^{(C)}_{(a,b)}=\lambda_A\gamma^{(A)}_a+\lambda_B\gamma^{(B)}_b
\end{equation}
for all possible pairs $(a,b)=c$. It may be reduced by eliminating those $\gamma^{(C)}_c$ for which the probability $w_c$ vanishes for all quantum states. Even if the number $\tilde M^{(C)}$ of different values $\gamma^{(C)}_c$ obeys $\tilde M^{(C)}\leq M$ as realized, for example, if $\tilde M^{(A)}\tilde M^{(B)}\leq M$, there is no guarantee that all $\gamma^{(C)}_c$ coincide with the eigenvalues of $\hat C$.

If $C^{(Q)}$ is a quantum observable its spectrum must coincide with the spectrum of $\hat C$. Furthermore, the classical product $C^{(Q)}\cdot C^{(Q)}$ must also be a quantum observable and obey
\begin{eqnarray}\label{Z2}
\langle C\cdot C\rangle&=&\lambda^2_A\langle A\cdot A\rangle+\lambda^2_B\langle B\cdot B\rangle +
2\lambda_A\lambda_B\langle A\cdot B\rangle\nonumber\\
&=&\textup{tr}(\hat C^2\rho)\\
&=&\lambda^2_A\textup{tr}(\hat A^2\rho)+\lambda^2_B\textup{tr}(\hat B^2\rho)+
\lambda_A\lambda_B\textup{tr}\Big(\{\hat A,\hat B\}\rho\Big).\nonumber
\end{eqnarray}
We find as a necessary condition that the classical product $A\cdot B$ must be computable in terms of $\rho_k$
\begin{equation}\label{Z3}
\langle A\cdot B\rangle=\frac12\textup{tr}  \Big(\{\hat A,\hat B\}\rho\Big).
\end{equation}
This has to hold for arbitrary $\rho$. If any (nontrivial) linear combination of $A^{(Q)}$ and $B^{(Q)}$ is a quantum observable one concludes that $A^{(Q)}\cdot B^{(Q)}$ must be a quantum observable with associated operator $A\cdot B\to \frac12\{\hat A,\hat B\}$. Similar restrictions arise for higher powers of $C$. 

The condition \eqref{Z3} is nontrivial. On the classical level we can derive from $\{p_\tau\}$ the probabilities $w_{(a,b)}$ that $A$ has the value $\gamma^{(A)}_a$ and $B$ takes the value $\gamma^{(B)}_b$. For quantum observables $A^{(Q)},B^{(Q)}$ the probabilities $w^{(A)}_a=\sum_b w_{(a,b)},w^{(B)}_b=\sum_a w_{(a,b)}$ can be computed from $\{\rho_k\}$. In general, the information contained in $\{\rho_k\}$ will not be sufficient to determine $w_{(a,b)}$, however. It will therefore often not be possible to express $\langle A\cdot B\rangle=\sum_{a,b}\gamma^{(A)}_a\gamma^{(B)}_b w_{(a,b)}$ in terms of $\{\rho_k\}$. Then $D=A\cdot B$ cannot be a quantum observable. On the other hand, if a linear relation between $\langle D\rangle$ and $\rho_k$ exists, $\langle D\rangle=e^{(D)}_0+e^{(D)}_k\rho_k$, we can write $\langle D\rangle=\textup{tr} (\hat D\rho)$ and eq. \eqref{Z3} implies $\hat D=\frac12\{\hat A,\hat B\}$. 

If $\lambda_AA^{(Q)}+\lambda_BB^{(Q)}$ is a quantum observable for arbitrary $\lambda_A,\lambda_B$, the associated operators $\hat A$ and $\hat B$ must commute, $[\hat A,\hat B]=0$. In order to show this, we first consider the case where a given 
$\gamma^{(C)}_{\bar c}=\lambda_A\gamma^{(A)}_{\bar a}+\lambda_B\gamma^{(B)}_{\bar b}$ corresponds to a unique combination $(\bar a,\bar b)$. If $C^{(Q)}$ is a quantum observable, there must exist probability distributions $\{p_\tau\}$ which are an ``eigenstate'' for the ``eigenvalue'' $\gamma^{(C)}_{\bar c}$. This implies $w_{\bar c}=1~,~w_{c\neq \bar c}=0$ or $w_{(\bar a,\bar b)}=1~,~w_{(a,b)}=0$ if $a\neq \bar a$ or $b\neq \bar b$ and therefore $w_{\bar a}=1~,~w_{a\neq\bar a}=0~,~w_{\bar b}=1~,~w_{b\neq\bar b}=0$. We conclude that this state is also a simultaneous eigenstate of the observables $A^{(Q)}$ and $B^{(Q)}$, with respective eigenvalues $\gamma^{(A)}_{\bar a}$ and $\gamma^{(B)}_{\bar b}$. In particular, we may consider a pure state $\psi_{\bar c}$ which is an eigenstate of $\hat C$ with eigenvalue $\gamma^{(C)}_{\bar c}$. It must obey $\hat A\psi_{\bar c}=\gamma^{(A)}_{\bar a}\psi_{\bar c}~,~\psi^T_{\bar c}\hat A=\gamma^{(A)}_{\bar a}\psi^T_{\bar c}$~,~$\hat B\psi_{\bar c}=\gamma^{(B)}_{\bar b}\psi_{\bar c}$~,~$\bar\psi^T_c\hat B=\gamma^{(B)}_{\bar b}\bar\psi^T_c$. We choose a basis where $\hat C$ is diagonal and $\psi^T_{\bar c}=\hat\psi^T_1=(1,0,\dots 0)$. In this basis $\hat A$ and $\hat B$ must be block-diagonal, $\hat A_{\alpha 1}=\gamma^{(A)}_{\bar a}\delta_{\alpha 1}$, $\hat A_{1\beta}=\gamma^{(A)}_{\bar a}\delta_{1\beta}$ and similar for $\hat B$. We can repeat this for other eigenvalues of $\hat C$. If for every eigenvalue $\gamma^{(C)}_c$ of $\hat C$ the composition out of eigenvalues of $A^{(Q)}$ and $B^{(Q)}$ is unique, we can infer that $\hat A$ and $\hat B$ must commute. Indeed, in the basis where $\hat C$ is diagonal both $\hat A$ and $\hat B$ must be simultaneously diagonal, and therefore $[\hat A,\hat B]=0$. 

In presence of multiple possibilities of composing $\gamma^{(C)}_{\bar c}$ from linear combinations of $\gamma^{(A)}_a$ and $\gamma^{(B)}_b$ the discussion is more involved. This case appears, however, only for particular coefficients $\lambda_A,\lambda_B$. If $C$ is a quantum observable for arbitrary $\lambda_A$ and $\lambda_B$ such degenerate cases can be avoided such that $\hat A$ and $\hat B$ must commute. Indeed, consider the case of a ``degenerate decomposition'' for a particular pair $(\lambda_A,\lambda_B)$. This occurs if there are two solutions $\lambda_A\gamma^{(A)}_{a_1}+\lambda_B\gamma^{(B)}_{b_1}=\gamma^{(C)}_{c_1}~,~\lambda_A\gamma^{(A)}_{a_2}+\lambda_B\gamma^{(B)}_{b_2}=\gamma^{(C)}_{c_2}$, with $\gamma^{(C)}_{c_1}=\gamma^{(C)}_{c_2}$~,~$\gamma^{(A)}_{a_1}\neq\gamma^{(A)}_{a_2}$. Performing an infinitesimal shift $\lambda_A\to\lambda_A+\delta_A$, while keeping $\lambda_B$ fixed, results in a separation of $\gamma^{(C)}_{c_1}$ and $\gamma^{(C)}_{c_2}$ , $\gamma^{(C)}_{c_2}-\gamma^{(C)}_{c_1}=\delta_A(\gamma^{(A)}_{a_2}-\gamma^{(A)}_{a_1})\neq 0$. Then $\gamma^{(C)}_{c_1}$ has a unique composition from $\gamma^{(A)}_{a_1}$ and $\gamma^{(B)}_{b_1}$. (We have discussed here the case of two-fold degeneracy where other eigenvalues of $C$ are separated from $\gamma^{(C)}_{c_1}~,~\gamma^{(C)}_{c_2}$ by a finite distance. Higher degeneracies can be treated similarly.)

\medskip\noindent
{\bf Comeasurable quantum observables}

Two classical quantum observables $A^{(Q)}$ and $B^{(Q)}$ are called ``{\em comeasurable quantum observables}'' if arbitrary linear combinations $\lambda_AA^{(Q)}+\lambda_BB^{(Q)}$ are also quantum observables. The operators $\hat A,\hat B$ associated to a pair of comeasurable observables must commute, $[\hat A,\hat B]=0$. Furthermore, the classical product of two comeasurable quantum observables is a quantum observable. The associated operators and probabilistic observables are given by the chain of maps
\begin{equation}\label{Z4}
A^{(Q)}\cdot B^{(Q)}\to \frac12\{\hat A,\hat B\}\to (A B)_s. 
\end{equation}
For comeasurable quantum observables the classical correlation is computable from the quantum system and equals the quantum correlation. 

We next consider general conditions for the classical product being a quantum observable, $D^{(Q)}=A^{(Q)}\cdot B^{(Q)}$. If for all eigenvalues of $D^{(Q)}$ the decompositions $\gamma^{(D)}_{\bar c}=\gamma^{(A)}_{\bar a}\gamma^{(B)}_{\bar b}$ are unique, the operators $\hat A$ and $\hat B$ must again commute. In this situation the quantum state specifies the probabilities $w_a$ for the observable $A$ to have the value $\gamma^{(A)}_a$, the analogue for $w_b$, and in addition the joint probability $w_c=w_{(a,b)}$ that a measurement of $A$ yields $\gamma^{(A)}_a$ and a measurement $B$ yields $\gamma^{(B)}_b$. Since the associated operators $\hat A$ and $\hat B$ commute we may choose a basis where both are diagonal. The probability for the operator $\hat A\hat B$ to take the value $\gamma^{(D)}_c=\gamma^{(A)}_a\gamma^{(B)}_b$, with $(a,b)=c$, is given by the corresponding diagonal element of the density matrix in this basis, $(\rho')_{\alpha\alpha}$. This must equal $w_c$, and we conclude that for all quantum states Tr$(\hat A\hat B\rho)=\sum_cw_c\gamma^{(D)}_c=\langle D\rangle=$Tr$(\hat D\rho)$ and therefore  eq. \eqref{Z4} applies. 

\medskip
Inversely, we cannot infer that for every pair of quantum observables $A^{(Q)},B^{(Q)}$, for which the associated operators commute, $[\hat A,\hat B]=0$, the classical product $A^{(Q)}\cdot B^{(Q)}$ must be a quantum observable. There is simply no guarantee that the joint probabilities $w_{(a,b)}$ find an expression in terms  of $\{\rho_k\}$. Furthermore, if $A^{(Q)}$ and $B^{(Q)}$ are comeasurable quantum observables and $A\cdot B=D$ is therefore a quantum observable, and if $F^{(Q)}$ is a quantum observable with associated operator $\hat F=\frac12\{\hat A,\hat B\}$, this does not imply that $(A\cdot B)_\tau$ equals $F_\tau$. There is a whole equivalence class of distinct classical quantum observables which are mapped to the operator $\hat F$, and it would not be clear with which one $(A\cdot B)_\tau$ should be identified. We only know that $(A\cdot B)_\tau$ belongs to the same equivalence class as $F_\tau$. 

One may understand this issue in more detail in the realization of four-state quantum mechanics discussed in sect. \ref{Simplequantum}. In general, even for two different $e_k$ for which the operators commute, as $e^{(A)}_k$ and $e^{(B)}_k$ in eq. \eqref{M22} (with $C$ replaced by $A$ here), the classical correlation
\begin{equation}\label{XA0}
\langle A\cdot B\rangle = \langle A\rangle \langle B\rangle+\sum_\tau(\delta p_e)_\tau A_\tau B_\tau
\end{equation}
will depend on the environment $\big\{(\delta p_e)_\tau\big\}$. We may, however, restrict $\big\{(\delta p_e)_\tau\big\}$ by imposing beyond the conditions \eqref{M14} the relations
\begin{equation}\label{XA1}
\langle (A\cdot B)^p\rangle=\text{tr}\big\{\rho(\hat A\hat B)^p\big\}
\end{equation}
for $p=1,2,3$. This enlarges the set of the possible system observables and the equivalence class corresponding to the operator $\hat F=\frac12\{\hat A,\hat B\}=\hat A\hat B$, such that $A\cdot B$ is now a quantum observable belonging to this equivalence class. The explicit construction of sect. \ref{Simplequantum} shows, however, that the classical observable $F=A(e^{(F)}_k)$, with $\hat F=e^{(F)}_kL_k$, is distinct from the observable $A\cdot B$. 

It is important that $A\cdot B$ and $F$ are in the same equivalence class, but not identical  classical observables. In particular, if there is another quantum observable $G$ which is comeasurable with $F$, such that  $F\cdot G$ is a quantum observable, this does not imply that $A\cdot B\cdot G$ is a quantum observable. The classical products of $G$ with two different representatives of a given equivalence class may be different. This is important in order to avoid contradictions for an implementation of several pairs of comeasurable observables. Assume for $M=4$ the operator representations $A\to L_3~,~B\to L_2~,~F\to L_1~,~G\to L_4$. While $A\cdot B$ can be a quantum observable in the same equivalence class as $F$, and $F$ and $G$ may be realized also as comeasurable observables, the classical product $A\cdot B\cdot G$ is not necessarily a quantum observable. If it would be one, also $A\cdot G\cdot B$ and $G\cdot A\cdot B$ would be identical quantum observables since the classical product is commutative. On the other hand, the operators $\hat B$ and $\hat G$ or $\hat A$ and $\hat G$ do not commute. Such a situation can lead to contradictions as we will see when we next discuss the notion of bit chains. Those are avoided if $A\cdot B\cdot G$ is not a quantum observable. We also recall that it is only an option to realize pairs of quantum observables which are associated to commuting operators as comeasurable observables. This requires additional constraints on $\big\{(\delta p_e)_\tau\big\}$ which need not to be imposed. 

As an upshot of this discussion we conclude  that the lack of a map ${\cal O}\to {\cal Q}$ leaves a lot of freedom in the choice and properties of the quantum observables on the classical level. Generically, linear combinations and classical products of quantum observables are not quantum observables themselves. 

\medskip
\noindent
{\bf Bit chains}

Finally, we discuss the special setting of ``bit chains''. Bit chains are sets of bits for which the probability for the outcome of ordered measurement sequences as $(+,-,+)$ for (bit 1, bit 2, bit 3) can be predicted from the knowledge of the state of the subsystem. The simplest bit chain is a set of three comeasurable two level observables which we take among the set of basis observables $A^{(k)}$. For the example $M=4$ we may consider the observables $T_1,T_2,T_3$ associated to the three commuting diagonal operators $L_1,L_2,L_3$. A bit chain arises if the expectation values of two  (or several) one-bit-observables as well as their (multiple) products can be determined simultaneously in a quantum system. 

Suppose that the first bit corresponds to $T_1$, the second to $T_2$. Each bit can take the two values $+1$ or $-1$. We can consider measurements of the two bits and construct a composite observable which takes the value $+1$ if the signs of measurements of $T_1$ and $T_2$ are equal, and $-1$ if they are opposite. If the information contained in the state of the system is sufficient in order to predict the outcome for the combined observable, such that its expectation value depends linearly on $\rho_k$, we can conclude that the combined observable must be a quantum observable. The combined observable should be in the same equivalence class as  $T_3$, with corresponding quantum operators obeying $L_1L_2=L_3$. 

From the expectation values $\rho_m=\langle T_m\rangle~,~m=1,2,3$, we can determine the probabilities $w_{\gamma,\epsilon}$ that a measurement of bit one finds the value $\gamma\pm 1$ and the measurement of bit 2 yields $\epsilon=\pm 1$,namely
\begin{equation}\label{130A}
w_{\gamma\epsilon}=\frac14(1+\gamma\rho_1+\epsilon\rho_2+\gamma\epsilon \rho_3).
\end{equation}
The order of the measurement does not matter here. A bit chain is closed in the sense that if $T_3$ is considered as a bit and $T_2$ as a second bit, the composite two level observable is now in the same equivalence class as $T_1$, according to $L_3L_2=L_1$. We can compute the probabilities for arbitrary sequences of measurements of $T_1$ and $T_2$ if we assume that the measurement correlations for the composite observable constructed from $T_1$ and $T_2$ are the same as for the bit $T_3$. For four-state quantum mechanics there are many different bit chains, associated to products of two commuting operators as $L_1L_4=L_6$.

\medskip\noindent
{\bf Comeasurable bit chains}

For a comeasurable bit chain the classical product $T_1\cdot T_2$ is also a quantum observable, in the same equivalence class as $T_3$. From the expectation values $\langle T_1\rangle~,~\langle T_2\rangle~,~\langle T_3\rangle$ we can determine all probabilities $p_{++}$ etc. for the four possibilities of values $(+,+)~,~(+,-)~,~(-,-)$ and $(-,+)$ for bits one and two \cite{CW1}. In summary, for $T_1, T_2$ and $T_3$ to form a comeasurable bit chain we restrict the classical probability distribution $p_\tau$ such that the expectation values obey
\begin{eqnarray}\label{144A}
\langle T_1\cdot T_2\rangle&=&\langle T_3\rangle~,~\langle T_1\cdot T_3\rangle=\langle T_2\rangle,\nonumber\\
\langle T_2\cdot T_3\rangle&=&\langle T_1\rangle~,~\langle T_1\cdot T_2\cdot T_3\rangle=1.
\end{eqnarray}

This can be easily generalized: for a comeasurable bit chain of quantum observables with $\tilde P$ members $T_i~,~j=1\dots \tilde P$, all mutual classical products are members of the bit chain, such that $T_i\cdot T_j$ is in the equivalence class of  $c_{ijk}T_k|(i\neq j)$, with $c_{ijk}=c_{jik}=1$ for one particular combination $(i,j,k)$ and zero otherwise. All associated operators $\hat T_j$ mutually commute, and $\hat T_i\hat T_j=c_{ijk}\hat T_k~,~\hat T^2_j=1$. This extends in a straightforward way to classical products of an arbitrary number of members of the bit chain. One infers that all linear combinations $\lambda_iT_i+\lambda_j T_j$ are quantum observables, represented by the operators $\lambda_i\hat T_i+\lambda_j\hat T_j$.

For given $M$ the maximal number of members of a bit chain is $\tilde P=M-1$ and we call such chains ``complete bit chains''. This restriction follows simply from the maximal number of mutually commuting operators. The presence of a bound for $\tilde P$ poses certain restrictions on the classical realizations of comeasurable bit chains associated to different sets of mutually commuting operators. 

As an example, consider the case $M=8$. A possible complete three bit chain with seven members can be associated to the operators $C_1\to (\tau_3\otimes 1\otimes 1)~,~C_2\to (1\otimes\tau_3\otimes 1)~, ~C_3\to (1\otimes 1\otimes\tau_3)~,~\tilde C_1\to (1\otimes\tau_3\otimes\tau_3)~,~\tilde C_2\to(\tau_3\otimes 1\otimes\tau_3)~,~\tilde C_3\to(\tau_3\otimes\tau_3\otimes 1)~,~{\stackrel \approx C}\to (\tau_3\otimes\tau_3\otimes\tau_3)$. For this ``$C$-chain'' one has for the operators (we omit hats here) $C_j\tilde C_j={\stackrel \approx C}=C_1C_2C_3$ for all $j=1,2,3~,~\tilde C_1 \tilde C_2=\tilde C_3$. Alternative candidates for complete three bit chains are the ``$A$-chain'' where $\tau_3$ is replaced by $\tau_1$, or the ``$B$-chain'' which obtains from the $C$-chain by the replacement $\tau_3\to \tau_2$, Further candidates are the ``$F$-chain'' $(C_1,A_2,A_3,\tilde F_1,\tilde F_2,\tilde F_3,{\stackrel \approx F})$, ``$G$-chain'' $(A_1,C_2,A_3,\tilde G_1,\tilde G_2,\tilde G_3,{\stackrel \approx G})$ or ``$H$-chain'' $(A_1,A_2,C_3,\tilde H_1,\tilde H_2,\tilde H_3,{\stackrel \approx H})$, with analogous multiplication structures given by the order of the elements in the list, i.e. ${\stackrel \approx F}\to (\tau_3\otimes\tau_1\otimes\tau_1)$ , ${\stackrel\approx G}\to(\tau_1\otimes\tau_3\otimes\tau_1)$, ${\stackrel \approx H}\to(\tau_1\otimes\tau_1\otimes\tau_3)$. Finally, we may consider a possible candidate ``$Q$-chain'' $({\stackrel \approx F},{\stackrel\approx G},{\stackrel\approx H},\tilde Q_1,\tilde Q_2.\tilde Q_3,{\stackrel\approx Q})$, with ${\stackrel\approx Q}\to-(\tau_3\otimes\tau_3\otimes\tau_3)$. 

If all these sets of observables are simultaneously realized as comeasurable bit chains, and if $A\cdot B~\widehat{=}~C~,~C\cdot D~\widehat{=}~E$ would imply $A\cdot B\cdot D~\widehat{=}~E$, we would run into contradiction. Here we denote by $A~\widehat{=}~B$ that $A$ and $B$ are in the same equivalence class, i.e. that the associated operators obey $\hat A=\hat B$. From the $Q$-chain we conclude ${\stackrel\approx F}\cdot{\stackrel\approx G}\cdot{\stackrel\approx H}~\widehat{=}~{\stackrel\approx Q}$. In turn, from the $F,G,H$ chains we infer $C_1\cdot A_2\cdot A_3~\widehat{=}~{\stackrel\approx F}$, $A_1\cdot C_2\cdot A_3~\widehat{=}~{\stackrel\approx G}$ , $A_1\cdot A_2\cdot C_3~\widehat{=}~{\stackrel\approx H}$. If this would imply ${\stackrel\approx Q}~\widehat{=}~C_1\cdot A_2\cdot A_3\cdot A_1
\cdot C_2\cdot A_3\cdot A_1\cdot A_2\cdot C_3=C_1\cdot C_2\cdot C_3~\widehat{=}~{\stackrel \approx C}$, we would find that the operators associated to ${\stackrel\approx C}$ and ${\stackrel\approx Q}$ have opposite sign, showing the contradiction. 

This clearly demonstrates that not every set of two level observables for which the associated operators mutually commute can be a comeasurable bit chain simultaneously. In our case this only poses a consistency condition for the possibilities of classical products of quantum observables being quantum observables themselves. If we had a map $\hat A\to A^{(Q)}$, with $f(\hat A)\to f(A^{(Q)})$ and $f(A^{(Q)})$ based on the classical product, one could show that for every pair of commuting operators $\hat A,\hat B$ the map implies $\hat A\hat B\to A\cdot B$. The resulting contradiction is a proof of the Kochen-Specker-theorem \cite{KS} - actually the above chains of observables correspond precisely to the elegant proof of this theorem by N. Straumann \cite{Str}.

The observation that not all ``candidate chains'' $C,F,G,H,Q$ can be simultaneously comeasurable bit chains does not mean that the associated sets of seven commuting operators are inequivalent. There is no problem to associate to each such operator set a comeasurable bit chain. Only the bit chain associated to the set $\hat Q_1=(\tau_3\otimes\tau_1\otimes\tau_1)~,~\hat Q_2=(\tau_1\otimes\tau_3\otimes\tau_1)~,~\hat Q_3=(\tau_1\otimes\tau_1\otimes\tau_3)~,~\hat Q_2\hat Q_3~,~\hat Q_3\hat Q_1~,~\hat Q_1\hat Q_2~,~\hat Q_1\hat Q_2\hat Q_3=-(\tau_3\otimes\tau_3\otimes\tau_3)$ should be a new comeasurable bit chain with quantum observables $(Q_1,Q_2,Q_3~,~Q_2\cdot Q_3~,~Q_3\cdot Q_1~,~Q_1\cdot Q_2~,~Q_1\cdot Q_2\cdot Q_3)$ which are different from the candidate $Q$-chain $({\stackrel \approx F}, {\stackrel \approx G},{\stackrel \approx H},\tilde Q_1,\tilde Q_2,\tilde Q_3,{\stackrel \approx Q})$ discussed above. This does not lead to any contradiction, since the map from quantum observables to operators is not invertible. Both the observables $Q_1$ and ${\stackrel \approx F}$ are mapped to the same operator $\hat Q_1$, but the classical product may be a quantum observable for $Q_1\cdot Q_2$ and not for ${\stackrel \approx F}\cdot Q_2$. In our explicit construction of a classical representation of observables we should not exclude $\hat Q_1$ from the set of independent operators with the argument that it can be obtained as the product of two commuting observables.

\section{Entanglement}
\label{Entanglement}

The violation of Bell's inequalities and entanglement are often considered as key features which distinguish quantum mechanics form classical statistical physics. In this section we demonstrate that the classical statistical ensembles which correspond to quantum systems can realize entanglement. We show that the conditional correlations which are appropriate for measurements in the subsystem indeed violate Bell's inequalities. We argue that the key ingredient for the violation of Bell's inequalities is statistical incompleteness.

\medskip
\noindent
{\bf Classical statistical entanglement and violation of Bell's inequalities}

Entanglement is a key feature of quantum mechanics. Its classical realization is best discussed in the context of bit chains. Consider $M=4$ and a bit chain of observables $T_1,T_2,T_3$, with corresponding commuting diagonal operators $L_1,L_2,L_3=L_1L_2$. We associate a first bit to $T_1$ and a second bit to $T_2$. The product of measurements of the two bits is then determined by $\rho_3=\langle T_3\rangle$. Let us concentrate on a state with $\rho_3=\langle T_3\rangle=-1$ , $\rho_1=\langle T_1\rangle=0~,~\rho_2=\langle T_2\rangle=0$. Depending on the other $\rho_k$ this may be a pure or mixed state, with purity $P=1+\sum_{k\geq 4}\rho^2_k$. From $\rho_1=0$ we infer an equal probability to find for the first bit the values $+1$ and $-1$, and similar for the second bit from $\rho_2=0$. On the other hand, $\rho_3=-1$ implies a maximal anticorrelation  between bits one and two. The probabilities vanish for all classical states for which both bit one and bit two have the same value, corresponding to $w_{++}=w_{--}=0$~,~$w_{-+}=w_{-+}=1/2$. 

We may assume that a first apparatus measures bit one, and a second one bit two. Whenever the first apparatus shows a positive result, the second apparatus will necessarily indicate a negative result, and vice versa. By itself, this anticorrelation does not yet indicate an entangled pure state. For example, it may be realized by a mixed state with $\rho_k=0$ for $k\geq 4$ , $P=1$, corresponding to a diagonal density matrix $\rho=(1/2)diag(0,1,1,0)$. 

We may compute the quantum or conditional correlation \eqref{16A}, \eqref{14} for measurements of a rotated spin observable $A(\vartheta)$ with associated operator $\hat A(\vartheta)=\cos\vartheta L_1+\sin\vartheta L_8$, together with a second rotated spin observable $B(\varphi)$ with $\hat B(\varphi)=\cos\varphi L_2+\sin \varphi L_4$. One finds for arbitrary $\rho_k$ 
\begin{eqnarray}\label{37}
\langle A(\vartheta)B(\varphi)\rangle&=&C(\vartheta,\varphi)=\frac12\textup{tr}  \big(\{\hat A(\vartheta),\hat B(\varphi)\}\rho\big)\nonumber\\
&=&\cos\vartheta\cos\varphi\rho_3+\cos\vartheta\sin\varphi\rho_6\\
&&+\sin\vartheta\cos\varphi\rho_{10}+\sin\vartheta\sin\varphi\rho_{12},\nonumber
\end{eqnarray}
where we recall the representations of the generators $L_k$ \eqref{FXA}. For all states with $\rho_3=\rho_{12}=-1~,~\rho_6=\rho_{10}=0$, one obtains the familiar quantum result for two spins with relative rotation
\begin{equation}\label{39}
\langle A(\vartheta)B(\varphi)\rangle=-\cos(\vartheta-\varphi)=\bar C(\vartheta-\varphi).
\end{equation}
Bell's inequality for local deterministic theories reads for this situation
\begin{equation}\label{40}
|C(\vartheta_1,0)-C(\vartheta_2,0)|\leq 1+C(\vartheta_1,\vartheta_2).
\end{equation}
With eq. \eqref{39} this reduces to $|\bar C(\vartheta_1)-\bar C(\vartheta_2)|\leq 1+\bar C(\vartheta_1-\vartheta_2)$. It is violated for $\vartheta_1=\pi/2~,~\vartheta_2=\pi/4$. This clearly shows that we have introduced a conditional correlation \eqref{16A} within a classical statistical setting which violates Bell's inequalities. 

We observe that the contribution $\sim\rho_{12}$ to the quantum correlation matters. For our choice $\vartheta_1=\pi/2~,~\vartheta_2=\pi/4$ the inequality \eqref{40} reads 
$|C(\pi/2,0)-C(\pi/4,0)|\leq 1+C(\pi/2,\pi/4)$. For $\rho_6=\rho_8=0~,~\rho_3=-1$ and general $\rho_{12}$ one finds $C(\pi/2,0)=0~,~C(\pi/4,0)=-1/\sqrt{2}$ and $C(\pi/2,\pi/4)=\rho_{12}/\sqrt{2}$. For $\rho_{12}=0$ Bell's inequality is now obeyed.

We conclude that the presence of off-diagonal elements in the density matrix (in a direct product basis for the two entangled spins) plays an important role for the coexistence of different complete bit chains. For our example with $M=4$, a second bit chain besides $T_1,T_2,T_3$ is given by $T_8,T_4,T_{12}$, with associated commuting operators $L_8,L_4,L_{12}=L_8L_4$. Not only the spins in one direction are maximally anticorrelated for $\rho_3=-1$, but also the spins in an orthogonal direction (represented by $L_8,L_4)$ are maximally anticorrelated for $\rho_{12}=-1$. The quantum state of the subsystem allows for a specification of several correlations by independent elements as $\rho_3$ and $\rho_{12}$. This possibility is closely connected to the use of quantum correlations for the calculation of the outcome of two measurements. In a setting where only the classical correlations are available, a simultaneous implementation of the two two-bit chains $(T_1,T_2,T_3)$ and $(T_8,T_4,T_{12})$ would require more than three mutually commuting objects and can therefore not be implemented for a state with purity $P\leq 3$.

We finally display the classical formulation for two particular entangled pure states. They are given by 
\begin{equation}\label{41}
\rho_3=\epsilon\rho_{12}=-\epsilon\rho_{14}=-1~,~\epsilon=\pm 1.
\end{equation}
The sign $\epsilon=+1$ corresponds to the rotation invariant spin singlet state with density matrix
\begin{equation}\label{42}
\rho=\frac14\big(1-(\tau_1\otimes\tau_1)-(\tau_2\otimes\tau_2)-(\tau_3\otimes\tau_3)\big),
\end{equation}
and wave function
\begin{equation}\label{43}
\psi=\frac{1}{\sqrt{2}}(\hat\psi_2-\hat\psi_3).
\end{equation}
For $\epsilon=-1$ the relative sign between $\hat\psi_2$ and $\hat\psi_3$ is positive. 

\medskip
\noindent
{\bf Probabilistic realism, locality and incompleteness}

It is often stated that Bell's inequalities imply that quantum mechanics has to abandon either realism or locality. We argue here that our implementation of quantum mechanics is compatible both with ``probabilistic realism'' and locality. What is not realized, however, is a notion of ``statistical completeness'' in the sense that joint probabilities for arbitrary pairs of observables are available and used for the measurement correlation. Statistical completeness is often implicitly assumed in the stochastic proofs of Bell's inequalities \cite{CC}. In contrast, our definition of the measurement correlation, which is based on conditional probabilities that can be determined by the state of the system alone without additional information from the environment, leads to ``incomplete statistics'' \cite{3}. This is the basic reason why measurement correlations violate Bell's inequalities. 

``Probabilistic realism'' starts from the premise that the most general fundamental description of reality is of statistical nature \cite{3}. ``Elements of reality'', which allow for definite predictions, correspond then to values of observables as well as to correlations. Let us consider the EPR case of two entangled spins, carried by spatially separated particles which originate from the decay of a spinless particle and therefore have total spin zero. In this case the element of reality is the maximal anticorrelation for all spin directions, rather than values of individual spins. This element of reality is revealed by measurements of both spins and has existed already before the first measurement. In contrast, the value of one of the spins is maximally undetermined before the first measurement and not an element of reality.

Due to the correlation, the two spins have to be considered as one system. Even for an arbitrarily large separation, such that signals cannot be exchanged any longer, we cannot divide the system into two independent subsystems, consisting of one of the spins each. The correlation between the two spins is then nonlocal. Nonlocal correlations are common in classical statistical systems, however. As an example we may take ferromagnetism where the mean value of the spins is ordered in domains with macroscopic size. ``Simultaneous measurements'' of the mean value in spatially separated regions within the domain will find the same mean value, even if no signals can be exchanged between the measurements in the two regions. The only condition for a causal local theory is in this case that the nonlocal correlation has been prepared in the past by local causal processes. This is precisely what happens for the EPR-spins. The maximal anticorrelation of the two spins has been ``prepared'' during the decay of the spinless particle, and persists later due to angular momentum conservation. (For the antiferromagnet, one may invoke that the mean value of the spins and not only the correlation could now correspond to an element of physical reality. However, we also could consider the system somewhat above the critical temperature, where the mean value vanishes but correlations persist for macroscopic distances.) We conclude that quantum mechanics shares the same properties of probabilistic realism and locality as any other classical statistical system.

What is different from many usual classical statistical systems as encountered, for example, in classical thermodynamics, is the property of ``incomplete statistics'' characterizing the quantum systems. We advocate that the general definition of a measurement correlation for a pair of observables cannot be based on joint probabilities for the two observables. We believe that this incompleteness holds, in principle, for all statistical systems. Complete statistics, where the measurement correlation is expressed in terms of joint probabilities, obtains only as a special limiting case of incomplete statistics.

The basic reason for statistical incompleteness is the observation that statistical completeness is not, in general, compatible with the notion of measurements in isolated statistical subsystems. Any isolated subsystem is characterized by system observables. They are a subset of all the observables of the larger ``total system'', which can be regarded as the subsystem and its environment. Isolation means that the probability for finding a given value from the spectrum of a system observable should be determined by the state of the subsystem alone. It should not involve additional properties of the environment. Furthermore, it should be possible to determine the state of the subsystem by a certain number of expectation values of system observables. If a measurement of a pair of two system observables respects the isolation of the subsystem, the outcome should again be determined by the state of the subsystem, without invoking further information from the environment.

These simple, rather compelling characterizations of the notion of an isolated statistical subsystem are not compatible, in general, with statistical completeness. The basic reason is that the mapping from the space of observables for the total system (including the environment) to the system observables is not invertible. Different observables of the total system are mapped to the same system observable. Their difference resides only in different properties of the environment, whereas from the point of view of the subsystem they are all equivalent. The system observables define equivalence classes. As we have shown in detail, the joint probabilities differ for different representatives of a given equivalence class. They are therefore not a property of the equivalence class alone. In other words, the joint probabilities are not properties of the system alone, but also involve detailed information about the environment. They cannot be computed from the information which characterizes the state of the subsystem. For this reason the joint probabilities cannot be used for predicting the outcome of measurements in an isolated statistical subsystem. A generic measurement correlation, which determines the outcome of measurements of pairs of system observables in terms of the state of the subsystem alone, therefore leads to incomplete statistics.

\section{Sequence of measurements in a subsystem}
\label{Sequenceofmeasurements}

We have constructed a consistent implementation of quantum mechanics within a classical statistical ensemble, which leads to the quantum laws for expectation values of observables, correlations between two measurement which may violate Bell's inequalities, and the unitary time evolution. The classical statistical description therefore reproduces all the surprising effects of quantum mechanics. Many of them are related to our consistent choice of a measurement correlation for isolated subsystems. In this section we discuss further properties of sequences of measurements and their close connection to the non-commutativity of quantum operators. We restrict the discussion of this section to two-level-observables. An extension to measurements of observables with a spectrum of more than two distinct values may need an appropriate generalization. 

\medskip\noindent
{\bf Measurement chains}

Consider first the classical statistical ensemble which describes the two-state quantum system $(M=2)$, and a chain of three measurements of two-level observables. For a sequence of measurements of first $C$, then $B$, and finally $A$ we compute the probability $w^{(ABC)}_{+++}$ that all observables are measured to have the value $+1$, or the probability $w^{(ABC)}_{-+-}$ that $A$ is found to have the value $-1$, $B$ the value $+1$ and $C$ the value $-1$, and similarly for other combinations. After a measurement of the second observable $B=1$ the system is projected to a density matrix $\rho_{B+}=\frac12(1+\hat B)$, independently of the first measurement of $C$, such that the conditional probabilities \eqref{9} depend only on $\langle AB\rangle_{m}=$tr$(\hat A\hat B)/2$. In consequence, one finds for $\gamma,\delta,\epsilon=\pm 1$
\begin{equation}\label{C1}
w^{(ABC)}_{\gamma\delta\epsilon}=\frac14(1+\gamma\delta\langle AB\rangle_{m})
(1+\delta\epsilon\langle BC\rangle_{m})w^C_{+,s}.
\end{equation}

In particular, we may consider the basis observables $A=C=A^{(1)},B=A^{(2)}$ and compute
\begin{equation}\label{C2}
w^{(ABA)}_{-\delta+}=\frac14w^A_{+,s}.
\end{equation}
This clearly demonstrates that a series of measurements which are compatible with the preservation of the isolation of the subsystem cannot be reduced to a consecutive elimination of states of the classical ensemble. For a classical elimination process all states $\tau$ for which $A_\tau=-1$ are eliminated if the first measurement yields $A=+1$. The ``classical probability'' of finding $A=-1$ in later measurements must therefore be zero, in contrast to the result \eqref{C2}. Our prescription for conditional probabilities reproduces the quantum mechanical feature that the second measurement of $A^{(2)}$ leaves after the measurement a state with equal probabilities to find $A^{(1)}=\pm 1$, independently of the preceeding history. This underlines the particular role of measurements of not only eliminating the states which contradict the measured value of the measured observable, but also reshuffling the probabilities for other observables that are not measured. The particular form of the modification of probabilities for $A^{(1)}$ as a consequence of a measurement of $A^{(2)}$, which results in eq. \eqref{C2}, is due to the requirement that after a ``good measurement'' the ensemble should still obey the purity constraint. One should find a new state of the subsystem for which future measurements should not depend on the environment. As well known, this can be experimentally verified by a sequence of three Stern-Gerlach measurements. ``Good measurements'' of properties of the subsystem act similar to polarization filters for electromagnetic waves and are closely related to the particle-wave duality in quantum mechanics.

\medskip\noindent
{\bf Sequence of measurements for $M$-state quantum systems}

We can use the measurement probabilities \eqref{C1} in order to define the three point correlation for a sequence of two-level observables with spectrum $\{+1,-1\}$ \cite{CW2}
\begin{equation}\label{52A}
\langle ABC\rangle_m=\sum_{\gamma,\delta,\epsilon}\gamma\delta\epsilon w^{(ABC)}_{\gamma\delta\epsilon}
\end{equation}
and find
\begin{equation}\label{C3}
\langle ABC\rangle_m=\frac14\text{tr}\Big(\big\{\{\hat A,\hat B\},\hat C\big\}\rho\Big).
\end{equation}
For $M=2$ one may use in eq. \eqref{C3} the identity $\{\hat A,\hat B\}=2\langle AB\rangle_{m}$ such that $\langle ABC\rangle_m=\langle AB\rangle_{m}\langle C\rangle$. We will next show that eq. \eqref{C3} holds for minimally destructive measurements for general $M$. For this purpose we will have to discuss the properties of the projection \eqref{13} in more detail. 

We first note that $\rho_{A+}$ becomes formally ill defined for states where $\langle A\rangle=-1$, due to the normalization factor $\sim (1+\langle A\rangle)^{-1}$ which appears for $M>2$ in the terms multiplying the $d_{klm}$-symbols
\begin{eqnarray}\label{C3A}
\rho_{A+}&=&\frac1M(1+\hat A)
\left(1+\frac{d_{klm}a_k\rho_lL_m}{1+\langle A\rangle}\right)\nonumber\\
&&+\frac{d_{klm}a_ka_lL_m(1-M\rho)}{2M(1+\langle A\rangle)}.
\end{eqnarray}
This divergence cancels for physical quantities - for example by multiplication with $w^A_{+,s}$ in eq. \eqref{16A}. For the conditional probability one obtains
\begin{eqnarray}\label{C4}
(w^B_+)^A_+&=&\frac12\text{tr}\big((1+\hat B)\rho_{A+}\big)\\
&=&\frac12+\frac{\text{tr}\big((\hat B+\hat A\hat B+\hat B\hat A+\hat A\hat B\hat A)\rho\big)}{4(1+\langle A\rangle)}\nonumber
\end{eqnarray}
and $(w^B_-)^A_+=1-(w^B_+)^A_+$. The boundedness and positivity of these expressions for $\langle A\rangle\to-1$ may not seem obvious. It is more convenient to work here with an unnormalized density matrix $\tilde\rho_{A+}$ after the first measurement, and perform the proper normalization at the end. We define $\tilde\rho_{A+}$ in terms of the projector $P_{A+}$
\begin{eqnarray}\label{F1}
\tilde\rho_{A+}&=&P_{A+}\rho P_{A+}~,~P_{A+}=\frac{1+\hat A}{2}~,~P^2_{A+}=P_{A+},\nonumber\\
\text{tr}\tilde\rho_{A+}&=&(1+\langle A\rangle)/2.
\end{eqnarray}
Associating the change after a measurement to a projection onto an unnormalized state has the advantage that a subsequent measurement can be described by the same procedure. In particular, a second measurement of $A$ does not change the state further. The map $\rho\to\tilde\rho_{A+}$ is now linear. With $(w^B_+)^A_++(w^B_-)^A_+=1$ we can express
\begin{equation}\label{F2}
(w^B_+)^A_+=\frac{1+R}{2}~,~R=
\frac{w^B_+)^A_+-(w^B_-)^A_+}{(w^B_+)^B_++(w^B_-)^A_+}.
\end{equation}
The normalization drops out of the ratio $R$,
\begin{equation}\label{F3}
R=\frac{\text{tr}(\hat B\tilde\rho_{A+})}{\text{tr}\tilde \rho_{A+}},
\end{equation}
and we obtain $0\leq(w^A_+)^B_+\leq 1$ provided $-1\leq R\leq 1$. Since $\hat B$ has eigenvalues $\pm 1$ we can infer
\begin{equation}\label{F4}
-\text{tr}\tilde\rho_{A+}\leq\text{tr}(\hat B\tilde\rho_{A+})\leq\text{tr}
\tilde\rho_{A+}
\end{equation}
using the positivity of all diagonal elements, $(\tilde\rho_{A+})_{\alpha\alpha}\geq 0$, in a basis where $\hat B$ is diagonal. (We can infer $(\tilde\rho_{A+})_{\alpha\alpha}\geq 0$ in a basis where $\hat A$ is diagonal and $\hat P_{A+}$ is a diagonal matrix with entries $1$ and $0$. A change of basis by unitary transformations preserves this property.) For any limiting sequence of states $\rho$ for which $\langle A\rangle$ approaches $-1$ and therefore tr$\tilde\rho_{A+}$ approaches zero from above, we can indeed conclude $|R|\leq 1$. Therefore $(w^B_+)^A_+$ is well defined, positive and smaller or equal one for $\langle A\rangle$ arbitrary close to $-1$. For states with $\langle A\rangle=-1$ the limit may not be unique, but this is irrelevant since for further measurements in such states only the conditional probabilities $(w^B_\pm)^A_-$ are needed.

For a product of two conditional probabilities
\begin{eqnarray}\label{F5}
(w^A_+)^B_+(w^B_+)^C_+&=&\text{tr}
\left(\frac{1+\hat A}{2}\frac{1+\hat B}{2}\rho_{C+}
\frac{1+\hat B}{2}\right)\nonumber\\
&&\times\frac{2}{1+\langle B\rangle_{\rho_{C+}}}(w^B_+)^C_+\\
&=&\frac18\text{tr}(1+\hat A)(1+\hat B)\rho_{C+}(1+\hat B)\nonumber
\end{eqnarray}
the normalization factor for $(w^A_+)^B_+$ involves the expectation value of $B$ evaluated for a density matrix $\rho_{C+}$. This follows since $A$ has to be evaluated for a density matrix $\rho_{B+}$, and $\rho_{B+}$ obtains from eq. \eqref{13} by replacing $\hat A\to\hat B$ and $\rho\to\rho_{C+}$. The normalization factor is exactly canceled by $(w^B_+)^C_+$. We can use eq. \eqref{F5} in order to derive the identity
\begin{equation}\label{F6}
(w^A_+)^B_+(w^B_+)^C_\pm +
(w^A_-)^B_-(w^B_-)^C_\pm =
(w^D_+)^C_\pm ~,~
\hat D=\frac12\{\hat A, \hat B\},
\end{equation}
where the observable $D$ is represented by the anticommutator of $\hat A$ and $\hat B$. Similarly, one has
\begin{equation}\label{F7}
(w^A_+)^B_-(w^B_-)^C_\pm +
(w^A_-)^B_+(w^B_+)^C_\pm=(w^D_-)^C_\pm.
\end{equation}

We may therefore define a ``combined observable'' $D=A\circ B$ for the measurement of $A$ after $B$, where the results of the measurements of $A$ and $B$ are multiplied. In other words, $D$ is again a two level observable, which takes the value $+1$ if $A$ and $B$ have the same value, whereas $D=-1$ if $A$ and $B$ have opposite values. The conditional probabilities $(w^D_+)^C_+$ etc. can be obtained if the observable $D$ is represented by the operator $\hat D=\{\hat A,\hat B\}/2$. One concludes that the order of the measurements of $A$ and $B$ does not matter. Eqs. \eqref{F6}, \eqref{F7} can directly be used for the measurement correlation \eqref{52A} for sequences of measurements,
\begin{eqnarray}\label{F8}
\langle ABC\rangle_m&=&\langle DC\rangle_m=\frac12\text{tr}
\Big(\{\hat D,\hat C\}\rho\Big)\nonumber\\
&=&\frac14\text{tr}
\Big(\big\{\{\hat A,\hat B\},\hat C\big\}\rho\Big),
\end{eqnarray}
and we recover eq. \eqref{52A}. The generalization to more than three measurements is obvious - one starts from the left to group pairs into combined observables. We emphasize that the order matters. For $\langle ABC\rangle_m$ we cannot group $BC$ in a combined observable, since the latter would involve a sum over $+1$ and $-1$ values of $B$, combined with the appropriate $+1$ and $-1$ values of $C$, while the combinations appearing in $\langle ABC\rangle_m$ cannot be factorized in this way.

\medskip\noindent
{\bf Quantum commutator}

The commutator between two quantum operators can be related to the issue of ordering of a sequence of three measurements. Indeed, we find for the difference between two measurement correlations for three measurements evaluated in a different order
\begin{eqnarray}\label{F9}
\langle ABC\rangle_m-\langle ACB\rangle_m&=&\frac14\text{tr}
\Big(\big[\hat A,[\hat B,\hat C]\big]\rho\Big),\nonumber\\
\langle ABC\rangle_m-\langle CBA\rangle_m&=&\frac14\text{tr}
\Big(\big[\hat B,[\hat A,\hat C]\big]\rho\Big),\nonumber\\
\langle ABC\rangle_m-\langle BAC\rangle_m&=&0.
\end{eqnarray}
The non-commutativity of $\langle ABC\rangle_m$ is deeply rooted in the use of conditional probabilities, since $AB$ can be combined to a ``composite observable'', but not $BC$. The classical correlation, where $\langle ABC\rangle=\langle ACB\rangle$, is a special case for which the commutator vanishes for all pairs of observables. We observe the non-commuting structure of the three point correlation even if relations like
\begin{eqnarray}\label{F10}
&&(w^A_+)^B_+(w^B_+)^C_++(w^A_-)^B_-(w^B_-)^C_+\nonumber\\
&=&(w^B_+)^A_+(w^A_+)^C_++(w^B_-)^A_-(w^A_-)^C_+
\end{eqnarray}
hold, as in our case, such that $\langle AB\rangle_m=\langle BA\rangle_m$. The measurement correlation that we propose for our classical statistical setting reproduces a sequence of quantum measurements. For example, if we consider $M=2$ and two orthogonal basis observables $A^{(1)}$ and $A^{(2)}$, we obtain
\begin{equation}\label{F11}
\langle A^{(1)}A^{(2)}A^{(1)}\rangle_m=0~,~
\langle A^{(1)}A^{(1)}A^{(2)}\rangle_m=\langle A^{(2)}\rangle.
\end{equation}

The commutator can be rooted directly in the conditional probabilities
\begin{eqnarray}\label{F12}
(w^A_\pm)^B_+w^B_{+,s}&=&\frac14(1\pm\langle A\rangle+\langle B\rangle\pm 
\langle AB\rangle_m)\nonumber\\
&&\pm \frac{1}{16}\text{tr}\Big(\big[[\hat B,\hat A],\hat B\big]\rho\Big).
\end{eqnarray}
For example, the probability of finding $A=+1$ after a measurement of $B$,
\begin{eqnarray}\label{F13}
(w^A_+)^B_+w^B_{+,s}+(w^A_+)^B_-w^B_{-,s}
=\frac12(1+\langle A\rangle)\nonumber\\
+\frac18\text{tr}
\Big(\big[[\hat B,\hat A],\hat B\big]\rho\Big),
\end{eqnarray}
differs by the last commutator term from the probability of finding $A=1$ without a measurement of $B$, is given by $w^A_{+,s}=(1+\langle A\rangle)/2$. This directly reflects the modifications of the probabilities for finding $A=\pm 1$ as a consequence of a measurement of $B$. As we have discussed above, this is a necessary property if the measurement of $B$ is compatible with the isolation of the subsystem, i.e. if predictions for the subsystem can be done without invoking knowledge of the environment. 

The commutator also characterizes the difference between the probabilities of first measuring $B=1$ and then $A=1$ or first $A=1$ and then $B=1$,
\begin{eqnarray}\label{F14}
(w^A_+)^B_+w^B_{+,s}-(w^B_+)^A_+w^A_{+,s}
=\frac{1}{16}\text{tr}\Big(\big[[\hat B,\hat A],(\hat B+\hat A)\big]\rho\Big).\nonumber\\
\end{eqnarray}
Similarly, one finds
\begin{eqnarray}\label{F15}
(w^A_+)^B_-w^B_{-,s}-(w^B_-)^A_+w^A_{+,s}
=\frac{1}{16}\text{tr}\Big(\big[[\hat B,\hat A],(\hat B-\hat A)\big]\rho\Big).\nonumber\\
\end{eqnarray}
The r.h.s of eqs. \eqref{F14}, \eqref{F15} changes sign if we switch the sign of all values of $A$ and $B$, leading to a commutative measurement correlation $\langle AB\rangle_m=\langle BA\rangle_m$. For commuting observables the order of the measurements does not matter for any pair of possible outcomes. This applies, in particular, for two observables with support in regions with spacelike separation, which have to commute in order to avoid contradictions with causality. 

\medskip\noindent
{\bf Sequence of four measurements and uncertainty relation}

Consider sequences of four measurements, where two distinct two-level-observables $A$ and $B$ $(A^2=B^2=1)$ are measured in a different order. With
\begin{equation}\label{157A}
\langle ABCD\rangle_m=\frac18 \text{tr}\big(\rho\{\{\{A, \hat B\},\hat C\},\hat D\}\big)
\end{equation}
we may compute the difference of measurement correlations
\begin{eqnarray}\label{157B}
&&\langle ABAB\rangle_m+\langle BABA\rangle_m-\langle A^2B^2\rangle_m-\langle B^2A^2\rangle_m=\nonumber\\
&&\quad\frac12\text{tr}\Big\{\rho\big([\hat A,\hat B]\big)^2\Big\},
\end{eqnarray}
where $\hat A$ and $\hat B$ are the quantum operators associated to $A$ and $B$. This relates the expectation value of the squared commutator directly to the difference of measurement sequences in different orders. Also recall Heisenberg's uncertainty relation
\begin{equation}\label{157C}
\Delta A^2\Delta B^2\geq \frac14\Big|\text{tr}\Big\{\rho\big([\hat A,\hat B]\big)\Big\}\Big|^2,
\end{equation}
which holds for pure states of our classical statistical ensemble. For $[\hat A,\hat B]$ proportional to the unit matrix this relates $\Delta A^2\Delta B^2$ directly to the sequence of measurements \eqref{157B}.

\medskip\noindent
{\bf Products of quantum observables}

The isomorphism between equivalence classes of quantum observables and the quantum operators allow for the introduction of two different products of quantum observables. A product of two quantum observables should again be a quantum observable and is therefore represented by a hermitean operator. We have already discussed before the symmetric product $(AB)_s=(BA)_s$ which is represented by the anticommutator
\begin{equation}\label{F16}
(AB)_s=(BA)_s\leftrightarrow\frac12\{\hat A,\hat B\}.
\end{equation}
The measurement correlation is described by the expectation value of this symmetric product,
\begin{equation}\label{F17}
\langle AB\rangle_m=\langle(AB)_s\rangle.
\end{equation}
Further, we can associate an antisymmetric product $(AB)_a$ to the commutator multiplied by $-i/2$,
\begin{equation}\label{F18}
(AB)_a=-(BA)_a\leftrightarrow-\frac i2[\hat A,\hat B].
\end{equation}
The antisymmetric product characterizes the difference between three point correlations in different order
\begin{equation}\label{F19}
\langle ABC\rangle_m-\langle ACB\rangle_m=-
\langle\big(A(BC)_a\big)_a\rangle.
\end{equation}

In this context we emphasize that the products $(AB)_s$ and $(AB)_a$ are the quantum observables which are associated to the operators $\frac12\{\hat A,\hat B\}$ and $-\frac i2[\hat A,\hat B]$. For example, if $\{\hat A,\hat B\}=0$, as for $M=2$ and $\hat A,\hat B$ representing two orthogonal spins, the product $(AB)_s$ is the zero-observable with all eigenvalues zero. This differs from the combined observable $D=A\circ B$ discussed before. The latter is a two level observable, with $D^2=1$, in contrast to $\big((AB)_s\big)^2=0$. The observable $A\circ B$ is mapped to the operator $\hat D=\frac12\{\hat A,\hat B\}$ - the same as for $(AB)_s$. It is, however, a system observable which is not a quantum observable. For example, $D^2=1$ is not represented by $\hat D^2$. We recall that different system observables can be mapped to the same operator. The inverse mapping from operators to probabilistic observables is only defined uniquely if the observable is specified to be a quantum observable. Examples for complete observable systems are classical realization off all possible observables for $M$-state quantum mechanics or of a maximal subset of commuting observables. 

If a classical ensemble can describe a subsystem which admits a complete set of classical quantum observables, all measurement correlations can be expressed as expectation values of appropriate observables. We note that the measurement correlations for the subsystem can be computed from the state of the subsystem even if the system of observables is incomplete, for example of $(AB)_s$ has no realization as a classical observable. It is interesting to observe, however, that for complete observable systems the measurement correlations have even formally the same states concerning ``reality'' as all other quantum observables.

\medskip\noindent
{\bf Complex structure}

All our discussion so far concerns only real physical quantities: the spectrum of observables comprises only real values, the probabilities for finding a value from this spectrum, the conditional probabilities for sequences, as well as expectation values and correlation functions are real. The state can be characterized by real numbers $\rho_k$, and the time evolution $\partial_t\rho_k=F_k$ involves real quantities $F_k$. Also the prodcuts $(AB)_s$ and $(AB)_a$ are ``real observables'' with a real spectrum. Nevertheless, we have found a convenient expression of all quantities in terms of complex $M\times M$ matrices.

The apparent complex structure in quantum mechanics can be related to the existence of two different product structures $(AB)_s$ and $(AB)_a$ for quantum observables. Indeed, we may define a complex product as
\begin{equation}\label{F20}
(AB)_c=(AB)_s+i(AB)_a.
\end{equation}
On the level of operators it is represented by the complex operator prodcut
\begin{equation}\label{F21}
(AB)_c\leftrightarrow\hat A\hat B,
\end{equation}
consistent with $\frac12\{\hat A,\hat B\}+i
\left(-\frac i2[\hat A,\hat B]\right)=\hat A\hat B$. This allows us to associate to $\hat A\hat B$ the two products of quantum observables $(AB)_s$ and $(AB)_a$. The isomorphism between the algebra of quantum operators and probabilistic quantum observables can be extended to operators which are no longer necessarily hermitean anymore.

\medskip\noindent
{\bf Complete observable systems}

We call a system of quantum observables ``complete'' if for arbitrary pairs $(A,B)$ of quantum observables their products $(AB)_s$ and $(AB)_a$ are also quantum observables of the system. Since $(AB)_s$ and $(AB)_a$ can always be defined as probabilistic quantum observables with the appropriate spectrum $(\gamma_a)$ and associated probabilities $w_a$, an incomplete system of observables may be completed by adding the quantum observables associated to the products $(AB)_{s,a}$. This is always possible on the level of probabilistic observables, but not necessarily on the classical level. Examples for complete observable systems are classical realizations of all possible observables for $M$-state quantum mechanics or of a maximal subset of commuting observables.

If a classical ensemble can describe a subsystem which admits a complete set of classical quantum observables, all measurement correlations can be expressed as expectation values of appropriate observables. We note that the measurement correlations for the subsystem can be computed from the state of the subsystem even if the subsystem of observables in incomplete, for example if $(AB)_s$ has no realization as a classical observable. It is interesting to observe, however, that for complete observable systems the measurement correlations have even formally the same status concerning ``reality'' as well as other quantum observables.

\section{Conclusions}
\label{Conclusions}

We have obtained all laws of quantum mechanics from classical statistics, including the concept of probability amplitudes $\psi$ and the associated superposition of states with interference and entanglement, as well as the unitary time evolution. Our classical statistical description is genuinely probabilistic and not a local deterministic model. It allows to predict probabilities for the outcome of a chain of measurements, but not a deterministic result of a given measurement in terms of some ``hidden variables''. Bell's inequalities can indeed be violated for our formulation of a correlation function which is based on conditional probabilities for a sequence of measurements. Since the mapping from classical observables to quantum operators is not invertible, no contradiction to the Kochen-Specker theorem arises.

Our setting can be extended to include observables like location and momentum by considering many two-level observables on a space-lattice and taking the limit of vanishing lattice spacing \cite{CWQP}. Both quantum particles and classical particles can be described by appropriate classical statistical ensembles. Even a continuous interpolation becomes possible. As a function of some continuous parameter $\gamma$ in the interval $[0,\frac{\pi}{2}]$, the classical ensembles describe for $\gamma=0$ a quantum particle - ``passing simultaneously'' through the two slits in a double slit experiment and producing an interference pattern - or for $\gamma=\pi/2$ a classical particle that passes only through one of the slits. A continuous interpolation becomes possible for intermediate values of $\gamma$ \cite{CWQP}. 

In our statistical mechanics setting quantum mechanics describes isolated subsystems of a larger ensemble that also includes the environment. Isolation does not mean that the subsystem can be described by classical probabilities for the states of the subsystem and sharp values of the observables in these states. It rather relates to a separated time evolution of the subsystem and to observables which can be described by quantities only associated to the subsystem, without explicit reference to the environment. The question why Nature shows a strong preference for subsystems that are described by quantum mechanics can now be addressed within the general framework of classical statistics which allows, in principle, also subsystems without the characteristic features of quantum physics. We conjecture that the answer is related to particular stability properties for the time evolution of subsystems with a quantum character \cite{CW2}.

The deep question if small deviations from quantum mechanics are possible, and of what nature they could be, finds in a classical statistical setting an appropriate framework to be addressed. The overall probabilistic description of the whole world (or the entire reality) involves an infinity of states or degrees of freedom both in the quantum and the classical statistical description. It is well conceivable that they can be mapped onto each other such that in this very general sense no deviations from quantum mechanics occur. The perhaps more interesting issue concerns the possible descriptions of isolated subsystems, as isolated atoms or isolated $M$-state quantum systems. For such systems it seems possible that quantum mechanics is only a very good approximation, but small deviations could occur. The following discussion shows that such deviations from simple $M$-state quantum systems or from the ideal quantum particle do actually occur in practice. They are often related to phenomena as decoherence or imperfect measurements, for which the quantum character of the description can only be restored if the environment is included in the description.

Consider a subsystem that is well approximated by $M$-state quantum mechanics, in the sense that only the expectation values $\rho_k$ are available for a description of the state of the system and that the quantum mechanical laws hold to a good approximation. There are several types of possible small deviations for an exact quantum behavior. (i) The time evolution may not be unitary. In particular, the purity may decrease (decoherence) or increase (syncoherence). (ii) The time evolution may not be linear. This happens if the Hamiltonian operator $H$, or more generally the matrix $T_{kl}$ and $D$ in eq. \eqref{16}, depend themselves on $\rho_k$. (iii) If the $\rho_k$ violate during the time evolution the purity constraint \eqref{54A}-\eqref{56A}, we expect phenomena as the violation of the quantum mechanical uncertainty relation. (iv) Deviations from quantum mechanics may occur if only a limited part of the infinitely many classical states necessary for a complete quantum description is accessible for a subsystem \cite{CW2}. (v) Our framework yields also a formalism for the description of imperfect measurements. In this case the measurement correlation \eqref{14}, which is based on minimally destructive measurements, has to be replaced by a different correlation, adapted to the ``imperfection'' of the measurement. 

We do not intend to enter here the debate if quantum mechanics or classical statistics are more fundamental -it is well known that classical statistics can be obtained as a limiting case of quantum mechanics. In our view classical statistics and quantum mechanics are two sides of the same medal. This may have far reaching consequences, as the possibility that the late time asymptotic state of a classical ensemble may be given by the equilibrium ensemble of quantum statistics, or that the classical statistical realization of certain steps in quantum computations can find a practical implementation. We find it remarkable that the conceptual foundations of quantum mechanics need not to go beyond the concepts of classical statistics.


\bigskip
\begin{thebibliography}{100}
\bibitem{Ze}D. Bouwmeester, J. W. Pan, K. Mattle, M. Eibl, H. Weinfurter, A. Zeilinger, Nature {\bf 390} (1997) 575
\bibitem{Zo}R. Feynman, Int. J. Theor. Phys. {\bf 21} (1982) 467;\\
D. Deutsch, Proc. R. Soc. London {\bf A400} (1985) 97;\\
J. I. Cirac, P. Zoller, Phys. Rev. Lett. {\bf 74} (1995) 4091
\bibitem{CW1}C. Wetterich, arXiv: 0809.2671
\bibitem{CW2}C. Wetterich, arXiv: 0810.0985
\bibitem{KS}S. Kochen, E. P. Specker, Journal of Mathematics and Mechanics {\bf 17} (1967), 59;\\
N. D. Mermin, Phys. Rev. Lett. {\bf 65} (1990) 3373;\\
A. Peres, J. Phys. A: Math. Gen. {\bf 24} (1991) L175
\bibitem{Str}N. Straumann, arXiv: 0801.4931 [quant-ph]
\bibitem{POB}A. S. Holevo, ``Probabilistic and Statistical Aspects of Quantum Theory'' (Amsterdam, North Holland) 1982; WS. T.  Ali, E. Prugovecki, J. Math. Phys. {\bf 18} (1977) 219; \\
M. Singer, W. Stulpe,  J. Math. Phys. {\bf 33} (1992) 131
\bibitem{BB}E. Beltrametti, S. Bugajski,  J. Phys. A: Math. Gen. {\bf 28} (1995) 3329; Int.~J.~Theor.~Phys. {\bf 34} (1995) 1221; \\
S. Bugajski, Int.~J.~Theor.~Phys. {\bf 35} (1996) 2229
\bibitem{Bell}J. S. Bell, Physica 1 (1964) 195
\bibitem{Z}D. Greenberger, M. Horne, A.~Zeilinger, in ``Bell's Theorem, Quantum Theory, and Conceptions of the Universe'', p69, ed. M.~Kafatos, Kluver, Dortrecht (1989), arXiv:0712.0921
\bibitem{CC}J. Clauser, M. Horne, A. Shimony, R. Holt, Phys. Rev. Lett. {\bf 23} (1969) 880;\\
J. Bell, ``Foundations of Quantum Mechanics'', ed. B. d'Espagnat (New York: Academic, (1971) p. 171;\\
J. Clauser, M. Horne, Phys. Rev. {\bf D10} (1974) 526;\\
J. Clauser, A. Shimony, Rep. Prog. Phys. {\bf 41} (1978) 1881
\bibitem{CWQP}C. Wetterich, arXiv: 0904.3048[quant-ph]
\bibitem{3}C. Wetterich, in ``Decoherence and Entropy in Complex Systems'', ed. T. Elze, p. 180, Springer Verlag 2004, arXiv: quant-ph/0212031
\bibitem{GenStat}C. Wetterich, Nucl. Phys. {\bf B314} (1989) 40; Nucl. Phys. 
{\bf B397} (1993) 299
\bibitem{4}C. Wetterich, Phys. Lett. {\bf B399}(1997) 123
\bibitem{DC}H. D. Zeh, Found. Phys. {\bf 1} (1970) 69;\\
E. Joos, H. D. Zeh, Z. Phys. {\bf B59} (1985) 273;\\
E. Joos, H. D. Zeh, C. Kiefer, D. Giulini, J. Kupsch, I.-O. Stamatescu,
``Decoherence and the appearance of the classical world'', Springer 2003; \\
W. Zurek, Rev. Mod. Phys. {\bf 75} (2003) 715
\end{thebibliography}
\end{document}